\documentclass[12pt]{article}

  \usepackage{graphicx}
  \usepackage{epsfig}
  \usepackage{amssymb}
  \usepackage{subfigure}
    \usepackage{feynmf}
\evensidemargin - 1.truecm
\begin{document}
\begin{fmffile}{vert}

\newcommand{\be}{\begin{equation}}
\newcommand{\ee}{\end{equation}}
\newcommand{\nn}{\nonumber}
\newcommand{\bea}{\begin{eqnarray}}
\newcommand{\eea}{\end{eqnarray}}
\newcommand{\bfig}{\begin{figure}}
\newcommand{\efig}{\end{figure}}
\newcommand{\bc}{\begin{center}}
\newcommand{\ec}{\end{center}}

\begin{titlepage}
\nopagebreak
{\flushright{
        \begin{minipage}{5cm}
        Freiburg-THEP 03/02\\
        {\tt hep-ph/0301170}\\
        \end{minipage}        }

}
\renewcommand{\thefootnote}{\fnsymbol{footnote}}
\vspace*{-1.5cm}
\vskip 3.5cm
\begin{center}
\boldmath
{\Large\bf Vertex diagrams for the QED\\[3mm]
form factors at the 2-loop level}\unboldmath
\vskip 1.cm
{\large  R.~Bonciani\footnote{Email:
Roberto.Bonciani@physik.uni-freiburg.de}
\footnote{This work was supported by the European Union under
contract HPRN-CT-2000-00149},}
\vskip .2cm
{\it Facult\"at f\"ur Mathematik und Physik, Albert-Ludwigs-Universit\"at
Freiburg, \\ D-79104 Freiburg, Germany} 
\vskip .2cm
{\large P.~Mastrolia\footnote{Email: Pierpaolo.Mastrolia@bo.infn.it}},
\vskip .2cm
{\it Dipartimento di Fisica dell'Universit\`a di Bologna, 
I-40126 Bologna, Italy} 
\vskip .1cm
{\it and Institut f\"ur Theoretische Teilchenphysik,
Universit\"at Karlsruhe, \\ D-76128 Karlsruhe, Germany}
\vskip .2cm
{\large E.~Remiddi\footnote{Ettore.Remiddi@bo.infn.it}}
\vskip .2cm
{\it Dipartimento di Fisica dell'Universit\`a di Bologna
and INFN Sezione di Bologna, I-40126 Bologna, Italy}
\end{center}
\vskip 1.2cm

\begin{abstract}
We carry out a systematic investigation of all the 2-loop integrals 
occurring in the electron vertex in QED in the continuous $D$-dimensional 
regularization scheme, for on-shell electrons, momentum transfer $t=-Q^2$ 
and finite squared electron mass $m_e^2=a$. We identify all the 
Master Integrals (MI's) of the problem and write the differential equations 
in $Q^2$ which they satisfy. The equations are expanded in powers of 
$\epsilon = (4-D)/2$ and solved by the Euler's method of the variation 
of the constants. As a result, we obtain the coefficients of the 
Laurent expansion in $\epsilon$ of the MI's up to zeroth order 
expressed in close analytic form in terms of Harmonic Polylogarithms. 

\end{abstract}
\vfill
\end{titlepage}

\section{Introduction \label{Intro}}

The QED electron form factors at two loops were considered 
in~\cite{2loop1}, for massive on-shell electrons and arbitrary 
momentum transfer within the Pauli-Villars regularization 
scheme and giving a fictitious small mass $\lambda$ to the photon 
for the parametrization of infrared divergences. The main results 
of~\cite{2loop1} are the analytic calculation of the imaginary parts 
of the form factors for arbitrary momentum transfer in terms of 
Nielsen's polylogarithms~\cite{Nielsen,Kolbig}, and 
of the charge slope of the electron at two loops (besides the check 
of the magnetic anomaly). 
The analytic evaluation of the real parts, expected to involve a class 
of functions wider than Nielsen's polylogarithms, was not attempted 
in~\cite{2loop1}. 

To our knowledge, the full analytic calculation of the real parts of 
the 2-loop QED form factors, for arbitrary momentum transfer and 
finite electron mass,  has not yet been carried out, despite 
a great number of papers dealing with a variety of kinematical 
configurations (neglecting typically the electron mass at large 
momentum transfer). 

In this paper we work out a systematic investigation of all the 
2-loop integrals occurring in the electron vertex in QED in the 
continuous $D$-dimensional regularization 
scheme~\cite{DimReg} (using the same $D$ for ultraviolet 
and infrared regularization) for on shell 
electrons of finite squared mass $m_e^2=a$ and arbitrary momentum 
transfer $t=-Q^2$. We identify all the Master Integrals (MI's) 
occurring in all the graphs and evaluate them analytically, in terms 
of Harmonic Polylogarithms~\cite{Polylog,Polylog3}, for arbitrary 
value of $Q^2$. We present the results for spacelike momentum transfer, 
i.e. $t<0$ or $Q^2>0$; the case of timelike $t$ can be obtained by 
standard analytic continuation. 

The extraction of the form factors and their expression in terms of 
the MI's will be carried out in a subsequent paper. 

The diagrams involved are those shown in Fig. (\ref{fig1}).

\bfig
\bc
\subfigure[]{
\begin{fmfgraph*}(35,35)
\fmfleft{i1,i2}
\fmfright{o}
\fmf{fermion}{i1,v1}
\fmf{fermion}{i2,v2}
\fmf{photon}{v5,o}
\fmflabel{$p_{2}$}{i1}
\fmflabel{$p_{1}$}{i2}
\fmflabel{$Q$}{o}
\fmf{fermion,tension=.3,label=$p_{1}-k_{1}$,label.side=left}{v2,v3}
\fmf{fermion,tension=.3,label=$p_{1}-k_{1}-k_{2}$,
label.side=left}{v3,v5}
\fmf{fermion,tension=.3,label=$p_{2}+k_{1}$,label.side=right}{v1,v4}
\fmf{fermion,tension=.3,label=$p_{2}+k_{1}+k_{2}$,
label.side=right}{v4,v5}
\fmf{photon,tension=0,label=$k_{1}$,label.side=right}{v2,v1}
\fmf{photon,tension=0,label=$k_{2}$,label.side=left}{v4,v3}
\end{fmfgraph*} }
%
%
%
\hspace{8mm}
\subfigure[]{
\begin{fmfgraph*}(35,35)
\fmfleft{i1,i2}
\fmfright{o}
\fmfforce{0.3w,0.6h}{v10}
\fmfforce{0.3w,0.4h}{v11}
\fmf{fermion}{i1,v1}
\fmf{fermion}{i2,v2}
\fmf{photon}{v5,o}
\fmflabel{$p_{2}$}{i1}
\fmflabel{$p_{1}$}{i2}
\fmflabel{$Q$}{o}
\fmflabel{$k_{1}$}{v10}
\fmflabel{$k_{2}$}{v11}
\fmf{fermion,tension=.3,label=$p_{1}-k_{1}$,label.side=left}{v2,v3}
\fmf{fermion,tension=.3,label=$p_{1} \! - \! k_{1} \! - \! k_{2}
$,label.side=left}{v3,v5}
\fmf{fermion,tension=.3,label=$p_{2}+k_{2}$,label.side=right}{v1,v4}
\fmf{fermion,tension=.3,label=$p_{2} \! + \! k_{1} \! + \! k_{2}
$,label.side=right,l.d=.02w}{v4,v5}
\fmf{photon,tension=0}{v2,v4}
\fmf{photon,tension=0}{v1,v3}
\end{fmfgraph*} }
%
%
\hspace{12mm}
\subfigure[]{
\begin{fmfgraph*}(35,35)
\fmfleft{i1,i2}
\fmfright{o}
\fmfforce{0.2w,0.93h}{v2}
\fmfforce{0.2w,0.07h}{v1}
\fmfforce{0.2w,0.5h}{v3}
\fmfforce{0.8w,0.5h}{v5}
\fmfforce{0.2w,0.3h}{v10}
\fmf{fermion}{i1,v1}
\fmf{fermion}{i2,v2}
\fmf{photon}{v5,o}
\fmflabel{$p_{2}$}{i1}
\fmflabel{$p_{1}$}{i2}
\fmflabel{$p_{2} \! + \! k_{2}$}{v10}
\fmfv{l=$p_{2} \! \! + \! \! k_{1} \! \! + \! \! k_{2}$,l.a=10,
l.d=.05w}{v3}
\fmflabel{$Q$}{o}
\fmf{fermion,tension=0,label=$p_{1}-k_{1}$,label.side=left}{v2,v5}
\fmf{fermion,tension=0}{v3,v4}
\fmf{photon,tension=.4,label=$k_{2}$,label.side=right}{v1,v4}
\fmf{fermion,tension=.4,label=$p_{2}+k_{1}$,label.side=right}{v4,v5}
\fmf{fermion,tension=0}{v1,v3}
\fmf{photon,tension=0,label=$k_{1}$,label.side=right}{v2,v3}
\end{fmfgraph*} } \\
%
%
\subfigure[]{
\begin{fmfgraph*}(35,35)
\fmfleft{i1,i2}
\fmfright{o}
\fmfforce{0.2w,0.93h}{v2}
\fmfforce{0.2w,0.07h}{v1}
\fmfforce{0.8w,0.5h}{v5}
\fmfforce{0.2w,0.4h}{v9}
\fmfforce{0.5w,0.45h}{v10}
\fmfforce{0.2w,0.5h}{v11}
\fmf{fermion}{i1,v1}
\fmf{fermion}{i2,v2}
\fmf{photon}{v5,o}
\fmflabel{$p_{2}$}{i1}
\fmflabel{$p_{1}$}{i2}
\fmflabel{$Q$}{o}
\fmfv{l=$k_{2}$,l.a=180,l.d=0.05w}{v10}
\fmfv{l=$k_{1}$,l.a=180,l.d=.06w}{v11}
\fmf{fermion,label=$p_{1} \! - \! k_{1}$,label.side=left}{v2,v3}
\fmf{photon,tension=.25,right}{v3,v4}
\fmf{fermion,tension=.25,label=$p_{1} \! - \! k_{1} \! - \! k_{2}
$,label.side=left}{v3,v4}
\fmf{fermion,label=$p_{1} \! - \! k_{1}$,label.side=left}{v4,v5}
\fmf{fermion,label=$p_{2} \! + \! k_{1}$,label.side=right}{v1,v5}
\fmf{photon}{v1,v2}
\end{fmfgraph*} }
%
%
\hspace{20mm}
\subfigure[]{
\begin{fmfgraph*}(35,35)
\fmfleft{i1,i2}
\fmfright{o}
\fmfforce{0.2w,0.93h}{v2}
\fmfforce{0.2w,0.07h}{v1}
\fmfforce{0.2w,0.3h}{v3}
\fmfforce{0.2w,0.7h}{v4}
\fmfforce{0.8w,0.5h}{v5}
\fmf{fermion}{i1,v1}
\fmf{fermion}{i2,v2}
\fmf{photon}{v5,o}
\fmflabel{$p_{2}$}{i1}
\fmflabel{$p_{1}$}{i2}
\fmflabel{$Q$}{o}
\fmf{fermion,label=$p_{1} \! - \! k_{1}$,label.side=left}{v2,v5}
\fmf{photon,label=$k_{1}$,label.side=left}{v1,v3}
\fmf{photon,label=$k_{1}$,label.side=right}{v2,v4}
\fmf{fermion,label=$p_{2} \! + \! k_{1}$,label.side=right}{v1,v5}
\fmf{fermion,right,label=$k_{1} \! + \! k_{2}$,label.side=right}{v4,v3}
\fmf{fermion,right,label=$k_{2}$,label.side=right}{v3,v4}
\end{fmfgraph*} }
%
%
\vspace*{8mm}
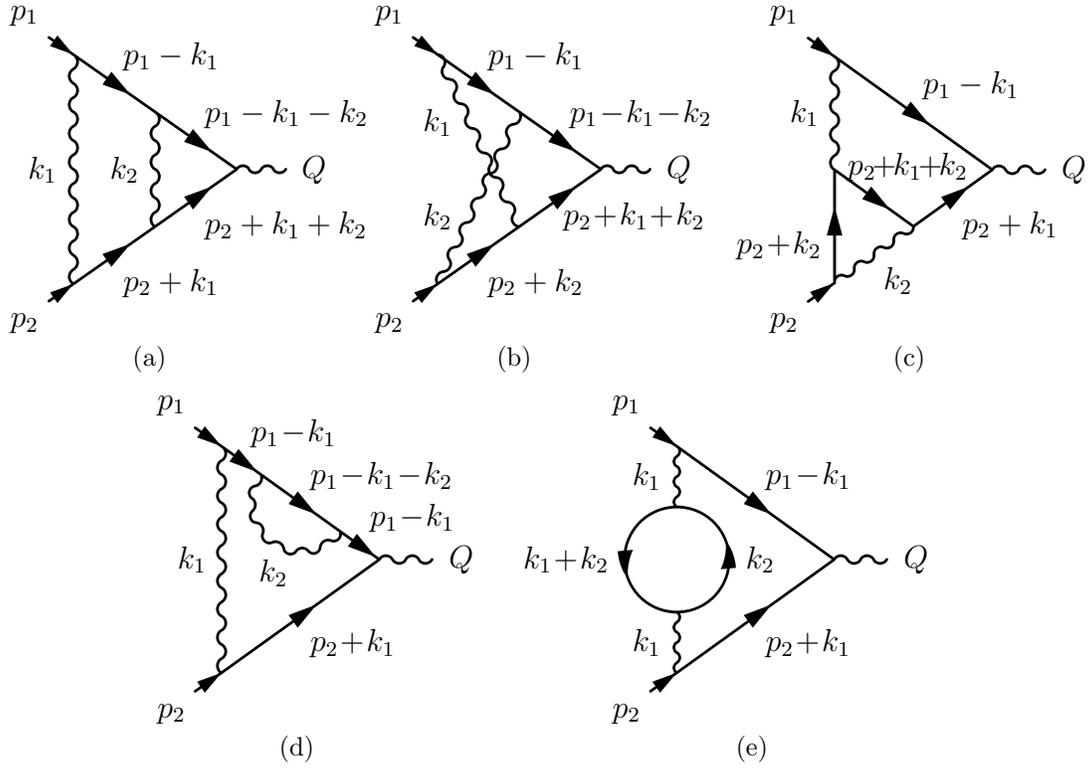
\caption{\label{fig1} 2-loop vertex diagrams for the QED form factor.
The fermionic external lines are on the mass-shell
$p_{1}^{2}=p_{2}^{2}=-a$, while the wavy line on the r.h.s. has momentum
$Q=p_{1}+p_{2}$, with $Q^{2}=-s$. }
\ec
\efig

Following a by now standard approach, we first express 
all the scalar integrals associated to each graph in terms of 
the Master Integrals (MI's) by using the integration by 
parts~\cite{Chet} 
and Lorentz invariance~\cite{Rem3} identities, then write the 
differential equations on the momentum transfer which are satisfied 
by the MI's, and expand the equations in powers of $\epsilon=(4-D)/2$ 
around $\epsilon=0$ ($D=4$) up to the required order. 
We obtain in that way a system of chained differential equations 
for the coefficients of the $\epsilon$-expansion of the MI's 
and finally solve the system for the coefficients 
by Euler's variation of constants method. As a result, we express
the coefficients in close analytic form in terms of 
Harmonic Polylogarithms~\cite{Polylog,Polylog3}.

Let us recall that the Euler's method requires the solution 
of the associated homogeneous equation. Even if general algorithms 
for the solution of differential equations are not available, 
it is to be stressed here that all the homogeneous equations 
which we had to solve came out to be essentially trivial (typically, 
first order homogeneous equations with rational coefficients). 

The present paper is structured as follows. In section \ref{Master} we
review briefly the techniques for reducing the calculation of 
generic multi-loop Feynman graph integrals to the calculation of 
the MI's. Integration by Parts, Lorentz Invariance and general symmetry 
relations are recalled and the application of this approach to our case
is discussed. In section \ref{DiffEqs} we review 
the method of differential equations for the calculation of the MI's. 
In section \ref{Exempla} we describe exhaustively the case of three
typical integrals, giving in some details the system of differential 
equations and the steps for obtaining the solution. 
In section \ref{Results} we present the results for all the MI's 
encountered in the calculation of the 2-loop vertex diagrams and in
section \ref{Results2} we give the results for the scalar 6-denominator 
vertex diagrams which are not MI's. Sections \ref{Q2grande} and 
\ref{Q2piccolo} contain respectively the expansions of the vertex 
6-denominator diagrams in the region of great and low momentum transfer.
Finally, after the summary, section \ref{Summa}, 
and appendix \ref{app1}, where we give the routing used for the 
explicit calculations,
 in appendix \ref{app2} 
we give the results of the 1-loop diagrams involved in our calculations
and in appendix \ref{app3} we list the results for all the reducible
diagrams appearing in the calculation.

\section{The reduction to master integrals \label{Master}}

The aim of this paper is the evaluation of all the possible scalar 
integrals which can occurr in the calculation of the Feynman diagrams of
Fig. (\ref{fig1}). 
They imply two loop momenta, $k_1$ and $k_2$, and three external 
momenta, $p_{1},p_{2},Q$; among them only two are independent, 
because of the momentum conservation law: $p_{1}+p_{2}=Q$. With 
two external momenta and two loop momenta, we can construct 
three Mandelstam invariant variables, $p_1^2, p_2^2$ and 
$Q^2=(p_1+p_2)^2$ 
(we use the Euclidean metric, so that the mass 
shell conditions are $p_1^2=p_2^2= -a$, where $a$ is the squared 
electron mass) and seven different scalar products involving 
the loop momenta, namely $(p_{1} \cdot k_{1}),\ $ 
$(p_{1} \cdot k_{2}),\ $ $(p_{2} \cdot k_{1}),\ $ 
$(p_{2} \cdot k_{2}),\ $ $(k_{1} \cdot k_{2}),\ $ 
$(k_{1}^{2})$ and $(k_{2}^{2})$. 

The graphs of Fig. (\ref{fig1}) involve up to 6 different propagators: 
more exactly, the graphs (a,b,c) have 6 different propagators, while 
graph (d) contains 2 equal electron propagators, graph (e) 
2 equal photon propagators, so that two graphs, (d,e), involve 
only 5 different propagators. In the following we will not consider 
anymore graphs but topologies: topologies will be drawn exactly 
as Feynman graphs, except that all propagators are different. 
To make an example, when a graph contains twice some propagator, as the 
two equal photon propagators of graph (e) above, the corresponding 
topology contains that propagator only once; indeed, the topology 
of graph (e) of Fig. (\ref{fig1}) is given by the topology (e) of 
Fig. (\ref{fig2}). 
Besides those topologies, we will also encounter all the subtopologies 
obtained by removing from the graphs one or more propagators in all 
possible ways. 

Let $t$ be the number of the propagators in any of the 
topologies or subtopologies; we can express $t$ of the 7 scalar products
containing the loop momenta in terms of the propagators, 
(the remaining $(7-t)$ scalar products will be called irreducible) 
and correspondingly the most general scalar integral associated to 
that topology or subtopology has the form 
\be
  I(p_1,p_2) \, = \, \int \{ d^{D}k_{1} \} \{ d^{D}k_{2} \} \, 
       \frac{S_{1}^{n_{1}} \cdots S_{q}^{n_{q}}}{{\mathcal D}_{1}^{m_{1}} 
       \cdots {\mathcal D}_{t}^{m_{t}}} \, , 
\label{b1} 
\ee 
where $\{ d^{D}k \}$ is the loop integration measure (its explicit 
expression, irrelevant here, will be given in section \ref{Results}), 
the integer $m_i, i=1,t$ are the powers of the $t$ propagators, 
with $m_1\ge 1$, and the integer $n_j, j=1,q,\ q=(7-t)$, with $n_j\ge0$,
are the powers of the irreducible scalar products. Let us further 
recall that the continuous 
dimensional regularization makes the definition meaningful for any 
values of the 7 integer $n_i, m_j$. 

We will denote with $I_{t,r,s}$ the family of the integrals with a 
same set of $t$ propagators, a total of $r = \sum_{i} (m_{i}-1)$ 
powers of the $t$ propagators and $s = \sum_{j} n_{j}$ powers of 
the $(7-t)$ irreducible scalar products. The number of the integrals 
contained in the family is 
\be
N \left[ I_{t,r,s} \right] \, = \, \pmatrix{r +t - 1 \cr t-1} \, 
                                   \pmatrix{s-t+6 \cr 6-t } \, .
\label{btrs}
\ee

As we will see more in detail in a moment, one can establish several 
identities involving integrals of the type of 
Eq. (\ref{b1}) with different sets of the 7 indices $m_i, n_j$. 
The identities can be written in the form of a sum of a finite number 
of terms set equal to zero, where each term is a polynomial 
(of finite order and with integer coefficients in the variable 
$D,a$ and the Mandelstam invariants) times an integral of the 
family, as will be seen explicitely in the example of next section.

The identities can be used to express as many as 
possible integrals of a given family in terms of as few as possible 
suitably choosen integrals of that family -- called the 
Master Integrals of that family. 

The identities will be generated by using Integration by Parts, 
Lorentz Invariance (or rotational invariance in $D$ dimensions) 
and symmetry considerations.

\subsubsection{Integration by Parts Identities}

Integration by Parts Identities (IBP-Id's) are among the most remarkable
properties of dimensionally regularized Feynman integrals \cite{Chet}. 
In our case, for each of the integrals defined in Eq. (\ref{b1}) 
one can write 
\bea
\int \{ d^{D}k_{1} \} \{ d^{D}k_{2} \} \, \frac{\partial}{
\partial k_{1}^{\mu}} \left\{ v^{\mu} \, \frac{S_{1}^{n_{1}} \cdots 
S_{q}^{n_{q}}}{{\mathcal D}_{1}^{m_{1}} \cdots {\mathcal D}_{t}^{m_{t}}} 
\right\} & = & 0 
\, , 
\label{b2} \\
\int \{ d^{D}k_{1} \} \{ d^{D}k_{2} \} \, \frac{\partial}{
\partial k_{2}^{\mu}} \left\{ v^{\mu} \, \frac{S_{1}^{n_{1}} \cdots 
S_{q}^{n_{q}}}{{\mathcal D}_{1}^{m_{1}} \cdots {\mathcal D}_{t}^{m_{t}}} 
\right\} & = & 0 
\, ,
\label{b3}
\eea
where thanks to the dimensional regularization everything is well 
defined and the identity holds trivially. 
In the above identities the vector $v^{\mu}$ can be 
any of the 4 independent vectors of the problem: $k_{1}$, $k_{2}$, 
$p_{1}$, or $p_{2}$, so that for each integrand there are 8 IBP-Id's. 
When evaluating explicitly the derivatives, one obtains a 
combination of integrands with a total 
power of the irreducible scalar products equal to $(s-1),\ $ $s$ and 
$(s+1)$ and total powers fo the propagators in the denominator 
equal to 
$(t+r)$ and $(t+r+1)$, therefore involving, 
besides the integrals of the family $I_{t,r,s}$, also 
the families $I_{t,r,s-1},\ $ $I_{t,r+1,s}$ and $I_{t,r+1,s+1}$. 

Simplifications between reducible 
scalar products and propagators in the denominator may also occur, 
giving lower powers of the propagators. It may happen that some 
propagator disappears at all in this process; the resulting term 
will then give an integral of a simpler family (or subtopology) with 
$(t-1)$ propagators. 

\bfig
\bc
\begin{fmfgraph*}(35,35)
\fmfleft{i1,i2}
\fmfright{o}
\fmfforce{0.25w,0.45h}{v11}
\fmf{fermion}{i1,v1}
\fmf{fermion}{i2,v2}
\fmf{photon}{v3,o}
\fmflabel{$p_{1}$}{i2}
\fmflabel{$p_{2}$}{i1}
\fmflabel{$Q$}{o}
\fmfv{l=$k_{1}$,l.a=0,l.d=.05w}{v11}
\fmf{fermion,tension=.3,label=$p_{1} \! + p_{2} \! - \! k_{1} \! - \! 
k_{2}$,label.side=left}{v2,v3}
\fmf{photon,tension=.3,label=$k_{2}$,label.side=right}{v1,v3}
\fmf{fermion,tension=0,label=$p_{2} \! - \! k_{2}$,
label.side=left}{v1,v2}
\fmf{fermion,tension=0,right=.5}{v2,v3}
\end{fmfgraph*}
\vspace*{8mm}
\caption{\label{figex} A 4-denominator topology. }
\ec
\efig

As an explicit example let us consider the case of the 4-denominator 
topology of Fig. (\ref{figex}). We have three irreducible scalar 
products, in this topology; we choose $(p_{1} \cdot k_{1})$, $(p_{2} 
\cdot k_{1})$ and $(k_{1} \cdot k_{2})$. Eq. (\ref{b2}), for generic
values of the indices $m_{i}$, $n_{i}$ and for a generic independent
vector $v_{\mu}$, reads:
\bea
\! \! \! \! \! \! \! \! \! \! \! \!
\int \{ \! d^{D} \! k_{1} \} \{ \! d^{D} \! k_{2} \} \frac{\partial}{
\partial k_{1}^{\mu}} \Biggl\{ \! \frac{v^{\mu} \, (p_{1} \cdot k_{1})^{
n_{1}} (p_{2} \cdot k_{1})^{n_{2}} (k_{1} \cdot k_{2})^{n_{3}}}{
[k_{1}^2 \! + \! a]^{m_{1}} \! [k_{2}^{2}]^{m_{2}} \! [(p_{2} \! - 
\! k_{2})^{2} \! + \! a]^{m_{3}} \! 
[(p_{1} \! \! + \! p_{2} \! \! - \! k_{1} \! \! - \! k_{2})^{2} \! + \! 
a]^{m_{4}}}  \! \! \Biggr\} \!  & & \nn\\
\! \! \! \! \! \! \! \! \! \! \! \! \! \! \! \! \! \! \! \! \! \! \! \! 
& & \! \! \! \! \! \! \! \! \! \! \! \! \! \! \! \! \! \! \! \! \! \! 
\! \!= 0 \, .
\label{ex1}
\eea

Let us take for simplicity $m_{1}= \cdots =m_{4}=1$, $n_{1}= \cdots
n_{3}=0$ and $v^{\mu}=p_{1}^{\mu}$. Performing the derivative with
respect to $k_{1}$ and simplifying the reducible scalar products
with the corresponding denominator, we write Eq. (\ref{ex1}) as follows:
\bea
\! \! \! \! \! \! \! \! \! \! \! \! 0 & = & - 2 
\parbox{15mm}{
\begin{fmfgraph*}(15,15)
\fmfleft{i1,i2}
\fmfright{o}
\fmfforce{0.4w,0.55h}{v5}
\fmf{plain}{i1,v1}
\fmf{plain}{i2,v2}
\fmf{photon}{v3,o}
\fmflabel{$(p_{1} \cdot k_{1})$}{o}
\fmf{plain,tension=.3}{v2,v3}
\fmf{photon,tension=.3}{v1,v3}
\fmf{plain,tension=0}{v2,v1}
\fmf{plain,tension=0,right=.5}{v2,v3}
\fmfv{decor.shape=circle,decor.filled=full,decor.size=.1w}{v5}
\end{fmfgraph*}} \qquad \qquad -2 
\parbox{15mm}{
\begin{fmfgraph*}(15,15)
\fmfleft{i1,i2}
\fmfright{o}
\fmfforce{0.48w,0.67h}{v5}
\fmf{plain}{i1,v1}
\fmf{plain}{i2,v2}
\fmf{photon}{v3,o}
\fmflabel{$(k_{1} \cdot k_{2})$}{o}
\fmf{plain,tension=.3}{v2,v3}
\fmf{photon,tension=.3}{v1,v3}
\fmf{plain,tension=0}{v2,v1}
\fmf{plain,tension=0,right=.5}{v2,v3}
\fmfv{decor.shape=circle,decor.filled=full,decor.size=.1w}{v5}
\end{fmfgraph*}} \qquad \qquad + 2 
\parbox{15mm}{
\begin{fmfgraph*}(15,15)
\fmfleft{i1,i2}
\fmfright{o}
\fmfforce{0.48w,0.67h}{v5}
\fmf{plain}{i1,v1}
\fmf{plain}{i2,v2}
\fmf{photon}{v3,o}
\fmflabel{$(p_{2} \cdot k_{1})$}{o}
\fmf{plain,tension=.3}{v2,v3}
\fmf{photon,tension=.3}{v1,v3}
\fmf{plain,tension=0}{v2,v1}
\fmf{plain,tension=0,right=.5}{v2,v3}
\fmfv{decor.shape=circle,decor.filled=full,decor.size=.1w}{v5}
\end{fmfgraph*}} \nn\\
\! \! \! \! \! \! \! \! \! \! \! \! & & + 
\parbox{15mm}{
\begin{fmfgraph*}(15,15)
\fmfleft{i1,i2}
\fmfright{o}
\fmf{plain}{i1,v1}
\fmf{plain}{i2,v2}
\fmf{photon}{v3,o}
\fmf{plain,tension=.3}{v2,v3}
\fmf{photon,tension=.3}{v1,v3}
\fmf{plain,tension=0}{v2,v1}
\fmf{plain,tension=0,right=.5}{v2,v3}
\end{fmfgraph*}} \, -  \, 
\parbox{15mm}{
\begin{fmfgraph*}(15,15)
\fmfleft{i}
\fmfright{o}
\fmfforce{0.5w,0.25h}{v5}
\fmf{photon}{i,v1}
\fmf{photon}{v2,o}
\fmf{plain,tension=.15,left}{v1,v2}
\fmf{photon,tension=.15}{v1,v2}
\fmf{plain,tension=.15,right}{v1,v2}
\fmfv{decor.shape=circle,decor.filled=full,decor.size=.1w}{v5}
\end{fmfgraph*} }  \, +  \, \frac{(1- \epsilon)}{a} \, \, 
\parbox{15mm}{
\begin{fmfgraph*}(15,15)
\fmfleft{i}
\fmfright{o}
\fmf{plain}{i,v1}
\fmf{plain}{v2,o}
\fmf{plain,tension=.22,left}{v1,v2}
\fmf{photon,tension=.22,right}{v1,v2}
\fmf{plain,right=45}{v2,v2}
\end{fmfgraph*} } \, ,
\eea
where a dot on a propagator line means that the propagator is squared
and irreducible scalar products left are explicitely written.

\subsubsection{Lorentz invariance identities}

Another class of identities can be derived from the fact that 
the integrals $I(p_i)$, Eq. (\ref{b1}), are Lorentz scalars (or rather 
$D$-dimensional rotational invariant) \cite{Rem3}. If we consider an 
infinitesimal Lorentz transformation on the external momenta, $p_{i} 
\rightarrow p_{i}+ \delta p_{i}$, where $\delta p_{i}^{\mu} = 
\epsilon^{\mu}_{\nu} p_{i}^{\nu}$, and $\epsilon^{\mu}_{\nu}$ is a 
completely antisymmetric tensor, we have 
\be
 I(p_{i}+ \delta p_{i}) \, = \, I(p_{i}) \, ,
\label{b5}
\ee
Because of the antisymmetry of $\epsilon^{\mu}_{\nu}$ and because
\bea
I(p_{i}+ \delta p_{i}) & = & I(p_{i}) + \sum_{n} 
\delta p_{n}^{\mu} \frac{\partial I(p_i)}{\partial p_{n}^{\mu}} 
\nn\\
& = & I(p_{i}) + \epsilon^{\mu}_{\nu} \, \left[ \sum_{n} 
p_{n}^{\nu} \frac{\partial I(p_i)}{\partial p_{n}^{\mu}} \right] 
\, ,
\label{b7}
\eea
we can write the following relation:
\be
\sum_{n} \left[ p_{n}^{\nu} \frac{\partial }{\partial p_{n}^{
\mu}} - p_{n}^{\mu} \frac{\partial }{\partial 
p_{n}^{\nu}} \right] I(p_i) \, = \, 0 \, .
\label{b8}
\ee

Eq. (\ref{b8}) can be contracted with all possible antisymmetric 
combination of the external momenta $p_{i}^{\mu}p_{j}^{\nu}$, to obtain
other identities for the considered integrals. 

In our case we have two external independent momenta and we can thus 
construct, besides Eqs. (\ref{b2},\ref{b3}), the further identity 
\begin{equation} 
    \left[ (p_{1} \cdot p_{2}) \left( 
p_{1}^{\mu} \frac{\partial }{\partial p_{1}^{\mu}} - 
p_{2}^{\mu} \frac{\partial }{\partial p_{2}^{\mu}} \right) + 
p_{2}^{2} \, \, p_{1}^{\mu} \frac{\partial }{\partial p_{2}^{\mu}} - 
p_{1}^{2} \, \, p_{2}^{\mu} \frac{\partial }{\partial p_{1}^{\mu}} 
    \right] I(p_i)  \, = \, 0 \, .
\label{b8a} 
\end{equation} 

Let us note that, in order to obtain non--trivial identities from 
Eq. (\ref{b8a}), the derivative with respect to the momentum $p_{i}$ has
to be intended as a differentiation under the loop integral in the 
definition, Eq. (\ref{b1}), i.e. directly on the integrand of 
$I_{t,r,s}$. Again, this is allowed by the dimensional regularization. 

As an explicit example let us consider the same topology as in the
previous section. If we take $n_{1}=n_{2}=n_{3}=0$ and
$m_{1}=m_{2}=m_{3}=m_{4}=1$, we have
\be
\! \! \! \! \! \! \! \! \! \! \! \! \! \! 
I(p_{1},p_{2}) \! = \! \int \! \{ d^{D}k_{1} \} \{ d^{D}k_{2} \} 
\frac{1}{[k_{1}^2 \! + \! a] k_{2}^{2} [(p_{2} \! - 
\! k_{2})^{2} \! + \! a] [(p_{1} \! + \! p_{2} \! - \! k_{1} \! - 
\! k_{2})^{2} \! + \! a]}  
,
\ee
and Eq. (\ref{b8a}) reads as follows:
\bea
\! \! \! \! \! \! \! \! \! 0 \! & = & \! 4a \Biggl\{
\parbox{15mm}{
\begin{fmfgraph*}(15,15)
\fmfleft{i1,i2}
\fmfright{o}
\fmfforce{0.23w,0.5h}{v5}
\fmf{plain}{i1,v1}
\fmf{plain}{i2,v2}
\fmf{photon}{v3,o}
\fmflabel{$(k_{1} \cdot k_{2})$}{o}
\fmf{plain,tension=.3}{v2,v3}
\fmf{photon,tension=.3}{v1,v3}
\fmf{plain,tension=0}{v2,v1}
\fmf{plain,tension=0,right=.5}{v2,v3}
\fmfv{decor.shape=circle,decor.filled=full,decor.size=.1w}{v5}
\end{fmfgraph*}} \qquad \qquad - 
\parbox{15mm}{
\begin{fmfgraph*}(15,15)
\fmfleft{i1,i2}
\fmfright{o}
\fmfforce{0.23w,0.5h}{v5}
\fmf{plain}{i1,v1}
\fmf{plain}{i2,v2}
\fmf{photon}{v3,o}
\fmflabel{$(p_{1} \cdot k_{1})$}{o}
\fmf{plain,tension=.3}{v2,v3}
\fmf{photon,tension=.3}{v1,v3}
\fmf{plain,tension=0}{v2,v1}
\fmf{plain,tension=0,right=.5}{v2,v3}
\fmfv{decor.shape=circle,decor.filled=full,decor.size=.1w}{v5}
\end{fmfgraph*}} \qquad \qquad - 
\parbox{15mm}{
\begin{fmfgraph*}(15,15)
\fmfleft{i1,i2}
\fmfright{o}
\fmfforce{0.23w,0.5h}{v5}
\fmf{plain}{i1,v1}
\fmf{plain}{i2,v2}
\fmf{photon}{v3,o}
\fmflabel{$(p_{2} \cdot k_{1})$}{o}
\fmf{plain,tension=.3}{v2,v3}
\fmf{photon,tension=.3}{v1,v3}
\fmf{plain,tension=0}{v2,v1}
\fmf{plain,tension=0,right=.5}{v2,v3}
\fmfv{decor.shape=circle,decor.filled=full,decor.size=.1w}{v5}
\end{fmfgraph*}} \qquad \qquad \Biggr\} \nn\\
\! \! \! \! \! \! \! \! \! & & -2as 
\parbox{15mm}{
\begin{fmfgraph*}(15,15)
\fmfleft{i1,i2}
\fmfright{o}
\fmfforce{0.23w,0.5h}{v5}
\fmf{plain}{i1,v1}
\fmf{plain}{i2,v2}
\fmf{photon}{v3,o}
\fmf{plain,tension=.3}{v2,v3}
\fmf{photon,tension=.3}{v1,v3}
\fmf{plain,tension=0}{v2,v1}
\fmf{plain,tension=0,right=.5}{v2,v3}
\fmfv{decor.shape=circle,decor.filled=full,decor.size=.1w}{v5}
\end{fmfgraph*}} \, + 2s \, \Biggl\{
\parbox{15mm}{
\begin{fmfgraph*}(15,15)
\fmfleft{i1,i2}
\fmfright{o}
\fmfforce{0.48w,0.67h}{v5}
\fmf{plain}{i1,v1}
\fmf{plain}{i2,v2}
\fmf{photon}{v3,o}
\fmflabel{$(k_{1} \cdot k_{2})$}{o}
\fmf{plain,tension=.3}{v2,v3}
\fmf{photon,tension=.3}{v1,v3}
\fmf{plain,tension=0}{v2,v1}
\fmf{plain,tension=0,right=.5}{v2,v3}
\fmfv{decor.shape=circle,decor.filled=full,decor.size=.1w}{v5}
\end{fmfgraph*}} \qquad \qquad - 2 
\parbox{15mm}{
\begin{fmfgraph*}(15,15)
\fmfleft{i1,i2}
\fmfright{o}
\fmfforce{0.48w,0.67h}{v5}
\fmf{plain}{i1,v1}
\fmf{plain}{i2,v2}
\fmf{photon}{v3,o}
\fmflabel{$(p_{2} \cdot k_{1})$}{o}
\fmf{plain,tension=.3}{v2,v3}
\fmf{photon,tension=.3}{v1,v3}
\fmf{plain,tension=0}{v2,v1}
\fmf{plain,tension=0,right=.5}{v2,v3}
\fmfv{decor.shape=circle,decor.filled=full,decor.size=.1w}{v5}
\end{fmfgraph*}} \qquad \qquad \Biggr\} \nn\\
\! \! \! \! \! \! \! \! \! & & - s^{2} 
\parbox{15mm}{
\begin{fmfgraph*}(15,15)
\fmfleft{i1,i2}
\fmfright{o}
\fmfforce{0.48w,0.67h}{v5}
\fmf{plain}{i1,v1}
\fmf{plain}{i2,v2}
\fmf{photon}{v3,o}
\fmf{plain,tension=.3}{v2,v3}
\fmf{photon,tension=.3}{v1,v3}
\fmf{plain,tension=0}{v2,v1}
\fmf{plain,tension=0,right=.5}{v2,v3}
\fmfv{decor.shape=circle,decor.filled=full,decor.size=.1w}{v5}
\end{fmfgraph*}} - [s-2a] \, \, 
\parbox{15mm}{
\begin{fmfgraph*}(15,15)
\fmfleft{i}
\fmfright{o}
\fmfforce{0.5w,0.5h}{v5}
\fmf{plain}{i,v1}
\fmf{plain}{v2,o}
\fmf{plain,tension=.15,left}{v1,v2}
\fmf{plain,tension=.15}{v1,v2}
\fmf{plain,tension=.15,right}{v1,v2}
\fmfv{decor.shape=circle,decor.filled=full,decor.size=.1w}{v5}
\end{fmfgraph*} } - s \, \, 
\parbox{15mm}{
\begin{fmfgraph*}(15,15)
\fmfleft{i}
\fmfright{o}
\fmfforce{0.5w,0.25h}{v5}
\fmf{plain}{i,v1}
\fmf{plain}{v2,o}
\fmf{plain,tension=.15,left}{v1,v2}
\fmf{plain,tension=.15}{v1,v2}
\fmf{plain,tension=.15,right}{v1,v2}
\fmfv{decor.shape=circle,decor.filled=full,decor.size=.1w}{v5}
\end{fmfgraph*} } \nn\\
\! \! \! \! \! \! \! \! \! & & + 2 s \, \, 
\parbox{15mm}{
\begin{fmfgraph*}(15,15)
\fmfleft{i}
\fmfright{o}
\fmfforce{0.5w,0.25h}{v5}
\fmf{photon}{i,v1}
\fmf{photon}{v2,o}
\fmf{plain,tension=.15,left}{v1,v2}
\fmf{photon,tension=.15}{v1,v2}
\fmf{plain,tension=.15,right}{v1,v2}
\fmfv{decor.shape=circle,decor.filled=full,decor.size=.1w}{v5}
\end{fmfgraph*} }  \, -  \, \frac{s[1- \epsilon]}{a} \, \, 
\parbox{15mm}{
\begin{fmfgraph*}(15,15)
\fmfleft{i}
\fmfright{o}
\fmf{plain}{i,v1}
\fmf{plain}{v2,o}
\fmf{plain,tension=.22,left}{v1,v2}
\fmf{photon,tension=.22,right}{v1,v2}
\fmf{plain,right=45}{v2,v2}
\end{fmfgraph*} } \, ,
\eea

\subsubsection{Symmetry relations}

In general further identities among Feynman graph integrals 
can arise when the Feynman graph has some symmetry. In such a case 
there can be a trasformation of the loop momenta which does not 
change the value of the integral, but transforms the integrand 
in a combination of different integrands. By imposing the 
identity of the initial integral to the combination 
of integrals resulting from the change of loop momenta one 
obtains further identities relating integrals corresponding to a 
same graph. 

As an example, let us consider again the topology of Fig. (\ref{figex}),
with generic indices on the numerator and on the denominator:
\bea
\! \! \! \! \! \! \! \! \! \! \! \! \! & & I(p_{1},p_{2}) 
= \nn\\
\! \! \! \! \! \! \! \! \! \! \! \! \! & & 
\int \{  \! d^{D} \! k_{1}  \! \} \{  \! d^{D} \! k_{2}  \! \} 
\! \frac{(p_{1} \cdot k_{1})^{n_{1}} (p_{2} \cdot k_{1})^{n_{2}}
(k_{1} \cdot k_{2})^{n_{3}}}{[k_{1}^2 \! + \! a]^{m_{1}} \, 
[k_{2}^{2}]^{m_{2}} [(p_{2} \! - \! k_{2})^{2} \, + \! a]^{m_{3}} \, 
[(p_{1} \! + p_{2} \! - \! k_{1} \! - \! k_{2})^{2} \! + \! a]^{m_{4}}}
.
\label{ex3}
\eea

The two propagators with momentum $k_{2}$ and $(p_{1}+p_{2}-k_{1}-
k_{2})$ have the same mass. The following redefinition of the 
integration momentum 
\be
  k_{1} = p_{1}+p_{2}-k_{1}^{'}-k_{2} \, , 
\label{kshift} 
\ee
that consists in the interchange of the two propagators in the closed 
electron loop, does not affect of course the value of the integral; 
nevertheless, the explicit form of the integrand can change, 
generating non-trivial identities. 

Taking for instance $n_{1}=n_{2}=n_{3}=0$ and $m_{1}=m_{2}=m_{4}=1$, 
$m_{3}=2$ in Eq.(\ref{ex3}), the substitution (\ref{kshift}) 
gives, for example, the following very simple relation: 
\be
0 \, = \, 
\parbox{15mm}{
\begin{fmfgraph*}(15,15)
\fmfleft{i1,i2}
\fmfright{o}
\fmfforce{0.4w,0.55h}{v5}
\fmf{plain}{i1,v1}
\fmf{plain}{i2,v2}
\fmf{photon}{v3,o}
\fmf{plain,tension=.3}{v2,v3}
\fmf{photon,tension=.3}{v1,v3}
\fmf{plain,tension=0}{v2,v1}
\fmf{plain,tension=0,right=.5}{v2,v3}
\fmfv{decor.shape=circle,decor.filled=full,decor.size=.1w}{v5}
\end{fmfgraph*}} \, \, -
\parbox{15mm}{
\begin{fmfgraph*}(15,15)
\fmfleft{i1,i2}
\fmfright{o}
\fmfforce{0.48w,0.67h}{v5}
\fmf{plain}{i1,v1}
\fmf{plain}{i2,v2}
\fmf{photon}{v3,o}
\fmf{plain,tension=.3}{v2,v3}
\fmf{photon,tension=.3}{v1,v3}
\fmf{plain,tension=0}{v2,v1}
\fmf{plain,tension=0,right=.5}{v2,v3}
\fmfv{decor.shape=circle,decor.filled=full,decor.size=.1w}{v5}
\end{fmfgraph*}} \, \, .
\ee

Taking $n_{1}=n_{2}=0$, $n_{3}=1$ and $m_{1}=m_{2}=m_{4}=1$, $m_{3}=2$,
we get the more complicated idetity 
\bea
\! \! \! \! \! \! \! \! \! 0 \! & = & \! 
\parbox{15mm}{
\begin{fmfgraph*}(15,15)
\fmfleft{i1,i2}
\fmfright{o}
\fmfforce{0.23w,0.5h}{v5}
\fmf{plain}{i1,v1}
\fmf{plain}{i2,v2}
\fmf{photon}{v3,o}
\fmflabel{$(k_{1} \cdot k_{2})$}{o}
\fmf{plain,tension=.3}{v2,v3}
\fmf{photon,tension=.3}{v1,v3}
\fmf{plain,tension=0}{v2,v1}
\fmf{plain,tension=0,right=.5}{v2,v3}
\fmfv{decor.shape=circle,decor.filled=full,decor.size=.1w}{v5}
\end{fmfgraph*}} \qquad \qquad + 
\parbox{15mm}{
\begin{fmfgraph*}(15,15)
\fmfleft{i1,i2}
\fmfright{o}
\fmfforce{0.23w,0.5h}{v5}
\fmf{plain}{i1,v1}
\fmf{plain}{i2,v2}
\fmf{photon}{v3,o}
\fmflabel{$(p_{1} \cdot k_{1})$}{o}
\fmf{plain,tension=.3}{v2,v3}
\fmf{photon,tension=.3}{v1,v3}
\fmf{plain,tension=0}{v2,v1}
\fmf{plain,tension=0,right=.5}{v2,v3}
\fmfv{decor.shape=circle,decor.filled=full,decor.size=.1w}{v5}
\end{fmfgraph*}} \qquad \qquad + 
\parbox{15mm}{
\begin{fmfgraph*}(15,15)
\fmfleft{i1,i2}
\fmfright{o}
\fmfforce{0.23w,0.5h}{v5}
\fmf{plain}{i1,v1}
\fmf{plain}{i2,v2}
\fmf{photon}{v3,o}
\fmflabel{$(p_{2} \cdot k_{1})$}{o}
\fmf{plain,tension=.3}{v2,v3}
\fmf{photon,tension=.3}{v1,v3}
\fmf{plain,tension=0}{v2,v1}
\fmf{plain,tension=0,right=.5}{v2,v3}
\fmfv{decor.shape=circle,decor.filled=full,decor.size=.1w}{v5}
\end{fmfgraph*}}  \nn\\
\! \! \! \! \! \! \! \! \! & & + \frac{s}{2} 
\parbox{15mm}{
\begin{fmfgraph*}(15,15)
\fmfleft{i1,i2}
\fmfright{o}
\fmfforce{0.23w,0.5h}{v5}
\fmf{plain}{i1,v1}
\fmf{plain}{i2,v2}
\fmf{photon}{v3,o}
\fmf{plain,tension=.3}{v2,v3}
\fmf{photon,tension=.3}{v1,v3}
\fmf{plain,tension=0}{v2,v1}
\fmf{plain,tension=0,right=.5}{v2,v3}
\fmfv{decor.shape=circle,decor.filled=full,decor.size=.1w}{v5}
\end{fmfgraph*}}  + \frac{1}{2} \, 
\parbox{15mm}{
\begin{fmfgraph*}(15,15)
\fmfleft{i}
\fmfright{o}
\fmfforce{0.5w,0.5h}{v5}
\fmf{plain}{i,v1}
\fmf{plain}{v2,o}
\fmf{plain,tension=.15,left}{v1,v2}
\fmf{plain,tension=.15}{v1,v2}
\fmf{plain,tension=.15,right}{v1,v2}
\fmfv{decor.shape=circle,decor.filled=full,decor.size=.1w}{v5}
\end{fmfgraph*} } \, \, .
\eea

\vskip 1truecm 

Summarizing, for each of the $N[I_{t,r,s}]$ integrals of the family 
$I_{t,r,s}$, Eq. (\ref{btrs}), we have the 9 identities,
Eqs. (\ref{b2},\ref{b3}) and Eq. (\ref{b8a}), involving integrals of 
the families up to $I_{t,r+1,s+1}$. For $(r=0,s=0)$ the number of 
all the integrals involved in the identities (for $t=6$ they are 14) 
exceeds the number of the equations obtained (which in this case is 9), 
but when writing systematically all the equations for increasing values 
of $r$ and $s$, $r=0,1,..\ ,\ s=0,1,...$, the number of the equations 
grows faster than the number of the integrals
\cite{Lap}, so that at some point one deals with more equations than 
involved integrals -- generating an apparently 
overconstrained set of linear equations for the unknown integrals. 
At this stage one can use the symmetry relations, somewhat 
reducing the number of the unknown integrals, after which 
one is left with the problem of solving the linear system of the 
identities. The problem is in principle trivial, but algebraically very 
lengthy, so that some organization is required for obtaining the 
solution. 

To that aim, one can order the integrals in some lexicographic order 
(which means giving a ``weight" to each integral; the weight can be 
almost any increasing function of the 
indices $m_i, n_j$, such that integrals with higher indices have 
bigger weights) and then solve the system by the Gauss substitution 
rule by considering one by one, in some order, the equations of 
the system and using each equation for expressing the integral with 
highest weight present in that equation 
in terms of the other integrals of lower weight, and then substituting 
in the rest of the system. The algorithm is straightforward, but its 
execution requires of course a great amount of algebra; indeed, 
it was implemented as a chain of programs, written in the computer 
language {\tt C}, which automatically runs programs written for the 
algebraic computer languages {\tt FORM} \cite{FORM} and {\tt Maple}
\cite{Maple}, reads the outputs and generates new input programs till 
all the equations are solved (and the solutions 
are written as a modulus of {\tt FORM} code). 
 
One finds that several equations are identically satisfied 
(the system is only apparently overconstrained), and all the appearing 
unknown integrals are expressed in terms of very few independent 
integrals, the Master Integrals (MI's) for that family of integrals. 
In so doing, the resulting MI's correspond 
to the integrals of lowest weight; but as the choice of the weight 
is to a large extent arbitrary, there is also some freedom in the choice
of the integrals to pick up as actual MI's (not in their number, of 
course!). Concerning in particular the calculation described in this 
paper, there are several cases in which two MI's are found for a given 
topology or subtopology, while sometimes 
only one MI is present. 
It may also happen that no MI for the considered topology is left -- 
i.e. all the integrals corresponding to the given $t$-propagator 
(sub)topology can be expressed in terms of MI's of its subtopolgies with
$(t-1)$ propagators. 

As a last remark, strictly speaking we are not able to prove that the 
MI's we find are really the minimal set of MI's, i.e. that they are 
all independent from each other (in the sense of the 
combination with polynomial factors described above); but in any case 
the number of the MI's which we find is small, so that reducing 
the several hundred of integrals occurring in the calculation of 
the vertex graphs form factors to a few (in fact 16 MI's, see 
section \ref{MIs}) is in any case a great 
progress. The (unlikely!) discovery that one of our MI's can be 
expressed as combination of the others would just simplify even further 
the calculation -- without spooling, however, the correctness of 
the already obtained results. 

Concerning the number of the subtopologies, a topology with 
$t$ propagators has $(t-1)$ subtopology with $(t-1)$ propagators, 
$(t-1)(t-2)$ subtopologies with $(t-2)$ propagators etc. It turns out, 
however, that most subtopologies are in fact equal due to symmetry 
relations, and the subtopologies coming from different graphs overlap 
to a great extent. For those reasons, the actual number of all the 
different subtopologies is relatively small. 

\bfig
\bc
\subfigure[]{
\begin{fmfgraph*}(23,23)
\fmfleft{i1,i2}
\fmfright{o}
\fmf{plain}{i1,v1}
\fmf{plain}{i2,v2}
\fmf{photon}{v5,o}
\fmf{plain,tension=.3}{v2,v3}
\fmf{plain,tension=.3}{v3,v5}
\fmf{plain,tension=.3}{v1,v4}
\fmf{plain,tension=.3}{v4,v5}
\fmf{photon,tension=0}{v2,v1}
\fmf{photon,tension=0}{v4,v3}
\end{fmfgraph*} }
%
%
%
\subfigure[]{
\begin{fmfgraph*}(23,23)
\fmfleft{i1,i2}
\fmfright{o}
\fmf{plain}{i1,v1}
\fmf{plain}{i2,v2}
\fmf{photon}{v5,o}
\fmf{plain,tension=.3}{v2,v3}
\fmf{plain,tension=.3}{v3,v5}
\fmf{plain,tension=.3}{v1,v4}
\fmf{plain,tension=.3}{v4,v5}
\fmf{photon,tension=0}{v2,v4}
\fmf{photon,tension=0}{v1,v3}
\end{fmfgraph*} }
%
%
\subfigure[]{
\begin{fmfgraph*}(23,23)
\fmfleft{i1,i2}
\fmfright{o}
\fmfforce{0.2w,0.93h}{v2}
\fmfforce{0.2w,0.07h}{v1}
\fmfforce{0.2w,0.5h}{v3}
\fmfforce{0.8w,0.5h}{v5}
\fmf{plain}{i1,v1}
\fmf{plain}{i2,v2}
\fmf{photon}{v5,o}
\fmf{plain,tension=0}{v2,v5}
\fmf{plain,tension=0}{v3,v4}
\fmf{photon,tension=.4}{v1,v4}
\fmf{plain,tension=.4}{v4,v5}
\fmf{plain,tension=0}{v1,v3}
\fmf{photon,tension=0}{v2,v3}
\end{fmfgraph*} } 
%
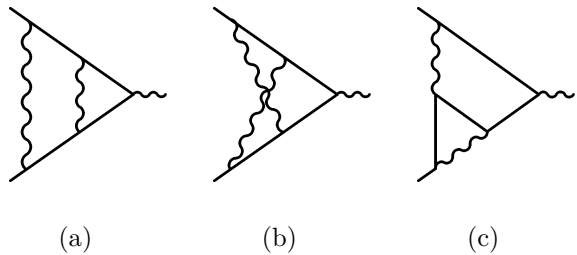
\caption{\label{fig1bis} The set of 3 independent 6-denominator 
topologies present in the graphs of Fig.(\ref{fig1}). }
\ec
\efig

We show in Figs. (\ref{fig1bis}--\ref{fig4}) all the different 
topologies (and subtopologies; we will refer to them as topologies as 
well). In all the figures, a straight external line stands for an 
electron on the mass-shell, while the wavy external line carries the 
momentum $Q$. 

There are 3 independent topologies involving 6 denominators, those of
Fig. (\ref{fig1bis}), corresponding to the original graphs (a,b,c) of 
Fig. (\ref{fig1}). One then finds the 8 independent topologies involving
5 denominators shown in Fig. (\ref{fig2}); note that the topologies 
(b,e) of Fig. (\ref{fig2}) correspond to the 6-propagator vertex graphs
(d,e) of Fig. (\ref{fig1}), the topology (f) of Fig. (\ref{fig2}) 
corresponds to a vacuum polarization graph, (g) is a constant (i.e. 
does not depend on the momentum transfer $Q^2$, but on a squared 
electron momentum on the mass-shell, $p^2=-a$) and (h) factorizes into 
two 1-loop topologies. 

Fig. (\ref{fig3}) contains all the 12 independent 4-denominator 
topologies; again, only the topologies (a,b,c,d) correspond to 
genuine 2-loop vertex topologies, while (e) corresponds to a vacuum 
polarization, (f,g,h) are constants and (i,j,k,l) factorize 
into two 1-loop topologies. 

Finally, Fig. (\ref{fig4}) contains all the 6 independent 3-denominator 
topologies; (a) corresponds to a vacuum polarization, the others 
are constants or factorizable.

At the 2-denominator level, the only topology giving a non-vanishing 
contribution corresponds to the product of two 1-loop tadpoles with
squared mass $a$ shown in Fig. (\ref{fig4}) (g).

\bfig
\bc
\subfigure[]{
\begin{fmfgraph*}(20,20)
\fmfleft{i1,i2}
\fmfright{o}
\fmf{plain}{i1,v1}
\fmf{plain}{i2,v2}
\fmf{photon}{v4,o}
\fmf{plain,tension=.4}{v2,v3}
\fmf{plain,tension=.2}{v3,v4}
\fmf{plain,tension=.15}{v1,v4}
\fmf{photon,tension=0}{v2,v1}
\fmf{photon,tension=0}{v1,v3}
\end{fmfgraph*} }
%
%
\hspace{3mm}
\subfigure[]{
\begin{fmfgraph*}(20,20)
\fmfleft{i1,i2}
\fmfright{o}
\fmf{plain}{i1,v1}
\fmf{plain}{i2,v2}
\fmf{photon}{v4,o}
\fmf{plain,tension=.4}{v2,v3}
\fmf{plain,tension=.2}{v3,v4}
\fmf{plain,tension=.15}{v1,v4}
\fmf{photon,tension=0}{v2,v1}
\fmf{photon,tension=0,left=.5}{v4,v3}
\end{fmfgraph*} }
%
%
\hspace{3mm}
\subfigure[]{
\begin{fmfgraph*}(20,20)
\fmfleft{i1,i2}
\fmfright{o}
\fmf{plain}{i1,v1}
\fmf{plain}{i2,v2}
\fmf{photon}{v4,o}
\fmf{photon,tension=.4}{v2,v3}
\fmf{plain,tension=.2}{v3,v4}
\fmf{plain,tension=.15}{v1,v4}
\fmf{plain,tension=0}{v2,v1}
\fmf{plain,tension=0}{v1,v3}
\end{fmfgraph*} }
%
%
\hspace{3mm}
\subfigure[]{
\begin{fmfgraph*}(20,20)
\fmfleft{i1,i2}
\fmfright{o}
\fmfforce{0.2w,0.93h}{v2}
\fmfforce{0.2w,0.07h}{v1}
\fmfforce{0.2w,0.5h}{v3}
\fmfforce{0.8w,0.5h}{v4}
\fmf{plain}{i1,v1}
\fmf{plain}{i2,v2}
\fmf{photon}{v4,o}
\fmf{photon,tension=0}{v1,v3}
\fmf{plain,tension=0}{v3,v4}
\fmf{photon,tension=0}{v2,v4}
\fmf{plain,tension=0}{v2,v3}
\fmf{plain,tension=0}{v1,v4}
\end{fmfgraph*} } \\
%
%
\subfigure[]{
\begin{fmfgraph*}(20,20)
\fmfleft{i1,i2}
\fmfright{o}
\fmfforce{0.2w,0.93h}{v2}
\fmfforce{0.2w,0.07h}{v1}
\fmfforce{0.2w,0.55h}{v3}
\fmfforce{0.2w,0.13h}{v5}
\fmfforce{0.8w,0.5h}{v4}
\fmf{plain}{i1,v1}
\fmf{plain}{i2,v2}
\fmf{photon}{v4,o}
\fmf{photon}{v2,v3}
\fmf{plain,left}{v3,v5}
\fmf{plain,right}{v3,v5}
\fmf{plain}{v1,v4}
\fmf{plain}{v2,v4}
\end{fmfgraph*} }
%
%
\hspace{3mm}
\subfigure[]{
\begin{fmfgraph*}(20,20)
\fmfforce{0.5w,0.2h}{v3}
\fmfforce{0.5w,0.8h}{v2}
\fmfforce{0.2w,0.5h}{v1}
\fmfforce{0.8w,0.5h}{v4}
\fmfleft{i}
\fmfright{o}
\fmf{photon}{i,v1}
\fmf{photon}{v4,o}
\fmf{plain,left=.4}{v1,v2}
\fmf{plain,right=.4}{v1,v3}
\fmf{plain,left=.4}{v2,v4}
\fmf{plain,right=.4}{v3,v4}
\fmf{photon}{v2,v3}
\end{fmfgraph*} } 
%
%
\hspace{3mm}
\subfigure[]{
\begin{fmfgraph*}(20,20)
\fmfforce{0.5w,0.2h}{v3}
\fmfforce{0.5w,0.8h}{v2}
\fmfforce{0.2w,0.5h}{v1}
\fmfforce{0.8w,0.5h}{v4}
\fmfleft{i}
\fmfright{o}
\fmf{plain}{i,v1}
\fmf{plain}{v4,o}
\fmf{plain,left=.4}{v1,v2}
\fmf{photon,right=.4}{v1,v3}
\fmf{photon,left=.4}{v2,v4}
\fmf{plain,right=.4}{v3,v4}
\fmf{plain}{v2,v3}
\end{fmfgraph*} } 
%
%
\hspace{3mm}
\subfigure[]{
\begin{fmfgraph*}(20,20)
\fmfleft{i1,i2}
\fmfright{o}
\fmf{plain}{i1,v1}
\fmf{plain}{i2,v2}
\fmf{photon}{v4,o}
\fmf{plain,tension=.3}{v2,v3}
\fmf{plain,tension=.3}{v1,v3}
\fmf{photon,tension=0}{v2,v1}
\fmf{plain,tension=.2,left}{v3,v4}
\fmf{plain,tension=.2,right}{v3,v4}
\end{fmfgraph*} }
%
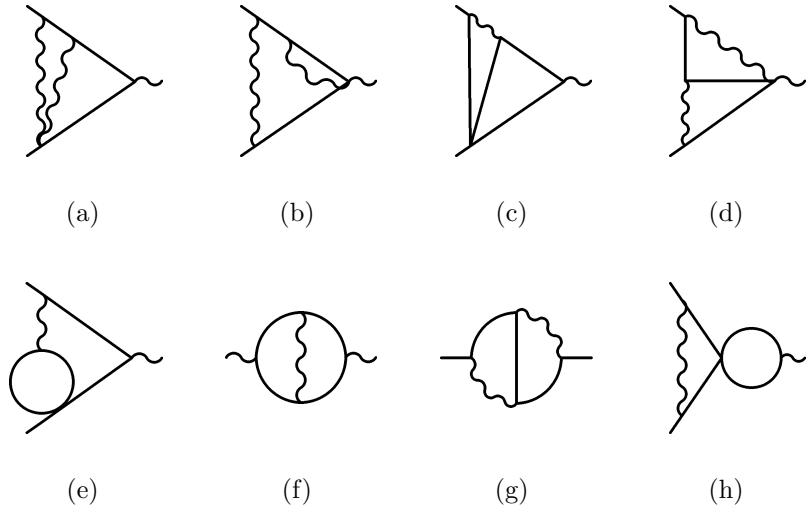
\caption{\label{fig2} The set of 8 independent 5-denominator topologies 
contained in the graphs of Fig.(\ref{fig1}). External fermion lines are
put on the mass-shell $p_{1}^{2}=p_{2}^{2}=-a$, while external wavy
lines carry an off-shell momentum $Q=p_{1}+p_{2}$. The topology (g) is
evaluated on the mass-shell. } 
\ec
\efig

%
%
\bfig
\bc
\subfigure[]{
\begin{fmfgraph*}(20,20)
\fmfleft{i1,i2}
\fmfright{o}
\fmf{plain}{i1,v1}
\fmf{plain}{i2,v2}
\fmf{photon}{v3,o}
\fmf{plain,tension=.3}{v2,v3}
\fmf{plain,tension=.3}{v1,v3}
\fmf{photon,tension=0,right=.5}{v2,v1}
\fmf{photon,tension=0,right=.5}{v1,v2}
\end{fmfgraph*} }
%
%
\subfigure[]{
\begin{fmfgraph*}(20,20)
\fmfleft{i1,i2}
\fmfright{o}
\fmf{plain}{i1,v1}
\fmf{plain}{i2,v2}
\fmf{photon}{v3,o}
\fmf{plain,tension=.3}{v2,v3}
\fmf{plain,tension=.3}{v1,v3}
\fmf{photon,tension=0}{v2,v1}
\fmf{photon,tension=0,right=.5}{v2,v3}
\end{fmfgraph*} }
%
%
\subfigure[]{
\begin{fmfgraph*}(20,20)
\fmfleft{i1,i2}
\fmfright{o}
\fmf{plain}{i1,v1}
\fmf{plain}{i2,v2}
\fmf{photon}{v3,o}
\fmf{plain,tension=.3}{v2,v3}
\fmf{plain,tension=.3}{v1,v3}
\fmf{plain,tension=0,right=.5}{v2,v1}
\fmf{plain,tension=0,right=.5}{v1,v2}
\end{fmfgraph*} }
%
%
\subfigure[]{
\begin{fmfgraph*}(20,20)
\fmfleft{i1,i2}
\fmfright{o}
\fmf{plain}{i1,v1}
\fmf{plain}{i2,v2}
\fmf{photon}{v3,o}
\fmf{plain,tension=.3}{v2,v3}
\fmf{photon,tension=.3}{v1,v3}
\fmf{plain,tension=0}{v2,v1}
\fmf{plain,tension=0,right=.5}{v2,v3}
\end{fmfgraph*} }
%
%
\subfigure[]{
\begin{fmfgraph*}(20,20)
\fmfforce{0.5w,0.2h}{v3}
\fmfforce{0.5w,0.8h}{v2}
\fmfforce{0.2w,0.5h}{v1}
\fmfforce{0.8w,0.5h}{v4}
\fmfleft{i}
\fmfright{o}
\fmf{photon}{i,v1}
\fmf{photon}{v4,o}
\fmf{plain,left=.4}{v1,v2}
\fmf{plain,right=.4}{v1,v3}
\fmf{plain,left=.4}{v2,v4}
\fmf{plain,right=.4}{v3,v4}
\fmf{photon,left=.6}{v3,v4}
\end{fmfgraph*} }
%
%
\subfigure[]{
\begin{fmfgraph*}(20,20)
\fmfforce{0.5w,0.2h}{v3}
\fmfforce{0.5w,0.8h}{v2}
\fmfforce{0.2w,0.5h}{v1}
\fmfforce{0.8w,0.5h}{v4}
\fmfleft{i}
\fmfright{o}
\fmf{plain}{i,v1}
\fmf{plain}{v4,o}
\fmf{photon,left=.4}{v1,v2}
\fmf{plain,right=.4}{v1,v3}
\fmf{photon,left=.4}{v2,v4}
\fmf{plain,right=.4}{v3,v4}
\fmf{photon,left=.6}{v3,v4}
\end{fmfgraph*} } \\
%
%
\subfigure[]{
\begin{fmfgraph*}(20,20)
\fmfforce{0.5w,0.2h}{v3}
\fmfforce{0.5w,0.8h}{v2}
\fmfforce{0.2w,0.5h}{v1}
\fmfforce{0.8w,0.5h}{v4}
\fmfleft{i}
\fmfright{o}
\fmf{plain}{i,v1}
\fmf{plain}{v4,o}
\fmf{plain,left=.4}{v1,v2}
\fmf{photon,right=.4}{v1,v3}
\fmf{plain,left=.4}{v2,v4}
\fmf{plain,right=.4}{v3,v4}
\fmf{plain,left=.6}{v3,v4}
\end{fmfgraph*} }
%
%
\subfigure[]{
\begin{fmfgraph*}(20,20)
\fmfforce{0.5w,0.2h}{v3}
\fmfforce{0.5w,0.8h}{v2}
\fmfforce{0.2w,0.5h}{v1}
\fmfforce{0.8w,0.5h}{v4}
\fmfleft{i}
\fmfright{o}
\fmf{plain}{i,v1}
\fmf{plain}{v4,o}
\fmf{plain,left=.4}{v1,v2}
\fmf{photon,right=.4}{v1,v3}
\fmf{plain,left=.4}{v2,v4}
\fmf{photon,right=.4}{v3,v4}
\fmf{photon,left=.6}{v3,v4}
\end{fmfgraph*} }
%
%
\subfigure[]{
\begin{fmfgraph*}(20,20)
\fmfleft{i1,i2}
\fmfright{o}
\fmf{plain}{i1,v1}
\fmf{plain}{i2,v2}
\fmf{photon}{v3,o}
\fmf{plain,tension=.3}{v2,v3}
\fmf{plain,tension=.3}{v1,v3}
\fmf{photon,tension=0}{v2,v1}
\fmf{plain,right=45}{v3,v3}
\end{fmfgraph*} }
%
%
\subfigure[]{
\begin{fmfgraph*}(20,20)
\fmfleft{i}
\fmfright{o}
\fmf{photon}{i,v1}
\fmf{photon}{v3,o}
\fmf{plain,tension=.2,left}{v1,v2}
\fmf{plain,tension=.2,right}{v1,v2}
\fmf{plain,tension=.2,left}{v2,v3}
\fmf{plain,tension=.2,right}{v2,v3}
\end{fmfgraph*} }
%
%
\subfigure[]{
\begin{fmfgraph*}(20,20)
\fmfbottom{v5}
\fmftop{v4}
\fmfleft{i}
\fmfright{o}
\fmf{plain}{i,v1}
\fmf{photon}{v3,o}
\fmf{plain}{v5,v2} 
\fmf{phantom}{v2,v4} 
\fmf{plain,tension=.2,left}{v1,v2}
\fmf{photon,tension=.2,right}{v1,v2}
\fmf{plain,tension=.2,left}{v2,v3}
\fmf{plain,tension=.2,right}{v2,v3}
\end{fmfgraph*} }
%
%
\subfigure[]{
\begin{fmfgraph*}(20,20)
\fmfbottom{v5}
\fmftop{v4}
\fmfleft{i}
\fmfright{o}
\fmf{plain}{i,v1}
\fmf{plain}{v3,o}
\fmf{plain}{v5,v2} 
\fmf{plain}{v2,v4} 
\fmf{plain,tension=.2,left}{v1,v2}
\fmf{photon,tension=.2,right}{v1,v2}
\fmf{photon,tension=.2,left}{v2,v3}
\fmf{plain,tension=.2,right}{v2,v3}
\end{fmfgraph*} }
%
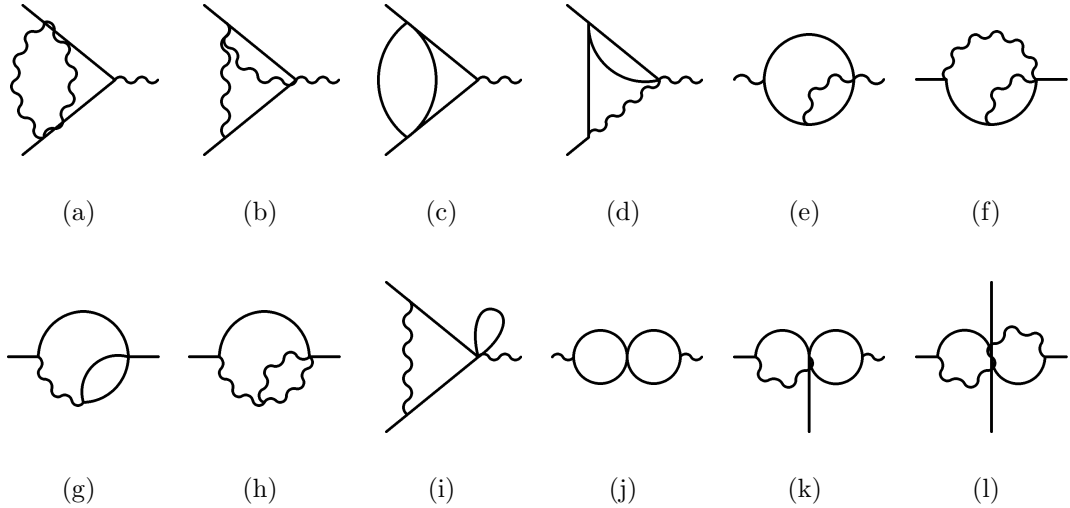
\caption{\label{fig3} The set of 12 independent 4-denominator 
(sub)topologies coming from the 5-denominator topologies of 
Fig.(\ref{fig2}).} 
\ec
\efig

%
%
\bfig
\bc
\subfigure[]{
\begin{fmfgraph*}(20,20)
\fmfleft{i}
\fmfright{o}
\fmf{photon}{i,v1}
\fmf{photon}{v2,o}
\fmf{plain,tension=.15,left}{v1,v2}
\fmf{photon,tension=.15}{v1,v2}
\fmf{plain,tension=.15,right}{v1,v2}
\end{fmfgraph*} } 
%
%
\subfigure[]{
\begin{fmfgraph*}(20,20)
\fmfleft{i}
\fmfright{o}
\fmf{plain}{i,v1}
\fmf{plain}{v2,o}
\fmf{plain,tension=.15,left}{v1,v2}
\fmf{photon,tension=.15}{v2,v1}
\fmf{photon,tension=.15,right}{v1,v2}
\end{fmfgraph*} } 
%
%
\subfigure[]{
\begin{fmfgraph*}(20,20)
\fmfleft{i}
\fmfright{o}
\fmf{photon}{i,v1}
\fmf{photon}{v2,o}
\fmf{plain,tension=.22,left}{v1,v2}
\fmf{plain,tension=.22,right}{v1,v2}
\fmf{plain,right=45}{v2,v2}
\end{fmfgraph*} } 
%
%
\subfigure[]{
\begin{fmfgraph*}(20,20)
\fmfleft{i}
\fmfright{o}
\fmf{plain}{i,v1}
\fmf{plain}{v2,o}
\fmf{plain,tension=.22,left}{v1,v2}
\fmf{photon,tension=.22,right}{v1,v2}
\fmf{plain,right=45}{v2,v2}
\end{fmfgraph*} }
%
%
\subfigure[]{
\begin{fmfgraph*}(20,20)
\fmfleft{i}
\fmfright{o}
\fmfforce{0.5w,0.1h}{v1}
\fmfforce{0.25w,0.62h}{v3}
\fmfforce{0.5w,0.9h}{v7}
\fmfforce{0.74w,0.62h}{v11}
\fmf{plain,left=.1}{v1,v3}
\fmf{plain,left=.5}{v3,v7}
\fmf{plain,left=.5}{v7,v11}
\fmf{plain,left=.1}{v11,v1}
\fmf{photon}{v1,v7}
\end{fmfgraph*} }
%
%
\subfigure[]{
\begin{fmfgraph*}(20,20)
\fmfleft{i}
\fmfright{o}
\fmf{plain}{i,v1}
\fmf{plain}{v2,o}
\fmf{plain,tension=.15,left}{v1,v2}
\fmf{plain,tension=.15}{v1,v2}
\fmf{plain,tension=.15,right}{v1,v2}
\end{fmfgraph*} }  \\
%
\subfigure[]{ 
\begin{fmfgraph*}(20,20)
\fmfleft{i}
\fmfright{o}
\fmf{phantom}{i,v1}
\fmf{phantom}{v1,o}
\fmf{plain,right=45}{v1,v1}
\fmf{plain,left=45}{v1,v1}
\end{fmfgraph*} } 
%
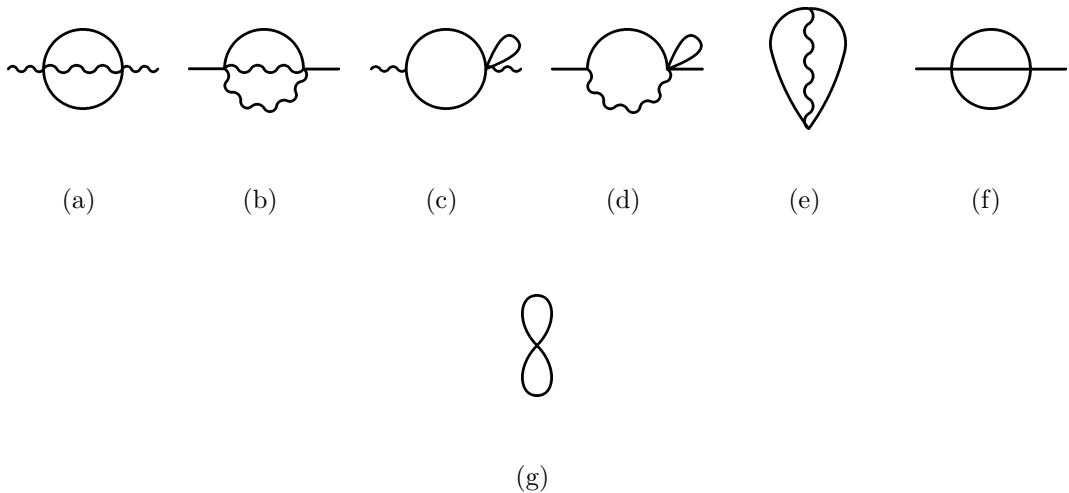
\caption{\label{fig4} The set of 6 independent 3-denominator 
(sub)topologies, (a)--(f), coming from the 4-denominator topologies of 
Fig.(\ref{fig3}) and the only non-vanishing topology at 2-denominator
level, (g), product of two tadpoles of squared mass $a$.} 
\ec
\efig

\subsection{The MI's. \label{MIs}}
For each independent topology we write the identities among the 
integrals of the associated family and solve them in terms of MI's 
according to the previous discussion, see section \ref{Master}. The 
resulting MI's are shown in Fig. (\ref{fig5}). The MI's are represented 
with a graph very much equal to the graph representing their topology, 
but occasional with some ``decoration". When no decoration is present, 
the corresponding MI is nothing but the corresponding full scalar graph
(first power of all the propagators, numerator equal to 1); that is the
case, for instance, of (a), (c), (e) etc. of Fig. (\ref{fig5}). When a 
propagator line is decorated with a dot, it appears squared in the MI, 
as in the case of (d),(h) of Fig. (\ref{fig5}). Finally, the decoration
can be a scalar product involving at least one loop momentum; the scalar
product appears in the numerator of the integrand of the corresponding 
MI, as in (b), (e) etc. of Fig. (\ref{fig5}).

As anticipated, there are topologies with two MI's, as the topology 
(b) of Fig. (\ref{fig1bis}) which has the two MI's (a), (b) of Fig. 
(\ref{fig5}), topologies with a single MI, as (a) of Fig. (\ref{fig3}) 
which has the MI (i) of Fig. (\ref{fig5}), and topologies without MI's,
i.e. topologies whose associated integrals can all be expressed in terms
of the MI's of their subtopologies; that is the case, for instance, of 
the topology (a) of Fig. (\ref{fig1bis}). 

As a last remark, let us recall that there is some arbitrariness on 
the actual scalar integrals to be choosen as MI's; in the case 
of Fig. (\ref{fig5}) some of the MI's correspond to graphs decorated 
with dots, other to graphs decorated by scalar products. The choice, 
by no means mandatory but rather somewhat accidental, was suggested by 
the convenience of later use. 

\bfig
\bc
\subfigure[]{
\begin{fmfgraph}(20,20) 
\fmfleft{i1,i2}
\fmfright{o}
\fmf{plain}{i1,v1}
\fmf{plain}{i2,v2}
\fmf{photon}{v5,o}
\fmf{plain,tension=.3}{v2,v3}
\fmf{plain,tension=.3}{v3,v5}
\fmf{plain,tension=.3}{v1,v4}
\fmf{plain,tension=.3}{v4,v5}
\fmf{photon,tension=0}{v2,v4}
\fmf{photon,tension=0}{v1,v3}
\end{fmfgraph}} 
%
%
\hspace{5mm}
\subfigure[]{
\begin{fmfgraph*}(20,20) 
\fmfleft{i1,i2}
\fmfright{o}
\fmf{plain}{i1,v1}
\fmf{plain}{i2,v2}
\fmf{photon}{v5,o}
\fmflabel{$(k_{1} \cdot k_{2})$}{o}
\fmf{plain,tension=.3}{v2,v3}
\fmf{plain,tension=.3}{v3,v5}
\fmf{plain,tension=.3}{v1,v4}
\fmf{plain,tension=.3}{v4,v5}
\fmf{photon,tension=0}{v2,v4}
\fmf{photon,tension=0}{v1,v3}
\end{fmfgraph*}}  
%
%
\hspace{15mm}
\subfigure[]{
\begin{fmfgraph}(20,20)
\fmfleft{i1,i2}
\fmfright{o}
\fmf{plain}{i1,v1}
\fmf{plain}{i2,v2}
\fmf{photon}{v4,o}
\fmf{plain,tension=.4}{v2,v3}
\fmf{plain,tension=.2}{v3,v4}
\fmf{plain,tension=.15}{v1,v4}
\fmf{photon,tension=0}{v2,v1}
\fmf{photon,tension=0}{v1,v3}
\end{fmfgraph} }
%
%
\hspace{5mm}
\subfigure[]{
\begin{fmfgraph}(20,20)
\fmfleft{i1,i2}
\fmfright{o}
\fmfforce{0.5w,0.3h}{v5}
\fmf{plain}{i1,v1}
\fmf{plain}{i2,v2}
\fmf{photon}{v4,o}
\fmf{plain,tension=.4}{v2,v3}
\fmf{plain,tension=.2}{v3,v4}
\fmf{plain,tension=.15}{v1,v4}
\fmf{photon,tension=0}{v2,v1}
\fmf{photon,tension=0}{v1,v3}
\fmfv{decor.shape=circle,decor.filled=full,decor.size=.1w}{v5}
\end{fmfgraph} } \\
%
%
\subfigure[]{
\begin{fmfgraph}(20,20)
\fmfleft{i1,i2}
\fmfright{o}
\fmf{plain}{i1,v1}
\fmf{plain}{i2,v2}
\fmf{photon}{v3,o}
\fmf{plain,tension=.3}{v2,v3}
\fmf{plain,tension=.3}{v1,v3}
\fmf{plain,tension=0,right=.5}{v2,v1}
\fmf{plain,tension=0,right=.5}{v1,v2}
\end{fmfgraph} }
%
%
\hspace{5mm}
\subfigure[]{
\begin{fmfgraph*}(20,20)
\fmfleft{i1,i2}
\fmfright{o}
\fmf{plain}{i1,v1}
\fmf{plain}{i2,v2}
\fmf{photon}{v3,o}
\fmflabel{$(p_{2} \cdot k_{1})$ }{o}
\fmf{plain,tension=.3}{v2,v3}
\fmf{plain,tension=.3}{v1,v3}
\fmf{plain,tension=0,right=.5}{v2,v1}
\fmf{plain,tension=0,right=.5}{v1,v2}
\end{fmfgraph*} }
%
%
\hspace{15mm}
\subfigure[]{
\begin{fmfgraph}(20,20)
\fmfleft{i1,i2}
\fmfright{o}
\fmf{plain}{i1,v1}
\fmf{plain}{i2,v2}
\fmf{photon}{v3,o}
\fmf{plain,tension=.3}{v2,v3}
\fmf{photon,tension=.3}{v1,v3}
\fmf{plain,tension=0}{v2,v1}
\fmf{plain,tension=0,right=.5}{v2,v3}
\end{fmfgraph} }
%
%
\hspace{5mm}
\subfigure[]{
\begin{fmfgraph}(20,20)
\fmfleft{i1,i2}
\fmfright{o}
\fmfforce{0.22w,0.5h}{v4}
\fmf{plain}{i1,v1}
\fmf{plain}{i2,v2}
\fmf{photon}{v3,o}
\fmf{plain,tension=.3}{v2,v3}
\fmf{photon,tension=.3}{v1,v3}
\fmf{plain,tension=0}{v2,v1}
\fmf{plain,tension=0,right=.5}{v2,v3}
\fmfv{decor.shape=circle,decor.filled=full,decor.size=.1w}{v4}
\end{fmfgraph}   } \\
%
%
\subfigure[]{
\begin{fmfgraph}(20,20)
\fmfleft{i1,i2}
\fmfright{o}
\fmf{plain}{i1,v1}
\fmf{plain}{i2,v2}
\fmf{photon}{v3,o}
\fmf{plain,tension=.3}{v2,v3}
\fmf{plain,tension=.3}{v1,v3}
\fmf{photon,tension=0,right=.5}{v2,v1}
\fmf{photon,tension=0,right=.5}{v1,v2}
\end{fmfgraph} }
%
%
\hspace{7mm}
\subfigure[]{
\begin{fmfgraph}(20,20)
\fmfleft{i1,i2}
\fmfright{o}
\fmf{plain}{i1,v1}
\fmf{plain}{i2,v2}
\fmf{photon}{v3,o}
\fmf{plain,tension=.3}{v2,v3}
\fmf{plain,tension=.3}{v1,v3}
\fmf{photon,tension=0}{v2,v1}
\fmf{photon,tension=0,right=.5}{v2,v3}
\end{fmfgraph} }
%
%
\hspace{7mm}
\subfigure[]{
\begin{fmfgraph}(20,20)
\fmfleft{i}
\fmfright{o}
\fmf{photon}{i,v1}
\fmf{photon}{v3,o}
\fmf{plain,tension=.2,left}{v1,v2}
\fmf{plain,tension=.2,right}{v1,v2}
\fmf{plain,tension=.2,left}{v2,v3}
\fmf{plain,tension=.2,right}{v2,v3}
\end{fmfgraph} }
%
%
\hspace{7mm}
\subfigure[]{
\begin{fmfgraph}(20,20)
\fmfleft{i}
\fmfright{o}
\fmf{plain}{i,v1}
\fmf{plain}{v2,o}
\fmf{plain,tension=.15,left}{v1,v2}
\fmf{plain,tension=.15}{v1,v2}
\fmf{plain,tension=.15,right}{v1,v2}
\end{fmfgraph} }  \\
%
%
\subfigure[]{
\begin{fmfgraph}(20,20)
\fmfleft{i}
\fmfright{o}
\fmf{photon}{i,v1}
\fmf{photon}{v2,o}
\fmf{plain,tension=.15,left}{v1,v2}
\fmf{photon,tension=.15}{v1,v2}
\fmf{plain,tension=.15,right}{v1,v2}
\end{fmfgraph} }
%
%
\hspace{5mm}
\subfigure[]{
\begin{fmfgraph*}(20,20)
\fmfleft{i}
\fmfright{o}
\fmf{photon}{i,v1}
\fmf{photon}{v2,o}
\fmflabel{$(k_{1} \cdot k_{2})$ }{o}
\fmf{plain,tension=.15,left}{v1,v2}
\fmf{photon,tension=.15}{v1,v2}
\fmf{plain,tension=.15,right}{v1,v2}
\end{fmfgraph*}  } 
%
%
\hspace{15mm}
\subfigure[]{
\begin{fmfgraph}(20,20)
\fmfleft{i}
\fmfright{o}
\fmf{plain}{i,v1}
\fmf{plain}{v2,o}
\fmf{plain,tension=.15,left}{v1,v2}
\fmf{photon,tension=.15}{v2,v1}
\fmf{photon,tension=.15,right}{v1,v2}
\end{fmfgraph} } 
%
%
\hspace{5mm}
\subfigure[]{
\begin{fmfgraph}(20,20)
\fmfleft{i}
\fmfright{o}
\fmf{photon}{i,v1}
\fmf{photon}{v2,o}
\fmf{plain,tension=.22,left}{v1,v2}
\fmf{plain,tension=.22,right}{v1,v2}
\fmf{plain,right=45}{v2,v2}
\end{fmfgraph} } 
%
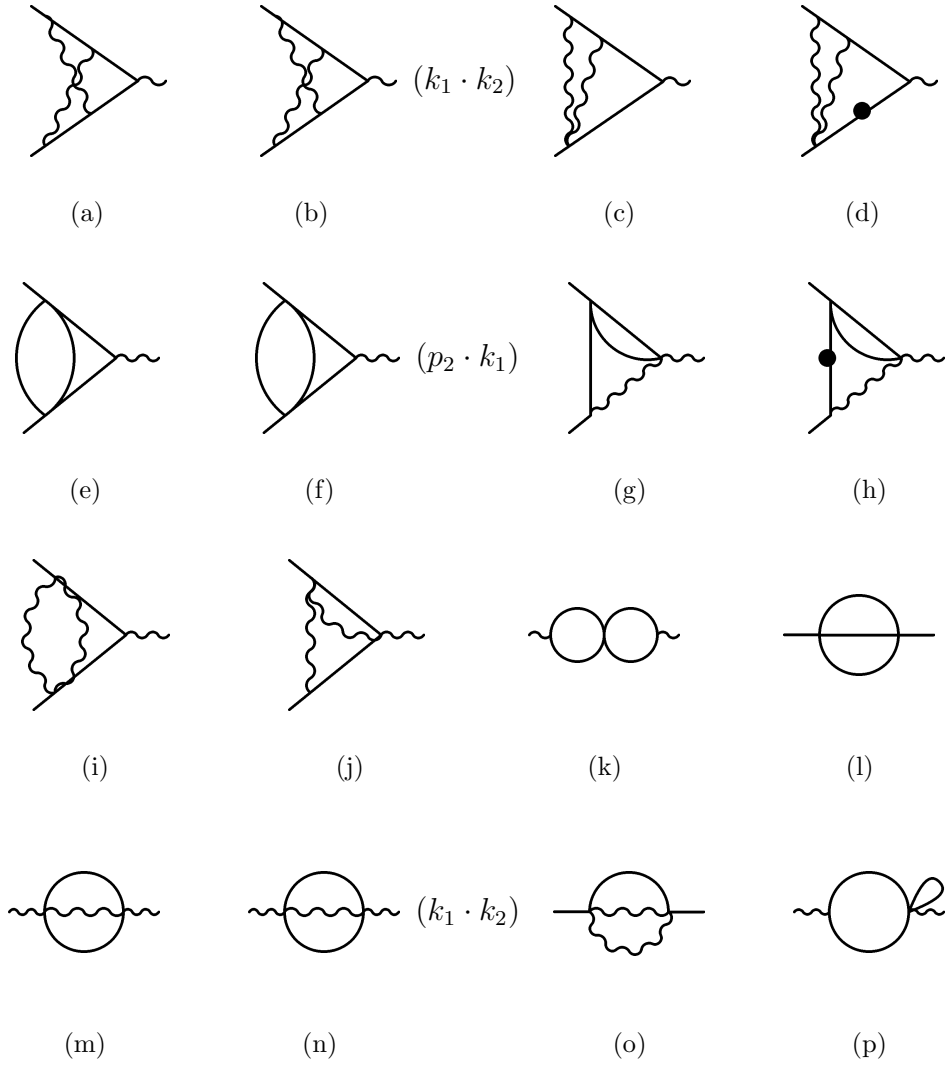
\caption{\label{fig5} The set of 16 Master Integrals (MIs). As explained
in section \ref{MIs}, the diagrams shown are a graphical representation
of the corresponding $D$-regularized integral. A dot on a propagator 
line means that the corresponding propagator is squared and 
an explicitly written scalar product means that the corresponding 
$D$-regularized integral has that scalar product in the numerator of 
the integrand.}
\ec
\efig

\section{Calculation of the MIs. The system of differential equations 
\label{DiffEqs}}

Once all the MIs of a given topology are obtained, the 
problem of their calculation arises. We will address the problem 
by the differential equations method, which 
turns out to be a really very powerful tool. 
The use of differential equations in one of the internal masses was 
first proposed out in \cite{Kot}, then extended to more general 
differential equations in any of Mandelstam variables 
in \cite{Rem1} and successively used in \cite{Rem2} for the MI's 
of the sunrise diagram with arbitrary internal masses. 
An application to the 4-point functions with 
massless internal propagators was worked out in \cite{Rem3} for the 
2-loop case and it brought to the complete evaluation of the master 
integrals for the planar \cite{Rem3,Rem4,Rem5} and non planar topologies
\cite{Rem6}. In this paper, we will write (and solve) the 
differential equations in the momentum transfer in the case of 3-point 
functions with massive fermionic propagators in QED, keeping the 
external electron legs on the mass shell. 

Let us summarize briefly the idea of the method. To begin with, 
consider any scalar integral $F(s_i)$, (we will be interested 
here in the MI's, but what follows applies to any scalar integral as 
well) defined as 
\be
F(s_i) \, = \, \int \{ d^{D}k_{1} \} \{ d^{D}k_{2} \} \, 
     \frac{ S_{1}^{n_{1}} \cdots S_{q}^{n_{q}} } 
          { {\mathcal D}_{1}^{m_{1}} \cdots {\mathcal D}_{t}^{m_{t}} } \, ;
\label{b21}
\ee 
$F(s_i)$ depends in general on the three external kinematical 
invariants $s_1=-p_1^2,\ s_2=-p_2^2$ and $s_3=-Q^2=-(p_1+p_2)^2$, 
where $p_1^2,\ p_2^2$ will be later constrained on the mass shell 
$p_1^2=p_2^2=-a$. 

Let us construct the following quantities: 
\be
O_{jk}(s_i) = p_{j}^{\mu} \frac{\partial }{\partial p_{k}^{\mu}} 
                         F(s_i) \, .
\label{b22}
\ee

As $F(s_i)$ depends on the Mandelstam invariants $s_i$, by the chain 
differentiation rule we have 
\be
  O_{jk}(s_i) = p_{j}^{\mu} \sum_{\xi} 
          \frac{\partial s_{\xi}}{\partial p_{k} ^{\mu}} 
          \frac{\partial }{\partial s_{\xi}} F(s_i) 
   =  \sum_{\xi} a_{\xi,jk}(s_{l}) \frac{\partial }{\partial s_{\xi}} 
                                             F(s_i)\, ,
\label{b23}
\ee
where the functions $a_{\xi,jk}(s_{l})$ are linear combinations of 
the Mandelstam invariants $s_i$.

As $j$ and $k$ take the two values $1,2$ we obtain in that way a system 
of 4 linear equations (not all linear independent), which we can solve 
for the three derivatives $ \frac{\partial}{\partial s_\xi}F(s_i) $; 
we have in particular 
\begin{equation} 
  \frac{\partial}{\partial Q^2} F(s_i) = \left[ 
     A \left( p_1^\mu \frac{\partial}{\partial p_1^\mu} 
             + p_2^\mu \frac{\partial}{\partial p_2^\mu} \right) 
   + B \left( p_1^\mu \frac{\partial}{\partial p_2^\mu} 
             + p_2^\mu \frac{\partial}{\partial p_1^\mu} \right) 
      \right] F(s_i) \, , 
\label{ddQ2} 
\end{equation} 
where 
\bea
A & = & \frac{1}{4} \left[ \frac{1}{Q^{2}} + \frac{1}{Q^{2}+4a} \right]
\, , \\
B & = & \frac{1}{4} \left[ \frac{1}{Q^{2}} - \frac{1}{Q^{2}+4a} \right]
\, .
\eea

Assume now that $F(s_i)$ is a master integral for some given topology. 
We can now substitute the right-hand side of Eq. (\ref{b21}) in the 
right-hand side of Eq. (\ref{ddQ2}) and perform the direct 
differentiation of the integrand. It is clear that we obtain a 
combination of several 
integrals, all belonging to the same topology as $F(s_i)$; therefore, 
we can use the solutions of the IBP and other identities for that 
topology and express everything in the r.h.s. of Eq. (\ref{ddQ2}) 
in terms of the MI's for the considered topology and its subtopologies. 
If there are several different MI's for that topology, the procedure 
can be repeated for all the other MI's as well. In so doing one obtains
a system of linear differential equations in $Q^2$ for $F(s_i)$ and the
other MI's (if any), expressing their $Q^2$-derivatives in terms of the
MI's of the considered topology and of its subtopologies; due to the 
presence of the MI's of the subtopologies the equations are in general 
non-homogeneous. 

At this point we can impose the mass-shell conditions, and the general 
structure of the system reads 
\bea 
\frac{\partial}{\partial Q^2} M_i(D,a,Q^2) & = & \sum_j A_j(D,a,Q^2) 
M_j(D,a,Q^2) \nn\\
& & \qquad \qquad + \sum_k B_k(D,a,Q^2) N_k(D,a,Q^2) \ , 
\label{Q2sys} 
\eea 
where the $M_i(D,a,Q^2)$ are the MI's of the topology with the electron 
legs on the mass-shell, $N_k(D,a,Q^2)$ the MI's of the subtopologies, we
have made explicit the dipendence on $D$ for later use (note that the 
above equations are exact in $D$) and the coefficients $A_j(D,a,Q^2), 
B_k(D,a,Q^2)$ are rational factors depending on $D, Q^2$ and the 
electron squared mass $a$. As will be apparent by the examples, the
singularities of $A_j(D,a,Q^2), B_k(D,a,Q^2)$ in the variable $Q^2$, 
such as $1/(Q^2+4a)$ and $1/Q^2$, and correspond to the thresholds and 
pseudothresholds of the corresponding Feynman graphs. 

It is clear that the procedure can be repeated in principle for the 
other Mandelstam variables as well. We are not interested in this 
further equations, as in the case we are considering the other 
Mandelstam variables are the invariant masses of the electrons, which 
we keep frozen on the mass-shell. 

As already observed, the Eqs. (\ref{Q2sys}) for the MI's $M_i(D,a,Q^2)$
of a given topology are not homogeneous, as they may involve the MI's 
$N_k(D,a,Q^2)$ of the subtopologies. It is therefore natural to proceed
bottom up, starting from the equations for the MI's of the simplest 
topologies (i.e. with less denominators), solving those equations and 
using the results within the equations for the MI's of the more 
complicated topologies with additional propagators, whose 
non-homogeneous part can then be considered as known.

\subsubsection{The boundary conditions \label{boundary}}

Some comments on boundary conditions.
As already observed, the coefficients of the differential equations 
Eq. (\ref{Q2sys}) are in general singular at $Q^2=0$ and $Q^2=-4a$; 
correspondingly, the solutions of the equations can have a singular 
behaviour in those points. But we know that the Vertex integrals are 
regular in $Q^2$ at $Q^2=0$ when the electron lines are on the 
mass shell, a qualitative result which can easily verified, when 
needed, by direct inspection of the very definition of the amplitudes as
loop integrals. It turns out that the qualitative information provided 
by the regularity behaviour (implying the absence, in the $Q^2\to0$ 
limit, of terms in $1/Q^2$ or $\ln{Q^2}$) is completely sufficient for 
the quantitative determination of the otherwise arbitrary integration 
constants which naturally arise when solving a system of differential 
equations.

\subsection{Laurent series expansion in ${\mathbf \epsilon}$
\label{Laurent}}

The system of differential equations Eq. (\ref{Q2sys}) is exact in $D$, 
but we are interested, in any case, on the Laurent expansion of the 
solutions in powers of $\epsilon=(4-D)/2$. 
It turns out that our 2-loop integrals have up to a double pole in
$\epsilon$ (which can be of ultraviolet or infrared origin), so that 
quite in general we will expand the two loop MI's as 
\begin{equation} 
M_i(D,a,Q^2) = \sum_{j=-2}^{n} \, {\epsilon^{j}} M_i^{(j)}(a,Q^2) 
+ {\mathcal O} (\epsilon^{(n+1)}) \ , 
\label{b40} 
\end{equation} 
where $n$ is the required order in $\epsilon$. We will present for all 
the considered integrals the coefficients of the $\epsilon$ expansion up
to the zeroth order included. In some cases, however, we had to expand 
intermediate results up to the term of fourth order in $\epsilon$. 
That depends on the fact that some of the MI's, which appears in the
non-homogeneous part of a system of differential equations for 
more complicated MI's, can be multiplied by coefficients which are also 
singular in $\epsilon$. 

When expanding systematically in $\epsilon$ all the MI's (including 
those appearing in the non-homogeneous part) and all the $D$-dependent 
coefficients of Eq. (\ref{Q2sys}), one obtains a system of chained 
equations formed by the equations valid for each power of $\epsilon$. 
The first equation corresponds to the coefficient of double pole in 
$\epsilon$ of the equation, and involves only the coefficients 
$M_i^{(-2)}(a,Q^2)$ as unknown; 
the next equation, corresponding to the single pole in $\epsilon$, 
involves the $M_i^{(-1)}(a,Q^2)$ as unknown, but can in general involve
$M_i^{(-2)}(a,Q^2)$ if some of the coefficients contains a power of 
$\epsilon$; such a term in $M_i^{(-2)}(a,Q^2)$ 
can be considered as known once the equation for the double pole 
has been solved. For the subsequent equations we have the same
structure: they involves the coefficient $M_i^{(j)}(a,Q^2)$ at the order
$j$ in $\epsilon$ as unknown and the previous coefficients as known 
non-homogeneous terms. 

Let us note that the homogeneous part of all the equations arising 
from the $\epsilon$ expansion of Eq. (\ref{Q2sys}) is always the same 
and obviously identical to the homogeneous part of Eq. (\ref{Q2sys}) at
$D=4$, i.e. $\epsilon=0$. It is natural to look for the solutions of the
chained equations by means of the Euler's method of the variation of the
constants, using repeatedly the solutions of the homogeneous equation,
as we will show in some examples in the next section. 

General algorithms for the solution of the homogeneous equations 
are not available; it turns out however that in all the cases considered
in this paper the homogeneous equations at $D=4$ have almost 
trivial solutions, so that Euler's formula can immediately be written. 
With the change of variable 
\begin{equation} 
 x = \frac{\sqrt{Q^2+4a} - \sqrt{Q^2} } 
          {\sqrt{Q^2+4a} + \sqrt{Q^2} } \ , 
\label{Q2tox} 
\end{equation} 
all integrations can further be carried out in closed analytic form, 
the result being a combination of the Harmonic Polylogarithms 
introduced in~\cite{Polylog} (see also \cite{Polylog3} for their 
numerical evaluation), a generalization of the already widely used 
Nielsen's Polylogarithms~\cite{Nielsen,Kolbig}. 

As a last remark some comments on the arbitrariness of the choice of 
the MI's. For the topologies we considered, we had at most 2
MI's, which means that we had to solve in principle a linear system of 
two equations or an equivalent second-order differential equation. 
However, the freedom in the explicit choice of the MI's can play an 
essential role in simplifying the calculation. It turns out, in fact, 
that if we choose the two MI's with a different leading singular 
behaviour in $\epsilon$, the system of the two coupled first-order 
linear differential equations does in fact decouple.\footnote{The 
decoupling can be, in some cases, exact in $\epsilon$, which means that 
a combination of integrals diagonalize the system without expanding it 
in powers of $\epsilon$. More in general, the homogeneous equation at 
$D=4$ is brought to acquire a triangular form, with the first 
differential equation which contains only one of the MI's and the 
second equation which involves both MI's.} As a result, instead of 
solving a second-order differential equation we can solve simply two 
first-order equations. 
We will show how to exploit this possibility 
in the solution of the systems considered in sections \ref{2exempl} and
\ref{3exempl}.

\section{Explicit calculations \label{Exempla}}

In this section the equations for three topologies and their solutions 
are discussed in some 
details. We chose a 4-, a 5- and a 6-dominator topology, shown 
respectively in Fig. (\ref{fig3}) (a), Fig. (\ref{fig4}) (a) and Fig. 
(\ref{fig1bis}) (b), to illustrate the algorithm for the solution of the
corresponding system of differential equations.

The 4-denominator topology is the simpler among the three cases, since 
it has only one MI. Correspondingly, the system in Eq. (\ref{Q2sys})
reduces to a single first-order linear differential equation -- whose 
solution is therefore trivial. 

The other two topologies are more difficult. They have both two MI's and
therefore, in both cases, we must solve in principle a system of two 
first-order coupled linear differential equations. 
As we have already remarked in section \ref{Laurent}, the explicit form 
of the system depends on the choice of the pair of MI's. 
Indeed, we choose in both cases two MI's with different leading 
behaviour in $\epsilon$, such that the system decouples order 
by order in $\epsilon$.

Before to go on, two remarks have to be done.

The first one is the following. We are interested in really 2-loop 
diagrams, but, as we have seen, in the reduction to the MI's we 
encountered topologies which factorize in the product of two 1-loop 
topologies.
In particular, two MI's have this structure; they are shown in Fig.
(\ref{fig5}) (k) and (p) and they consist respectively in a product of 
two bubbles in the $Q$-channel and of one bubble in the $Q$-channel and
a Tadpole. The same algorithm explained in sections 
\ref{Master}--\ref{DiffEqs} was applied, therefore, to the 1-loop 
problem and the results are shown in appendix \ref{app2}, where we
discuss in particular the solution of the differential equation for the
bubble with two massive propagators with equal squared mass $a$.

The second remark concerns the 3-denominator MI's. They constitute the
simplest non-trivial 2-loop topologies of the entire pyramid of MI's, 
and thus they are present in the non-homogeneous part of the systems of 
differential equations for all the other topologies. For the two MI's of
the Sunrise with two equal-mass and one mass-less propagator, Fig.
(\ref{fig5}) (m) and (n), a system of two non-homogeneous first-order
differential equations can be established, the non-homogeneous part
consisting essentially on the product of two massive 1-loop Tadpoles, 
times a ratio of polynoms in $Q^{2}$, $a$ and $\epsilon$. But, if we
consider the Sunrise with two mass-less propagators, analogous to that
one in Fig. (\ref{fig5}) (o), but with the external leg off-shell, this 
is no longer possible. The resulting system is, in fact, homogeneous, as
we can understand from the fact that contracting a propagator line we 
have at least a product with a mass-less Tadpole, which vanishes in 
dimensional regularization. In this case the conditions of regularity of
the integrals in $Q^{2}=0$ are not sufficient to determine the boundary
conditions, as explained in section \ref{boundary}. In this situation we
are forced to evaluate the integrals by direct integration, as we did 
for the MI in Fig. (\ref{fig5}) (o); but that is not a problem, given 
the simplicity of the integrals.

\subsection{The full calculation for the topology in Fig. (\ref{fig3}) 
(a)}

The topology in Fig. (\ref{fig3}) (a) has only one MI. We choose the
simpler one, i.e. the scalar integral itself, Fig. (\ref{fig5}) (e):
\be
F(\epsilon,a,Q^{2}) = 
\parbox{15mm}{\begin{fmfgraph*}(15,15)
\fmfleft{i1,i2}
\fmfright{o}
\fmf{plain}{i1,v1}
\fmf{plain}{i2,v2}
\fmf{photon}{v3,o}
\fmf{plain,tension=.3}{v2,v3}
\fmf{plain,tension=.3}{v1,v3}
\fmf{photon,tension=0,right=.5}{v2,v1}
\fmf{photon,tension=0,right=.5}{v1,v2}
\end{fmfgraph*} } = \mu_{0}^{2(4-D)} 
\int \{ d^{D}k_{1} \} \{ d^{D}k_{2} \}
\frac{1}{{\mathcal D}_{1} {\mathcal D}_{2} {\mathcal D}_{14} {\mathcal D}_{15}} \, ,
\label{d19}
\ee
where the exlicit expressions of the denominators ${\mathcal D}_{i}$ are 
given in appendix \ref{app1}.

The first-order linear differential equation in the variable $Q^{2}$, 
which we obtain by the methods described in the previous sections, 
reads 
\bea
\frac{dF(\epsilon,a,Q^{2})}{dQ^{2}} \! & = & - \frac{1}{2} \left[ 
\frac{1}{Q^{2}} \! - \frac{(1 - 4 \epsilon)}{(Q^{2}+4a)} \right] 
F (\epsilon,a,Q^{2})\nn\\
& & - \frac{(2 - 3 \epsilon)}{4a} 
\left[ \frac{1}{Q^{2}} - \frac{1}{(Q^{2}+4a)} \right] \,
\parbox{15mm}{\begin{fmfgraph*}(15,15)
\fmfleft{i}
\fmfright{o}
\fmf{plain}{i,v1}
\fmf{plain}{v2,o}
\fmf{plain,tension=.15,left}{v1,v2}
\fmf{photon,tension=.15}{v2,v1}
\fmf{photon,tension=.15,right}{v1,v2}
\end{fmfgraph*}} \, ,
\label{d20}
\eea
where the 3-denominator diagram on the non-homogeneous part of the
equation is the MI of Fig. (\ref{fig5}) (o), function only of the
squared mass $a$; its expansion in Laurent series of $\epsilon$ is given
in Eqs. (\ref{e10}--\ref{e12}).

As we can notice, from Eq. (\ref{d20}) we see that its solutions can be 
singular at $Q^2=-4a$ and $Q^2=0$. The integral we are considering, 
Eq. (\ref{d19}), is indeed singular at the physical threshold $Q^2=-4a$,
but regular at the pseudothreshold $Q^2=0$. This allows us to get the 
initial condition directely from the differential equation. In fact, 
multiplying Eq. (\ref{d20}) for $Q^2$ and taking the limit 
$Q^{2} \rightarrow 0$, the left-hand side simply vanishes; the 
right-hand side gives us:
\be
F(\epsilon,a,Q^2=0) = 
- \frac{(2- 3 \epsilon)}{2a} \, \, 
\parbox{20mm}{\begin{fmfgraph*}(15,15)
\fmfleft{i}
\fmfright{o}
\fmf{plain}{i,v1}
\fmf{plain}{v2,o}
\fmf{plain,tension=.15,left}{v1,v2}
\fmf{photon,tension=.15}{v2,v1}
\fmf{photon,tension=.15,right}{v1,v2}
\end{fmfgraph*}}
\label{d21}
\ee

Even if in this particular case it is possible to find a solution of Eq.
(\ref{d20}) exact in $D=4-2 \epsilon$, \cite{PhDThesis}, in this section
we look for a solution expanded in Laurent series of $\epsilon$:
\be
F(\epsilon,a,Q^{2}) = \sum_{i=-2}^{0} \epsilon^{i} F_{i}(a,Q^{2})
 + {\mathcal O} \left( \epsilon \right) \, ,
\label{d22}
\ee

The homogeneous equation at $D=4$, i.e. $\epsilon=0$, is 
\be
\frac{df(a,y)}{dy} = - \frac{1}{2} \left[ \frac{1}{y} 
- \frac{1}{(y+4a)} \right] \, f(a,y) \, ,
\label{d23}
\ee
whose solution is 
\be
f(a,y) = k \sqrt{1 + \frac{4a}{y}} \, , 
\label{d24}
\ee
where $k$ is a normalization constant. 

Order by order in $\epsilon$, we obtain the solution of the 
non-homogeneous equation by means of the method of the variation of 
the constant $k$ (Euler's method). Substituting, into Eq. (\ref{d20}), 
Eq. (\ref{d22}) and the expansion in $\epsilon$ of the 3-denominator 
integral on the non-homogeneous part, Eq. (\ref{e9}), the result reads 
as follows:
\bea
F_{i}(a,Q^{2}) & = & \sqrt{1 + \frac{4a}{Q^{2}}} \, \, \,  
\Biggl\{ \int^{Q^{2}} \frac{dy}{\sqrt{1 + \frac{4a}{y}}} \, \biggl[ - 
\frac{1}{y+4a} F_{i-1}(a,y) \nn\\
& & - \frac{1}{2a} \left( \frac{1}{y} -
\frac{1}{y+4a} \right) C_{i} + \frac{3}{4a} \left( \frac{1}{y} -
\frac{1}{y+4a} \right) C_{i-1} \biggr] + k_{i} \Biggr\} \, ,
\label{d25}
\eea
where the explicit values of the constants $C_{i}$ are given in 
Eqs.(\ref{e10}--\ref{e12}).

The determination of the constants $k_{i}$ is made by imposing that the
solution, Eq. (\ref{d25}) satisfies the initial condition, Eq. 
(\ref{d21}), or, which is the same thing, imposing the regularity of the
solution at $Q^{2}=0$.

It is then useful to express the result in terms of the variable $x$, 
defined in Eq. (\ref{Q2tox}). The explicit form of the solution up to 
the zeroth order in $\epsilon$ is given in Eqs. (\ref{e26}--\ref{e28}).

\subsection{The full calculation for the topology in Fig. (\ref{fig2}) 
(a) \label{2exempl}}

The topology in Fig. (\ref{fig2}) (a) has two MI's. In order to decouple
the system of differential equations order by order in $\epsilon$, we
choose the couple of MI's given by Fig. (\ref{fig5}) (c) and (d), i.e. 
the fully scalar integral and the scalar integral with a squared 
propagator:
\bea
F_{1}(\epsilon,a,Q^{2}) & = & 
\parbox{15mm}{\begin{fmfgraph*}(15,15)
\fmfleft{i1,i2}
\fmfright{o}
\fmf{plain}{i1,v1}
\fmf{plain}{i2,v2}
\fmf{photon}{v4,o}
\fmf{plain,tension=.4}{v2,v3}
\fmf{plain,tension=.2}{v3,v4}
\fmf{plain,tension=.15}{v1,v4}
\fmf{photon,tension=0}{v2,v1}
\fmf{photon,tension=0}{v1,v3}
\end{fmfgraph*} } = \mu_{0}^{2(4-D)} 
\int \{ d^{D}k_{1} \} \{ d^{D}k_{2} \}
\frac{1}{{\mathcal D}_{1} {\mathcal D}_{2} {\mathcal D}_{9} {\mathcal D}_{14}
{\mathcal D}_{15}} \, , 
\label{d26} \\
F_{2}(\epsilon,a,Q^{2}) & = & 
\parbox{15mm}{\begin{fmfgraph*}(15,15)
\fmfleft{i1,i2}
\fmfright{o}
\fmfforce{0.5w,0.3h}{v5}
\fmf{plain}{i1,v1}
\fmf{plain}{i2,v2}
\fmf{photon}{v4,o}
\fmf{plain,tension=.4}{v2,v3}
\fmf{plain,tension=.2}{v3,v4}
\fmf{plain,tension=.15}{v1,v4}
\fmf{photon,tension=0}{v2,v1}
\fmf{photon,tension=0}{v1,v3}
\fmfv{decor.shape=circle,decor.filled=full,decor.size=.1w}{v5}
\end{fmfgraph*} } = \mu_{0}^{2(4-D)} 
\int \{ d^{D}k_{1} \} \{ d^{D}k_{2} \}
\frac{1}{{\mathcal D}_{1} {\mathcal D}_{2} {\mathcal D}_{9} {\mathcal D}_{14}
{\mathcal D}^{2}_{15}} \, .
\label{d27} 
\eea

The corresponding system of first-order linear differential equations 
in the variable $Q^{2}$ reads 
\bea
\! \! \! \! 
\frac{dF_{1}(\epsilon,a,Q^{2})}{dQ^{2}} \! & = \! & 
   - \frac{1}{2} \left[ \frac{1}{Q^{2}} +
     \frac{(1-2 \epsilon)}{(Q^{2}+4a)} \right] \, 
     F_{1}(\epsilon,a,Q^{2}) 
     \nonumber\\ 
     &-& F_{2}(\epsilon,a,Q^{2}) + \Omega^{(1)}(\epsilon,a,Q^{2}) \, , 
\label{d28} \\
\! \! \! \! 
\frac{dF_{2}(\epsilon,a,Q^{2})}{dQ^{2}} \! & = \! & 
    \frac{\epsilon^{2}}{2a} \left[ \frac{1}{Q^{2}} - 
    \frac{1}{(Q^{2}+4a)} 
    \right] \, F_{1}(\epsilon,a,Q^{2}) \nonumber\\ 
   &-& \left[ \frac{1}{Q^{2}} + \frac{(1+2 \epsilon)}{(Q^{2}+4a)} 
   \right] 
   \, F_{2}(\epsilon,a,Q^{2}) + \Omega^{(2)}(\epsilon,a,Q^{2}) \, , 
\label{d29}
\eea
where the functions $\Omega^{(i)}(\epsilon,a,Q^{2})$ are the following
combinations of MI's:
\bea
\Omega^{(1)}(\epsilon,a,Q^{2}) & = & \frac{1}{2aQ^{2}}  
\frac{(1-3 \epsilon)(1-2 \epsilon)}{(1-4 \epsilon)} \, \, 
\parbox{15mm}{\begin{fmfgraph*}(15,15)
\fmfleft{i1,i2}
\fmfright{o}
\fmf{plain}{i1,v1}
\fmf{plain}{i2,v2}
\fmf{photon}{v3,o}
\fmf{plain,tension=.3}{v2,v3}
\fmf{plain,tension=.3}{v1,v3}
\fmf{photon,tension=0}{v2,v1}
\fmf{photon,tension=0,right=.5}{v2,v3}
\end{fmfgraph*} } + \frac{1}{2 \epsilon} \Biggl\{ \frac{(1-2 
\epsilon)^{2} (3-4 \epsilon)}{16a^{2}(1-4 \epsilon)Q^{2}} \nn\\
& & - \frac{(3 - 20 \epsilon^{2} + 16 \epsilon^{3})}{16a^{2}(1-4 
\epsilon)(Q^{2}+4a)} - \frac{(11 - 108 \epsilon + 256 \epsilon^{2} 
- 168 \epsilon^{3})}{4a(1-4 \epsilon)(Q^{2}+4a)^{2}} \nn\\
& & + \frac{3(5 - 16 \epsilon + 12 \epsilon^{2})}{(Q^{2}+4a)^{3}} 
\Biggr\} \, \, 
\parbox{15mm}{\begin{fmfgraph*}(15,15)
\fmfleft{i}
\fmfright{o}
\fmf{photon}{i,v1}
\fmf{photon}{v2,o}
\fmf{plain,tension=.15,left}{v1,v2}
\fmf{photon,tension=.15}{v1,v2}
\fmf{plain,tension=.15,right}{v1,v2}
\end{fmfgraph*} } - \frac{1}{\epsilon} \Biggl\{ 
\frac{3(1-2 \epsilon)^{2}(1- \epsilon)}{16a^{3}(1-4 \epsilon) Q^{2}}
\nn\\
& & - \frac{3(1-2 \epsilon)^{2}(1- \epsilon)}{16a^{3}(1-4 \epsilon) 
(Q^{2}+4a)} - \frac{3(1-\epsilon -4 \epsilon^{2} +4 \epsilon^{3})}{
4a^{2}(1-4 \epsilon) (Q^{2}+4a)^{2}} \nn\\
& & - \frac{9(1-3 \epsilon +2 \epsilon^{2})}{a(Q^{2}+4a)^{3}} \Biggr\}
\, \,
\parbox{15mm}{\begin{fmfgraph*}(15,15)
\fmfleft{i}
\fmfright{o}
\fmf{photon}{i,v1}
\fmf{photon}{v2,o}
\fmflabel{$(k_{1} \cdot k_{2})$ }{o}
\fmf{plain,tension=.15,left}{v1,v2}
\fmf{photon,tension=.15}{v1,v2}
\fmf{plain,tension=.15,right}{v1,v2}
\end{fmfgraph*} } \qquad \quad \quad - \Biggl\{ 
\frac{(3-5 \epsilon + 2 \epsilon^{2})}{16a^{3}(1-4 \epsilon) Q^{2}} 
\nn\\
& & - \frac{(3-5 \epsilon + 2 \epsilon^{2})}{16a^{3}(1-4 \epsilon)
(Q^{2}+4a)} - \frac{(2-7 \epsilon + 5 \epsilon^{2})}{4a^{2}(1-4 
\epsilon)(Q^{2}+4a)^{2}} \nn\\
& & - \frac{(3-9 \epsilon + 6 \epsilon^{2})}{a \epsilon
(Q^{2}+4a)^{3}} \Biggr\} \, T^{2}(\epsilon,a) \, , 
\label{d30} \\
\Omega^{(2)}(\epsilon,a,Q^{2}) & = & \frac{\epsilon ( 1-2 \epsilon)(1-3
\epsilon)}{2a^{2}(1-4 \epsilon)} \left[ \frac{1}{Q^{2}} -
\frac{1}{(Q^{2}+4a)} \right] \, \, 
\parbox{15mm}{\begin{fmfgraph*}(15,15)
\fmfleft{i1,i2}
\fmfright{o}
\fmf{plain}{i1,v1}
\fmf{plain}{i2,v2}
\fmf{photon}{v3,o}
\fmf{plain,tension=.3}{v2,v3}
\fmf{plain,tension=.3}{v1,v3}
\fmf{photon,tension=0}{v2,v1}
\fmf{photon,tension=0,right=.5}{v2,v3}
\end{fmfgraph*} 
} \nn\\
& & + \frac{( 1 - 2 \epsilon)(1 - 3
\epsilon)}{4a} \Biggl[ \frac{1}{4aQ^{2}} - \frac{1}{4a(Q^{2}+4a)} \nn\\
& & - \frac{1}{(Q^{2} \! + \! 4a)^{2}} \Biggr]
\parbox{15mm}{\begin{fmfgraph*}(15,15)
\fmfleft{i1,i2}
\fmfright{o}
\fmf{plain}{i1,v1}
\fmf{plain}{i2,v2}
\fmf{photon}{v3,o}
\fmf{plain,tension=.3}{v2,v3}
\fmf{plain,tension=.3}{v1,v3}
\fmf{photon,tension=0,right=.5}{v2,v1}
\fmf{photon,tension=0,right=.5}{v1,v2}
\end{fmfgraph*} 
}  + \frac{( 1 \! - \! 2 \epsilon)(1 \! - \! 3
\epsilon)(2 \! - \! 3 \epsilon)}{8a^{2}(1-4 \epsilon)} \Biggl[ \! 
\frac{1}{4aQ^{2}} \nn\\
& & - \frac{1}{4a(Q^{2}+4a)} - \frac{1}{(Q^{2}+4a)^{2}} \Biggr] \, \,
\parbox{15mm}{\begin{fmfgraph*}(15,15)
\fmfleft{i}
\fmfright{o}
\fmf{plain}{i,v1}
\fmf{plain}{v2,o}
\fmf{plain,tension=.15,left}{v1,v2}
\fmf{photon,tension=.15}{v2,v1}
\fmf{photon,tension=.15,right}{v1,v2}
\end{fmfgraph*} 
}  \nn\\
& & + \frac{( 1 \! -  \! \epsilon)(1 \! - \! 2 \epsilon)}{4a^{2}} 
\Biggl[ \frac{1}{4aQ^{2}} \! -  \! \frac{1}{4a(Q^{2} \! +  \! \! 4a)} 
\! - \! \frac{1}{(Q^{2} \! + \! 4a)^{2}} \! \Biggr]
\parbox{15mm}{\begin{fmfgraph*}(15,15)
\fmfleft{i}
\fmfright{o}
\fmf{photon}{i,v1}
\fmf{photon}{v2,o}
\fmf{plain,tension=.22,left}{v1,v2}
\fmf{plain,tension=.22,right}{v1,v2}
\fmf{plain,right=45}{v2,v2}
\end{fmfgraph*} } \nn\\
& & - \Biggl\{ \! \frac{( 1 \! - \! 2 \epsilon)(1 \! - 
\! 11 \epsilon \! + \! 12 \epsilon^{2})}{32a^{3} (1 \! - \! 4 \epsilon)}
\Biggl[ \frac{1}{Q^{2}} - \frac{1}{(Q^{2}+4a)} \Biggr] \nn\\
& & - \frac{(1-2 \epsilon)}{4 \epsilon} 
\Biggl[ \frac{(3-4 \epsilon)(1-4 \epsilon)}{16a^{2}Q^{2}} 
- \frac{(3-36 \epsilon+ 296 \epsilon^{2} - 288 \epsilon^{3})}{
16a^{2} (1-4 \epsilon)(Q^{2}+4a)^{2}} \nn\\
& & - \frac{3(1-16 \epsilon + 20 
\epsilon^{2})}{2a (1-4 \epsilon)(Q^{2}+4a)^{3}} + 
\frac{3(5-6 \epsilon)(1-6 \epsilon)}{
(Q^{2}+4a)^{4}} \Biggl] \Biggl\}
\parbox{15mm}{\begin{fmfgraph*}(15,15)
\fmfleft{i}
\fmfright{o}
\fmf{photon}{i,v1}
\fmf{photon}{v2,o}
\fmf{plain,tension=.15,left}{v1,v2}
\fmf{photon,tension=.15}{v1,v2}
\fmf{plain,tension=.15,right}{v1,v2}
\end{fmfgraph*} 
}  \nn\\
& & - \Biggl\{ \! \frac{3( 1 \! - \! \epsilon)(1 \! - 
\! 2 \epsilon )(1+4 \epsilon)}{64a^{4} (1 \! - \! 4 \epsilon)}
\Biggl[ \frac{1}{Q^{2}} - \frac{1}{(Q^{2}+4a)} \Biggr] \nn\\
& & + \frac{3(1- \epsilon)(1-2 \epsilon)}{2a \epsilon} 
\Biggl[ \frac{(1-4 \epsilon)}{16a^{2}Q^{2}} - \frac{(1-6 \epsilon + 24 
\epsilon^{2})}{16a^{2} (1-4 \epsilon)(Q^{2}+4a)^{2}} \nn\\
& & - \frac{(1-7 \epsilon + 36 \epsilon^{2})}{2a (1-4 \epsilon)
(Q^{2}+4a)^{3}} + \frac{3(1-6 \epsilon)}{
(Q^{2}+4a)^{4}} \Biggl] \Biggl\}
\parbox{15mm}{\begin{fmfgraph*}(15,15)
\fmfleft{i}
\fmfright{o}
\fmf{photon}{i,v1}
\fmf{photon}{v2,o}
\fmflabel{$(k_{1} \cdot k_{2})$ }{o}
\fmf{plain,tension=.15,left}{v1,v2}
\fmf{photon,tension=.15}{v1,v2}
\fmf{plain,tension=.15,right}{v1,v2}
\end{fmfgraph*} 
} \nn\\
& & - \Biggl\{ \! \frac{( 1 \! - \! \epsilon)(3 \! - 
\! 31 \epsilon + 122 \epsilon^{2} - 104 \epsilon^{3})}{128a^{4} (1 \! - 
\! 4 \epsilon)}
\Biggl[ \frac{1}{Q^{2}} - \frac{1}{(Q^{2}+4a)} \Biggr] \nn\\
& & - \frac{(1- \epsilon)}{2a \epsilon} 
\Biggl[ \frac{\epsilon (1-4 \epsilon)}{16a^{2}Q^{2}} 
+ \frac{(3-32 \epsilon+ 130 \epsilon^{2} - 120 \epsilon^{3})}{
16a^{2} (1-4 \epsilon)(Q^{2}+4a)^{2}} \nn\\
& & + \frac{(3 \! - \! 29 \epsilon \! + \! 134 
\epsilon^{2} \! \! - \! \! 168 \epsilon^{3})}{2a (1 \! - \! 4 \epsilon)
(Q^{2} \! + \! 4a)^{3}} \! + \! \frac{3(1 \! - \! 2 \epsilon)(1 \! - \! 
6 \epsilon)}{(Q^{2}+4a)^{4}} \Biggl] \Biggl\} T^{2}(\epsilon,a) 
\label{d31} , 
\eea 
where $T(\epsilon,a)$ stands for the tadpole 
explicitly definited in Eq. (\ref{defT}). 

For what initial conditions are concerned, we know that 
$F_{1}(\epsilon,a,Q^{2})$ is analytic 
in $Q^{2}=0$. For Euclidean momenta, the limit $Q^2 \rightarrow 0$ 
(which implies $Q \rightarrow 0$) can be recovered by the limit $p_{2} 
\rightarrow -p_{1}$. Taking this limit directly within the integrand of 
Eq. (\ref{d26}) we obtain 
\bea
\! \! \! \!
F_{1}(\epsilon,a,Q^{2}=0) \! & = & \, \,  
\parbox{15mm}{\begin{fmfgraph*}(15,15)
\fmfforce{0.5w,0.2h}{v3}
\fmfforce{0.5w,0.8h}{v2}
\fmfforce{0.2w,0.5h}{v1}
\fmfforce{0.8w,0.5h}{v4}
\fmfforce{0.3w,0.3h}{v9}
\fmfleft{i}
\fmfright{o}
\fmf{plain}{i,v1}
\fmf{plain}{v4,o}
\fmf{photon,left=.4}{v1,v2}
\fmf{plain,right=.4}{v1,v3}
\fmf{photon,left=.4}{v2,v4}
\fmf{plain,right=.4}{v3,v4}
\fmf{photon,left=.6}{v3,v4}
\fmfv{decor.shape=circle,decor.filled=full,decor.size=.1w}{v9}
\end{fmfgraph*}} \, , \nn\\
\! \! \! \! \! & = & \! - \frac{(2 \! - \! 3 \epsilon)(1 \! - \! 3 
\epsilon)}{2a^{2}(1-4 \epsilon)} \,
\parbox{15mm}{\begin{fmfgraph*}(15,15)
\fmfleft{i}
\fmfright{o}
\fmf{plain}{i,v1}
\fmf{plain}{v2,o}
\fmf{plain,tension=.15,left}{v1,v2}
\fmf{photon,tension=.15}{v2,v1}
\fmf{photon,tension=.15,right}{v1,v2}
\end{fmfgraph*}} - \frac{(1 \! - \! \epsilon)^{2}(1 \! - \! 2 
\epsilon)}{2a^{3}(1-2 \epsilon)} T^{2}(\epsilon,a) .
\label{d32}
\eea

Once we have the initial condition for $F_{1}(\epsilon,a,Q^{2})$, we 
can calculate the initial condition for the second MI, 
$F_{2}(\epsilon,a,Q^{2})$, directly from Eq. (\ref{d28}). In fact, 
because of the analyticity of $F_{1}(\epsilon,a,Q^{2})$ in 
$Q^{2}=0$, we can multiply Eq. (\ref{d28}) by $Q^{2}$ and take the 
limit $Q^{2} \rightarrow 0$. The left-hand side vanishes and 
we find the following relation:
\bea
\! \! \! \! \! \! 
F_{2}(\epsilon,a,Q^{2}=0) \! & = & \! -
\frac{3 \epsilon (1-3 \epsilon)(2-3 \epsilon)}{8a^{3}(1-4 \epsilon)} \,
\parbox{15mm}{\begin{fmfgraph*}(15,15)
\fmfleft{i}
\fmfright{o}
\fmf{plain}{i,v1}
\fmf{plain}{v2,o}
\fmf{plain,tension=.15,left}{v1,v2}
\fmf{photon,tension=.15}{v2,v1}
\fmf{photon,tension=.15,right}{v1,v2}
\end{fmfgraph*}
} \nn\\
\! \! \! \! \! \! 
& & \qquad \quad + \frac{(16 \epsilon^{4}-16 \epsilon^{3}-76 
\epsilon^{2}+226 \epsilon -171)}{32a^{4}} T^{2}(\epsilon,a) \, .
\label{d33}
\eea

We look for a solution of the system of Eqs. (\ref{d28}--\ref{d29}) 
in terms of the coefficients of the Laurent series in $\epsilon$: 
\bea
F_{1}(\epsilon,a,Q^{2}) & = & \sum_{i=-2}^{0} 
\epsilon^{i} F^{(1)}_{i}(a,Q^{2}) + 
{\mathcal O} \left( \epsilon \right) \, , 
\label{d34} \\
F_{2}(\epsilon,a,Q^{2}) & = & \sum_{i=-2}^{0} 
\epsilon^{i} F^{(2)}_{i}(a,Q^{2}) + 
{\mathcal O} \left( \epsilon \right) \, ,
\label{d35}
\eea

As immediately seen by direct inspection of Eqs. (\ref{d28},\ref{d29}), 
the systematic expansion in powers of $\epsilon$ 
gives a triangular system in
which the second equation, at order $i$ in the $\epsilon$ expansion, 
consists of a homogeneous part with 
$F^{(2)}_i(a,Q^{2})$ only, and a non-homogeneous part which 
contains expansion terms of order lower than $i$. 
The first equation for $F^{(1)}_i(a,Q^{2})$, on 
the contrary, contains both $F^{(1)}_i(a,Q^{2})$ and 
$F^{(2)}_i(a,Q^{2})$ in the homogeneous part; but as a first step we 
can solve the second equation for $F^{(2)}_i(a,Q^{2})$, substitute 
the result in the first equation and split again the first equation 
in a new homogeneous part, which now involves only 
$F^{(1)}_i(a,Q^{2})$, and a non-homogeneous part which is known. 
As a result, the original system splits into two homogeneous decoupled 
equations, which are 
\bea
\frac{df_{1}(a,y)}{dy} & = & - \frac{1}{2} \left[ 
\frac{1}{y} + \frac{1}{(y+4a)} \right] \, f_{1} (a,y) \, , 
\label{d36} \\
\frac{df_{2}(a,y)}{dy} & = & - \left[ \frac{1}{y} + 
\frac{1}{(y+4a)} \right] \, f_{2}(a,y) \, , 
\label{d37}
\eea
with solutions 
\bea
f_{1}(a,y) & = & \frac{k_{1}}{\sqrt{y(y+4a)}}  \, , 
\label{d38} \\
f_{2}(a,y) & = & \frac{k_{2}}{y(y+4a)} \, .
\label{d39} 
\eea

By means of the Euler method we can find, order by order in $\epsilon$,
the solution of the non-homogeneous system, solving the two first order
differential equations in the order of the two quadrature formulas: 
\bea
\! \! \! \! \! \! 
F^{(2)}_{i}(a,Q^{2}) \! & = & \! \frac{1}{Q^{2}(Q^{2}+4a)} 
\Biggl\{
\int^{Q^{2}} \! dy \, y(y+4a) \biggl[ \frac{1}{2a} \left( \frac{1}{y} -
\frac{1}{(y+4a)} \right) F^{(1)}_{i-2}(a,y) \nn\\
\! \! \! \! \! \! \! \! \!  & & \! \! \! - \frac{2}{(y+4a)}
F^{(2)}_{i-1} + \Omega^{(2)}_{i}(a,y) 
\biggr] \! + \! k^{(2)}_{i} 
\Biggr\} \, , 
\label{d40}
\\
\! \! \! \! \! \! 
F^{(1)}_{i}(a,Q^{2}) \! & = & \! 
\frac{1}{\sqrt{Q^{2}(Q^{2}+4a)}} \Biggl\{
\int^{Q^{2}} \! dy \sqrt{y(y+4a)} \biggl[ \frac{1}{(y+4a)} 
F^{(1)}_{i-1}
- F^{(2)}_{i}(a,y)  \nn\\
\! \! \! \! \! \! \! \! \!  & & \! \! \! + 
\Omega^{(1)}_{i}(a,y) \biggr] \! + \! k^{(1)}_{i} \! \Biggr\} ,
\label{d41}
\eea
where, for simplicity, we put:
\bea
\Omega^{(1)}(\epsilon,a,Q^2) & = & \sum_{i=-2}^{0} 
\epsilon^{i} \Omega^{(1)}_{i}(a,Q^{2}) + 
{\mathcal O} \left( \epsilon \right) \, , 
\label{d42} \\
\Omega^{(2)}(\epsilon,a,Q^{2}) & = & \sum_{i=-2}^{0} 
\epsilon^{i} \Omega^{(2)}_{i}(a,Q^{2}) + 
{\mathcal O} \left( \epsilon \right) \, ,
\label{d43} 
\eea

The determination of the constants $k^{(1)}_{i}$ and $k^{(2)}_{i}$ is
made imposing the initial conditions Eqs. (\ref{d32}, \ref{d33}).
The solution, expressed in terms of the variable $x$, is given in Eqs.
(\ref{e51}-\ref{e53}).

\subsection{The full calculation for the topology in Fig. 
(\ref{fig1bis}) (b) \label{3exempl}}

The topology in Fig. (\ref{fig1bis}) (b) has two MI's. We choose 
the MI's corresponding to Fig. (\ref{fig5}) (a) and (b), i.e. the fully 
scalar integral and the scalar integral with the scalar product 
$(k_1\cdot k_2)$ in the numerator of the integrand. Also in this case 
the choice of the MI's diagonalizes 
the system in the limit $\epsilon \rightarrow 0$.
\bea
\! \! \! \! \! \! \! \! \! \! 
F_{1}(\epsilon,a,Q^{2}) \! & = & \! \! \! \! \! \! 
\parbox{15mm}{\begin{fmfgraph*}(15,15)
\fmfleft{i1,i2}
\fmfright{o}
\fmf{plain}{i1,v1}
\fmf{plain}{i2,v2}
\fmf{photon}{v5,o}
\fmf{plain,tension=.3}{v2,v3}
\fmf{plain,tension=.3}{v3,v5}
\fmf{plain,tension=.3}{v1,v4}
\fmf{plain,tension=.3}{v4,v5}
\fmf{photon,tension=0}{v2,v4}
\fmf{photon,tension=0}{v1,v3}
\end{fmfgraph*} } = \mu_{0}^{2(4-D)} \! \! \! 
\int \{ \! d^{D} \! k_{1} \! \} \{ \! d^{D} \! k_{2} \! \}
\frac{1}{{\mathcal D}_{1} {\mathcal D}_{2} {\mathcal D}_{9} 
{\mathcal D}_{11} {\mathcal D}_{14} {\mathcal D}_{15} }  \, , 
\label{d44} \\
\! \! \! \! \! \! \! \! \! \! 
F_{2}(\epsilon,a,Q^{2}) \! & = & \! \! \! \! \! \! 
\parbox{15mm}{\begin{fmfgraph*}(15,15)
\fmfleft{i1,i2}
\fmfright{o}
\fmf{plain}{i1,v1}
\fmf{plain}{i2,v2}
\fmf{photon}{v5,o}
\fmflabel{$(k_{1} \cdot k_{2})$}{o}
\fmf{plain,tension=.3}{v2,v3}
\fmf{plain,tension=.3}{v3,v5}
\fmf{plain,tension=.3}{v1,v4}
\fmf{plain,tension=.3}{v4,v5}
\fmf{photon,tension=0}{v2,v4}
\fmf{photon,tension=0}{v1,v3}
\end{fmfgraph*} } \qquad \qquad \! = \! \mu_{0}^{2(4-D)}  \! \! \! 
\int \{ \! d^{D} \! k_{1} \! \} \{ \! d^{D} \! k_{2} \! \}
\frac{k_{1} \cdot k_{2}}{{\mathcal D}_{1} {\mathcal D}_{2} {\mathcal D}_{9} 
{\mathcal D}_{11} {\mathcal D}_{14} {\mathcal D}_{15} } .
\label{d45} 
\eea

The system of first-order linear differential equations is the
following:

\bea
\! \! \! \! \! \! 
\frac{dF_{1}(\epsilon,a,Q^{2})}{dQ^{2}} & = & - (1+2 \epsilon) \left[ 
  \frac{1}{Q^{2}} + \frac{(1-2 \epsilon)}{(Q^{2}+4a)} \right] \, 
       F_{1}(\epsilon,a,Q^{2}) \nonumber\\ 
  & & - \frac{2\epsilon}{a}  \left[ \frac{1}{Q^{2}} 
    - \frac{(1-2 \epsilon)}{(Q^{2}+4a)}\right] \, 
    F_{2}(\epsilon,a,Q^{2}) 
    + \Omega^{(1)}(\epsilon,a,Q^{2}) \, , 
\label{d46} \\
\! \! \! \! \! \! 
   \frac{dF_{2}(\epsilon,a,Q^{2})}{dQ^{2}} & = & \epsilon \, \, 
   F_{1}(\epsilon,a,Q^2) \nonumber\\ 
 & & - \frac{(1 \, - \, 2 \epsilon)}{2} \left[ \frac{1}{Q^{2}} \! + 
\! \frac{1}{(Q^{2} \! + \! 4a)}
\right] \! F_{2}(\epsilon,a,Q^{2}) \! + \! 
\Omega^{(2)}(\epsilon,a,Q^{2}) , 
\label{d47}
\eea
where the functions $\Omega^{(i)}(\epsilon,a,Q^{2})$ are defined as 
follows:
\bea
\Omega^{(1)}(\epsilon,a,Q^{2}) & = & \frac{\epsilon}{a} \left[
\frac{1}{Q^{2}} - \frac{1}{(Q^{2}+4a)} \right] \, \,
\parbox{15mm}{\begin{fmfgraph*}(15,15)
\fmfleft{i1,i2}
\fmfright{o}
\fmf{plain}{i1,v1}
\fmf{plain}{i2,v2}
\fmf{photon}{v4,o}
\fmf{plain,tension=.4}{v2,v3}
\fmf{plain,tension=.2}{v3,v4}
\fmf{plain,tension=.15}{v1,v4}
\fmf{photon,tension=0}{v2,v1}
\fmf{photon,tension=0}{v1,v3}
\end{fmfgraph*} 
} - \frac{3(1-2 \epsilon)}{8a^{2}} \Biggl[ 
\frac{4a}{Q^{4}} - \frac{1}{Q^{2}} \nonumber\\ 
& & + \frac{1}{(Q^{2}+4a)} \Biggr] \, \,
\parbox{15mm}{\begin{fmfgraph*}(15,15)
\fmfleft{i1,i2}
\fmfright{o}
\fmf{plain}{i1,v1}
\fmf{plain}{i2,v2}
\fmf{photon}{v3,o}
\fmf{plain,tension=.3}{v2,v3}
\fmf{plain,tension=.3}{v1,v3}
\fmf{plain,tension=0,right=.5}{v2,v1}
\fmf{plain,tension=0,right=.5}{v1,v2}
\end{fmfgraph*} 
} + \frac{3(2-3 \epsilon)}{4a^{3}} \Biggl[ 
\frac{4a}{Q^{4}} - \frac{1}{Q^{2}} \nonumber\\ 
& & + \frac{1}{(Q^{2}+4a)} \Biggr] \, \,
\parbox{15mm}{\begin{fmfgraph*}(15,15)
\fmfleft{i1,i2}
\fmfright{o}
\fmf{plain}{i1,v1}
\fmf{plain}{i2,v2}
\fmf{photon}{v3,o}
\fmflabel{$(p_{2} \cdot k_{1})$ }{o}
\fmf{plain,tension=.3}{v2,v3}
\fmf{plain,tension=.3}{v1,v3}
\fmf{plain,tension=0,right=.5}{v2,v1}
\fmf{plain,tension=0,right=.5}{v1,v2}
\end{fmfgraph*} } \qquad \qquad + \frac{5(1-3 \epsilon)}{4a^{2}} 
\Biggl[ \frac{1}{Q^{2}} \nonumber\\ 
& & - \frac{1}{(Q^{2}+4a)} \Biggr] 
\parbox{15mm}{\begin{fmfgraph*}(15,15)
\fmfleft{i1,i2}
\fmfright{o}
\fmf{plain}{i1,v1}
\fmf{plain}{i2,v2}
\fmf{photon}{v3,o}
\fmf{plain,tension=.3}{v2,v3}
\fmf{photon,tension=.3}{v1,v3}
\fmf{plain,tension=0}{v2,v1}
\fmf{plain,tension=0,right=.5}{v2,v3}
\end{fmfgraph*} } 
+ \frac{5}{2a} \left[ \frac{1}{Q^{2}} - \frac{1}{(Q^{2}+4a)} \right] 
\parbox{15mm}{\begin{fmfgraph*}(15,15)
\fmfleft{i1,i2}
\fmfright{o}
\fmfforce{0.22w,0.5h}{v4}
\fmf{plain}{i1,v1}
\fmf{plain}{i2,v2}
\fmf{photon}{v3,o}
\fmf{plain,tension=.3}{v2,v3}
\fmf{photon,tension=.3}{v1,v3}
\fmf{plain,tension=0}{v2,v1}
\fmf{plain,tension=0,right=.5}{v2,v3}
\fmfv{decor.shape=circle,decor.filled=full,decor.size=.1w}{v4}
\end{fmfgraph*} } \nn\\
& & - \frac{3(1-2 \epsilon)(1-3 \epsilon)}{4a^{2}(1-4 \epsilon)} 
\left[ \frac{1}{Q^{2}} - \frac{1}{(Q^{2}+4a)} \right] \, \,
\parbox{15mm}{\begin{fmfgraph*}(15,15)
\fmfleft{i1,i2}
\fmfright{o}
\fmf{plain}{i1,v1}
\fmf{plain}{i2,v2}
\fmf{photon}{v3,o}
\fmf{plain,tension=.3}{v2,v3}
\fmf{plain,tension=.3}{v1,v3}
\fmf{photon,tension=0}{v2,v1}
\fmf{photon,tension=0,right=.5}{v2,v3}
\end{fmfgraph*} 
} \nn\\
& & + \frac{(1 \! - \! 2 \epsilon)(1 \! - \! 3 \epsilon)}{16a^{2} 
\epsilon} \! \left[ \frac{1}{Q^{2}} \! - \! \frac{1}{(Q^{2}+4a)} \! - \!
\frac{4a}{(Q^{2}+4a)^{2}} \right] \,
\parbox{15mm}{\begin{fmfgraph*}(15,15)
\fmfleft{i1,i2}
\fmfright{o}
\fmf{plain}{i1,v1}
\fmf{plain}{i2,v2}
\fmf{photon}{v3,o}
\fmf{plain,tension=.3}{v2,v3}
\fmf{plain,tension=.3}{v1,v3}
\fmf{photon,tension=0,right=.5}{v2,v1}
\fmf{photon,tension=0,right=.5}{v1,v2}
\end{fmfgraph*} 
} \nn\\
& &  - \Biggl\{ \frac{(17-116 \epsilon +248 \epsilon^{2} - 128 
\epsilon^{3})}{64a^{3} \epsilon (1-4 \epsilon)} \left[ 
\frac{1}{Q^{2}} - \frac{1}{(Q^{2}+4a)} \right] \nn\\
& & + \frac{(3-4 \epsilon)(1-4 \epsilon)(1-22 \epsilon)}{64a^{2} 
\epsilon^{2}} \frac{1}{Q^{4}} \nn\\
& & + \frac{(3-162 \epsilon +968 \epsilon^{2} 
- 2240 \epsilon^{3} + 1536 \epsilon^{4})}{64a^{2} \epsilon^{2} (1-4 
\epsilon)} \frac{1}{(Q^{2}+4a)^{2}} \nn\\
& & + \frac{3(1-2 \epsilon)(1+34 \epsilon -140 
\epsilon^{2} +120 \epsilon^{3}}{8a \epsilon^{2} (1-4 \epsilon)}
\frac{1}{(Q^{2}+4a)^{3}} \nn\\
& & - \frac{3(1-2 \epsilon)(5-6 \epsilon)(1 -6 
\epsilon)}{4 \epsilon^{2}} \frac{1}{(Q^{2}+4a)^{4}} \Biggr\} \, \,
\parbox{15mm}{\begin{fmfgraph*}(15,15)
\fmfleft{i}
\fmfright{o}
\fmf{photon}{i,v1}
\fmf{photon}{v2,o}
\fmf{plain,tension=.15,left}{v1,v2}
\fmf{photon,tension=.15}{v1,v2}
\fmf{plain,tension=.15,right}{v1,v2}
\end{fmfgraph*} 
}  \nn\\
& &  + \Biggl\{ \frac{3(1- \epsilon)(9 -2 \epsilon - 112 \epsilon^{2})}{
64a^{4} \epsilon (1-4 \epsilon)} \left[ 
\frac{1}{Q^{2}} - \frac{1}{(Q^{2}+4a)} \right] \nn\\
& & - \frac{3(1- \epsilon)(1-4 \epsilon)(1-22 \epsilon)}{32a^{3} 
\epsilon^{2}} \frac{1}{Q^{4}} \nn\\
& & + \frac{3(1- \epsilon)(1-48 \epsilon +196 \epsilon^{2} -
128 \epsilon^{3})}{32a^{3} \epsilon^{2} (1-4 \epsilon)}
\frac{1}{(Q^{2}+4a)^{2}} \nn\\
& & + \frac{3(1- \epsilon)(1-2 \epsilon)(1-37 
\epsilon + 96 \epsilon^{2})}{4a^{2} \epsilon^{2} (1-4 \epsilon)}
\frac{1}{(Q^{2}+4a)^{3}} \nn\\
& & + \frac{9(1 \! -  \! \epsilon)(1 \! - \! 2 \epsilon)(1  \! - \! 6 
\epsilon)}{2a \epsilon^{2}} \frac{1}{(Q^{2} \! + \! 4a)^{4}} \Biggr\} 
\parbox{15mm}{\begin{fmfgraph*}(15,15)
\fmfleft{i}
\fmfright{o}
\fmf{photon}{i,v1}
\fmf{photon}{v2,o}
\fmflabel{$(k_{1} \! \cdot \! k_{2})$ }{o}
\fmf{plain,tension=.15,left}{v1,v2}
\fmf{photon,tension=.15}{v1,v2}
\fmf{plain,tension=.15,right}{v1,v2}
\end{fmfgraph*} 
}   \nn\\
& &  + \frac{(1-2 \epsilon)(1-3 \epsilon)(2-3 \epsilon)}{32a^{3} 
\epsilon (1-4 \epsilon)} \Biggl[ 
\frac{1}{Q^{2}} - \frac{1}{(Q^{2}+4a)} \nn\\
& & - \! \frac{4a}{(Q^{2} \! + \! 4a)^{2}}  \! 
\Biggr] \! 
\parbox{15mm}{\begin{fmfgraph*}(15,15)
\fmfleft{i}
\fmfright{o}
\fmf{plain}{i,v1}
\fmf{plain}{v2,o}
\fmf{plain,tension=.15,left}{v1,v2}
\fmf{photon,tension=.15}{v2,v1}
\fmf{photon,tension=.15,right}{v1,v2}
\end{fmfgraph*}}  \! 
-  \! \frac{3(2 \! - \! 3 \epsilon)}{16a^{3}} 
\Biggl[ \frac{4a}{Q^{4}}  \! -  \! \frac{1}{Q^{2}}  \! 
 + \! \frac{1}{(Q^{2} \! + \! 4a)} \! \Biggr] \times \nn\\
& & \times
\parbox{15mm}{\begin{fmfgraph*}(15,15)
\fmfleft{i}
\fmfright{o}
\fmf{plain}{i,v1}
\fmf{plain}{v2,o}
\fmf{plain,tension=.15,left}{v1,v2}
\fmf{plain,tension=.15}{v1,v2}
\fmf{plain,tension=.15,right}{v1,v2}
\end{fmfgraph*}} 
- \frac{3(2 \! - \! 3 \epsilon)}{32a^{3}} \! \left[ 
\! \frac{4a}{Q^{4}} \! - \! \frac{1}{Q^{2}} \! + \! \frac{1}{(Q^{2} \! 
+ \! 4a)} \! \right] \, 
\parbox{15mm}{\begin{fmfgraph*}(15,15)
\fmfleft{i}
\fmfright{o}
\fmf{photon}{i,v1}
\fmf{photon}{v2,o}
\fmf{plain,tension=.22,left}{v1,v2}
\fmf{plain,tension=.22,right}{v1,v2}
\fmf{plain,right=45}{v2,v2}
\end{fmfgraph*}}  \nn\\
& &  - \Biggl\{ \frac{(1 \! -  \! \epsilon)(3 \! - \! 89 \epsilon \! + 
\! 436 \epsilon^{2} \! - \! 576 \epsilon^{3} \! + \! 16 \epsilon^{4})}{
128a^{4} \epsilon^{2} (1-2 \epsilon)(1-4 \epsilon)} \! \left[ 
\frac{1}{Q^{2}} \! - \! \frac{1}{(Q^{2} \! + \! 4a)}  \! \right] \nn\\
& & - \frac{(1- \epsilon)(1-38 \epsilon +100 \epsilon^{2})}{32a^{3} 
\epsilon (1-2 \epsilon)} \frac{1}{Q^{4}} \nn\\
& & - \frac{(1- \epsilon)(3-84 \epsilon +310 \epsilon^{2} -208 
\epsilon^{3})}{32a^{3} \epsilon^{2} (1-4 \epsilon)} 
\frac{1}{(Q^{2}+4a)^{2}} \nn\\
& & - \frac{(1- \epsilon)(3-89 \epsilon +374 \epsilon^{2} -408 
\epsilon^{3})}{8a^{2} \epsilon^{2} (1-4 \epsilon)} 
\frac{1}{(Q^{2}+4a)^{3}} \nn\\
& & - \frac{3(1- \epsilon)(1-2 \epsilon)(1 -6 \epsilon)}{2a 
\epsilon^{2}} \frac{1}{(Q^{2}+4a)^{4}} \Biggr\} \, \, T^{2}(\epsilon,a)
\, , 
\label{d48} \\
\Omega^{(2)}(\epsilon,a,Q^{2}) & = & \epsilon \, 
\parbox{15mm}{\begin{fmfgraph*}(15,15)
\fmfleft{i1,i2}
\fmfright{o}
\fmf{plain}{i1,v1}
\fmf{plain}{i2,v2}
\fmf{photon}{v4,o}
\fmf{plain,tension=.4}{v2,v3}
\fmf{plain,tension=.2}{v3,v4}
\fmf{plain,tension=.15}{v1,v4}
\fmf{photon,tension=0}{v2,v1}
\fmf{photon,tension=0}{v1,v3}
\end{fmfgraph*} } + \Biggl\{
\frac{(1-2 \epsilon)(1-4 \epsilon)}{2 \epsilon} \Biggl[ 
\frac{4a}{Q^{4}}
+ \frac{1}{(Q^{2}+4a)} \Biggr] \nn\\
& & - \frac{(1-2 \epsilon)}{8a \epsilon}
\frac{1}{Q^{2}} \Biggr\} \,
\parbox{15mm}{\begin{fmfgraph*}(15,15)
\fmfleft{i1,i2}
\fmfright{o}
\fmf{plain}{i1,v1}
\fmf{plain}{i2,v2}
\fmf{photon}{v3,o}
\fmf{plain,tension=.3}{v2,v3}
\fmf{plain,tension=.3}{v1,v3}
\fmf{plain,tension=0,right=.5}{v2,v1}
\fmf{plain,tension=0,right=.5}{v1,v2}
\end{fmfgraph*} } 
+ \Biggl\{ \frac{(2-3 \epsilon)(1-4 \epsilon)}{a \epsilon} 
\frac{1}{Q^{4}}  \nn\\
& & - \frac{(1-2 \epsilon)(2-3 \epsilon)}{4a^{2} \epsilon}
\Biggl[ \frac{1}{Q^{2}} - \frac{1}{(Q^{2}+4a)} \Biggr] \, 
\parbox{15mm}{\begin{fmfgraph*}(15,15)
\fmfleft{i1,i2}
\fmfright{o}
\fmf{plain}{i1,v1}
\fmf{plain}{i2,v2}
\fmf{photon}{v3,o}
\fmflabel{$(p_{2} \cdot k_{1})$ }{o}
\fmf{plain,tension=.3}{v2,v3}
\fmf{plain,tension=.3}{v1,v3}
\fmf{plain,tension=0,right=.5}{v2,v1}
\fmf{plain,tension=0,right=.5}{v1,v2}
\end{fmfgraph*} } \nn\\
& & - \frac{(1-3 \epsilon)}{a} \frac{1}{(Q^{2}+4a)} \, 
\parbox{15mm}{\begin{fmfgraph*}(15,15)
\fmfleft{i1,i2}
\fmfright{o}
\fmf{plain}{i1,v1}
\fmf{plain}{i2,v2}
\fmf{photon}{v3,o}
\fmf{plain,tension=.3}{v2,v3}
\fmf{photon,tension=.3}{v1,v3}
\fmf{plain,tension=0}{v2,v1}
\fmf{plain,tension=0,right=.5}{v2,v3}
\end{fmfgraph*} } - \Biggl[ \frac{(1-2 \epsilon)}{2 \epsilon}
\frac{1}{Q^{2}} \nn\\
& & + \frac{(1 \! - \! 6 \epsilon)}{2 \epsilon}
\frac{1}{(Q^{2} \! + \! 4a)} \Biggr] \! \! 
\parbox{15mm}{\begin{fmfgraph*}(15,15)
\fmfleft{i1,i2}
\fmfright{o}
\fmfforce{0.22w,0.5h}{v4}
\fmf{plain}{i1,v1}
\fmf{plain}{i2,v2}
\fmf{photon}{v3,o}
\fmf{plain,tension=.3}{v2,v3}
\fmf{photon,tension=.3}{v1,v3}
\fmf{plain,tension=0}{v2,v1}
\fmf{plain,tension=0,right=.5}{v2,v3}
\fmfv{decor.shape=circle,decor.filled=full,decor.size=.1w}{v4}
\end{fmfgraph*} } + \! \frac{(1 \! - \! 2 \epsilon)(1 \! - \! 3 
\epsilon)}{2a(1-4 \epsilon)} \frac{1}{Q^{2}} \! \!
\parbox{15mm}{\begin{fmfgraph*}(15,15)
\fmfleft{i1,i2}
\fmfright{o}
\fmf{plain}{i1,v1}
\fmf{plain}{i2,v2}
\fmf{photon}{v3,o}
\fmf{plain,tension=.3}{v2,v3}
\fmf{plain,tension=.3}{v1,v3}
\fmf{photon,tension=0}{v2,v1}
\fmf{photon,tension=0,right=.5}{v2,v3}
\end{fmfgraph*} } \nn\\
& &  + \Biggl\{ \frac{(1-2 \epsilon)(3-4 \epsilon)(1-4 \epsilon)}{
16a \epsilon^{2}} \frac{1}{Q^{4}} \nn\\
& & + 
\frac{(3-18 \epsilon +4 \epsilon^{2}+136 \epsilon^{3}-160 \epsilon^{4}
)}{64a^{2} \epsilon^{2}(1-4 \epsilon)} \frac{1}{Q^{2}} \nn\\
& & + \frac{3(1-2 \epsilon)(5-6 \epsilon)(1-6 \epsilon)}{
4 \epsilon^{2}} \frac{1}{(Q^{2}+4a)^{3}} \nn\\
& & - 
\frac{(3-74 \epsilon +410 \epsilon^{2}-816 \epsilon^{3}+504 \epsilon^{4}
)}{8a \epsilon^{2}(1-4 \epsilon)} \frac{1}{(Q^{2}+4a)^{2}} \nn\\
& & -
\frac{(3-18 \epsilon +36 \epsilon^{2}+40 \epsilon^{3}-96 \epsilon^{4}
)}{64a^{2} \epsilon^{2}(1-4 \epsilon)} \frac{1}{(Q^{2}+4a)} \Biggr\} \,
\parbox{15mm}{\begin{fmfgraph*}(15,15)
\fmfleft{i}
\fmfright{o}
\fmf{photon}{i,v1}
\fmf{photon}{v2,o}
\fmf{plain,tension=.15,left}{v1,v2}
\fmf{photon,tension=.15}{v1,v2}
\fmf{plain,tension=.15,right}{v1,v2}
\end{fmfgraph*} }  \nn\\
& &  - \frac{3(1- \epsilon)}{2a\epsilon^{2}} \Biggl\{ 
\frac{(1-2 \epsilon)(1-4 \epsilon)}{
4a} \frac{1}{Q^{4}} + 
\frac{(1-32 \epsilon^{2}+72 \epsilon^{3})}{16a^{2} (1-4 \epsilon)} 
\Bigl( \frac{1}{Q^{2}} \nn\\
& & -  \frac{1}{(Q^{2}+4a)} \Biggr) - 3(1-2 \epsilon)(1-6 
\epsilon) \frac{1}{(Q^{2}+4a)^{3}} \nn\\
& & - 
\frac{(1-5 \epsilon +4 \epsilon^{2}+12 \epsilon^{3})}{2a (1-4 \epsilon)}
\frac{1}{(Q^{2}+4a)^{2}} \Biggr\} \,
\parbox{15mm}{\begin{fmfgraph*}(15,15)
\fmfleft{i}
\fmfright{o}
\fmf{photon}{i,v1}
\fmf{photon}{v2,o}
\fmflabel{$(k_{1} \cdot k_{2})$ }{o}
\fmf{plain,tension=.15,left}{v1,v2}
\fmf{photon,tension=.15}{v1,v2}
\fmf{plain,tension=.15,right}{v1,v2}
\end{fmfgraph*} }   \nn\\
& & - \frac{(1-2 \epsilon)(2 -3 \epsilon)(1-3 \epsilon)}{16a^{2}
\epsilon (1-4 \epsilon)} \Biggl[ \frac{1}{Q^{2}} -
\frac{1}{(Q^{2}+4a)^{2}} \Biggr] \,
\parbox{15mm}{\begin{fmfgraph*}(15,15)
\fmfleft{i}
\fmfright{o}
\fmf{plain}{i,v1}
\fmf{plain}{v2,o}
\fmf{plain,tension=.15,left}{v1,v2}
\fmf{photon,tension=.15}{v2,v1}
\fmf{photon,tension=.15,right}{v1,v2}
\end{fmfgraph*}}  \nn\\
& &  - \Biggl\{ \frac{(2-3 \epsilon)(1-4 \epsilon)}{8a \epsilon}
\frac{1}{Q^{4}} -  \frac{(1-2 \epsilon)(2-3 \epsilon)}{8a \epsilon}
\Biggl[ \frac{1}{Q^{2}} \nn\\
& & -
\frac{1}{(Q^{2}+4a)^{2}} \Biggr] \Biggr\} \, 
\parbox{15mm}{\begin{fmfgraph*}(15,15)
\fmfleft{i}
\fmfright{o}
\fmf{plain}{i,v1}
\fmf{plain}{v2,o}
\fmf{plain,tension=.15,left}{v1,v2}
\fmf{plain,tension=.15}{v1,v2}
\fmf{plain,tension=.15,right}{v1,v2}
\end{fmfgraph*}} 
+ \Biggl\{ \frac{(1- \epsilon)(1-4 \epsilon)}{2a 
\epsilon} \frac{1}{Q^{4}} \nn\\
& &  +  \frac{(1- \epsilon)(1-2 \epsilon)}{8a^{2} \epsilon}
\Biggl[ \frac{1}{Q^{2}} -
\frac{1}{(Q^{2}+4a)^{2}} \Biggr] \Biggr\} \, 
\parbox{15mm}{\begin{fmfgraph*}(15,15)
\fmfleft{i}
\fmfright{o}
\fmf{photon}{i,v1}
\fmf{photon}{v2,o}
\fmf{plain,tension=.22,left}{v1,v2}
\fmf{plain,tension=.22,right}{v1,v2}
\fmf{plain,right=45}{v2,v2}
\end{fmfgraph*}}  \nn\\
& &  - \! \frac{(1 \! - \! \epsilon)}{2a} \Biggl\{ \frac{(1 \! - \! 4 
\epsilon)}{4a (1 \! - \! 2 \epsilon)} \frac{1}{Q^{4}} \! + \! 
\frac{(3 \! - \! 25 \epsilon  \! - \! 71 \epsilon^{2} \! + \! 58 
\epsilon^{3} \! + \! 32 \epsilon^{4})}{16a^{2} \epsilon^{2}(1 \! - \! 2 
\epsilon)(1 \! - \! 4 \epsilon)} \Biggl( \frac{1}{Q^{2}} \nn\\
& & - \frac{1}{(Q^{2}+4a)} \Biggr) - \frac{3(1-2 \epsilon)(1-6 
\epsilon)}{\epsilon^{2}} \frac{1}{(Q^{2}+4a)^{3}} \nn\\
& & - 
\frac{3(1-7 \epsilon +14 \epsilon^{2})}{4a \epsilon^{2}(1-4 \epsilon)} 
\frac{1}{(Q^{2}+4a)^{2}}  \Biggr\} \, T^{2}(\epsilon,a) \, .
\label{d49} 
\eea

For what concerns initial conditions, we know that 
$F_{1}(\epsilon,a,Q^2)$ 
is analytic in $Q^{2}=0$. For Euclidean momenta, 
the limit $Q^2 \rightarrow 0$ (which implies $Q \rightarrow 0$) can be 
recovered by the limit $p_{2} 
\rightarrow -p_{1}$. Taking this limit directly within the integrand of 
Eq. (\ref{d44}) we have:
\bea
\! \! \! \! \! \! \! \!
F_{1}(\epsilon,a,Q^{2}=0) & = & 
\parbox{15mm}{\begin{fmfgraph*}(15,15)
\fmfforce{0.5w,0.2h}{v3}
\fmfforce{0.5w,0.8h}{v2}
\fmfforce{0.2w,0.5h}{v1}
\fmfforce{0.8w,0.5h}{v4}
\fmfforce{0.5w,0.5h}{v9}
\fmfleft{i}
\fmfright{o}
\fmf{plain}{i,v1}
\fmf{plain}{v4,o}
\fmf{plain,left=.4}{v1,v2}
\fmf{photon,right=.4}{v1,v3}
\fmf{photon,left=.4}{v2,v4}
\fmf{plain,right=.4}{v3,v4}
\fmf{plain}{v2,v3}
\fmfv{decor.shape=circle,decor.filled=full,decor.size=.1w}{v9}
\end{fmfgraph*}} \, , \nn\\
\! \! \! \! \! \! \! \! & = & - \frac{3 \epsilon (2 \! - \! 3 \epsilon)
(1 \! - \! 3 \epsilon)}{4a^{3}(1 \! - \! 4 \epsilon)(1 \! + \! 2 
\epsilon)} 
\parbox{15mm}{\begin{fmfgraph*}(15,15)
\fmfleft{i}
\fmfright{o}
\fmf{plain}{i,v1}
\fmf{plain}{v2,o}
\fmf{plain,tension=.15,left}{v1,v2}
\fmf{photon,tension=.15}{v2,v1}
\fmf{photon,tension=.15,right}{v1,v2}
\end{fmfgraph*}} 
\! + \! \frac{3 (2 \! - \! 3 \epsilon)(1 \! - \! 3 \epsilon)}{64
a^{3} \epsilon}
\parbox{15mm}{\begin{fmfgraph*}(15,15)
\fmfleft{i}
\fmfright{o}
\fmf{plain}{i,v1}
\fmf{plain}{v2,o}
\fmf{plain,tension=.15,left}{v1,v2}
\fmf{plain,tension=.15}{v1,v2}
\fmf{plain,tension=.15,right}{v1,v2}
\end{fmfgraph*}} \nn\\
\! \! \! \! & & + \frac{(1- \epsilon)^{2}(9+3 \epsilon - 160
\epsilon^{2} -196 \epsilon^{3})}{
64a^{4} \epsilon (1-2 \epsilon) (1+2 \epsilon)} \, \, T^{2}(\epsilon,a)
.
\label{d50}
\eea

We can find the initial condition for $F_{2}$ from Eq. (\ref{d46}),
multiplying by $Q^{2}$ and taking the limit $Q^{2} \rightarrow 0$, or
performing the limit directely inside the integral of Eq. (\ref{d45}).
What we get is the following expression:

\bea
\! \! \! \!
F_{2}(\epsilon,a,Q^{2}=0) & = & 
\frac{(2-3 \epsilon)(1-3 \epsilon)}{8
a^{2} \epsilon} \, \, 
\parbox{15mm}{\begin{fmfgraph*}(15,15)
\fmfleft{i}
\fmfright{o}
\fmf{plain}{i,v1}
\fmf{plain}{v2,o}
\fmf{plain,tension=.15,left}{v1,v2}
\fmf{photon,tension=.15}{v2,v1}
\fmf{photon,tension=.15,right}{v1,v2}
\end{fmfgraph*}
} + \frac{(2-3 \epsilon)(1-3 \epsilon)}{32
a^{2} \epsilon^{2}} \, \, 
\parbox{15mm}{\begin{fmfgraph*}(15,15)
\fmfleft{i}
\fmfright{o}
\fmf{plain}{i,v1}
\fmf{plain}{v2,o}
\fmf{plain,tension=.15,left}{v1,v2}
\fmf{plain,tension=.15}{v1,v2}
\fmf{plain,tension=.15,right}{v1,v2}
\end{fmfgraph*}} \nn\\
\! \! \! \! & & + \frac{(1- \epsilon)^{2}(3-15 \epsilon + 16
\epsilon^{2})}{32a^{3} \epsilon^{2} (1-2 \epsilon)} \, \, 
T^{2}(\epsilon,a) .
\label{d51}
\eea

We look for a solution of the system of Eqs. (\ref{d46}--\ref{d47}) 
expanded in Laurent series of $\epsilon$:
\bea
F_{1}(\epsilon,a,Q^{2}) & = & \sum_{i=-2}^{0} 
\epsilon^{i} F^{(1)}_{i}(a,Q^{2}) + 
{\mathcal O} \left( \epsilon \right) \, , 
\label{d52} \\
F_{2}(\epsilon,a,Q^{2}) & = & \sum_{i=-2}^{0} 
\epsilon^{i} F^{(2)}_{i}(a,Q^{2}) + 
{\mathcal O} \left( \epsilon \right) \, ,
\label{d53}
\eea

According to the previous remarks, the solution can be built, 
order by order, by the method of the variation of the constants 
of the associated homogeneous system, which reads 
\bea
\frac{df_{1}(a,y)}{dy} & = & - \left[ 
\frac{1}{y} + \frac{1}{(y+4a)} \right] \, f_{1} (a,y) \, , 
\label{d54} \\
\frac{df_{2}(a,y)}{dy} & = & - \frac{1}{2} \left[ \frac{1}{y} + 
\frac{1}{(y+4a)} \right] \, f_{2}(a,y) \, , 
\label{d55}
\eea

As we can see, the homogeneous system is completely diagonalized in the
limit $\epsilon \rightarrow 0$. The solution of the system is the 
following:
\bea
f_{1}(a,y) & = & \frac{k_{1}}{y(y+4a)}  \, , 
\label{d56} \\
f_{2}(a,y) & = & \frac{k_{2}}{\sqrt{y(y+4a)}} \, .
\label{d57} 
\eea

By means of the Euler's method we can find, order by order in 
$\epsilon$, the solution of the non-homogeneous system:
\bea
\! \! 
F^{(1)}_{i}(a,Q^{2}) & = & \frac{1}{Q^{2}(Q^{2}+4a)} \Biggl\{
\int^{Q^{2}} dy \, y(y+4a) \biggl[ - 2 \left( \frac{1}{y} \! + \! 
\frac{1}{(y+4a)} \right) F^{(1)}_{i-1}(a,y) \nn\\
\! \! 
& & + \frac{4}{(y+4a)} F^{(1)}_{i-2}(a,y) - \frac{2}{a} \left( 
\frac{1}{y} - 
\frac{1}{(y+4a)} \right) F^{(2)}_{i-1}(a,y) 
\nn\\
\! \! 
& & - \frac{4}{a(y+4a)} F^{(2)}_{i-2}(a,y) + \Omega^{(1)}_{i}(a,y) 
\biggr] \! + \! k^{(1)}_{i} \Biggr\} ,
\label{d58} \\
\! \! 
F^{(2)}_{i}(a,Q^{2}) & = & \frac{1}{\sqrt{Q^{2}(Q^{2}+4a)}} \Biggl\{
\int^{Q^{2}} dy \, \sqrt{y(y+4a)} \biggl[ F^{(1)}_{i-1}(a,y) \nn\\
\! \! 
& & + \left( \frac{1}{y} - \frac{1}{(y+4a)} \right) F^{(2)}_{i-1}(a,y) 
+ \Omega^{(2)}_{i}(a,y) \biggr] \! + \! k^{(2)}_{i} 
\Biggr\} \, , 
\label{d59}
\eea

where, for simplicity, we put:
\bea
\Omega^{(1)}(\epsilon,a,Q^2) & = & \sum_{i=-2}^{0} 
\epsilon^{i} \Omega^{(1)}_{i}(a,Q^{2}) + 
{\mathcal O} \left( \epsilon \right) \, , 
\label{d60} \\
\Omega^{(2)}(\epsilon,a,Q^{2}) & = & \sum_{i=-2}^{0} 
\epsilon^{i} \Omega^{(2)}_{i}(a,Q^{2}) + 
{\mathcal O} \left( \epsilon \right) \, ,
\label{d61} 
\eea

The determination of the constants $k^{(1)}_{i}$ and $k^{(2)}_{i}$ is
made imposing the initial conditions Eqs. (\ref{d50}, \ref{d51}).
The solution, expressed in terms of the variable $x$, is given in Eqs.
(\ref{e56}-\ref{e58}).

\section{Results for the MI's \label{Results}}

We list in this section all the MI's necessary for the calculation of 
the 2-loop vertex diagrams of Fig. (\ref{fig1}).

We give them as a Laurent series in $\epsilon = (4-D)/2$ and we express 
the coefficients of the series in terms of harmonic polylogarithms of 
the variable $x$, already introduced in Eq. (\ref{Q2tox}) 
$$  x = \frac{\sqrt{Q^2+4a} - \sqrt{Q^2} }
          {\sqrt{Q^2+4a} + \sqrt{Q^2} } \ . $$ 

For brevity we present the results only up to the zeroth order in 
$\epsilon$, but of course the method
allows us to calculate any order in $\epsilon$.
All the coefficients of the $\epsilon$ expansion depend of course on 
$a$ and $Q^2$ (or the above variable $x$); for reducing the size of the 
formulas, we will not 
write anymore the dependence of the coefficients on those variables. 
The analytic results are expressed in terms of Harmonic Polylogarithms 
of argument $x$. Definition, notation and properties of the Harmonic 
Polylogarithms can be found in~\cite{Polylog,Polylog3}.

The scale $\mu_{0}$ is the regularization scale and $a=m_{e}^{2}$ is 
the only mass involved (in our case the mass of the electron).

The explicit values of the ${\mathcal D}_i$ is given in appendix \ref{app1}.

For what concerns the normalization of our integrals, we define the 
1-loop Tadpole with mass $a$ as 
\be
T(\epsilon,a) = \mu_{0}^{2(4-D)} \int \{ d^{D}k \} \frac{1}{k^{2}+a}
\, .
\ee

We further define $\{ d^{D}k \}$ in order to have\footnote{
With this normalization the tadpole $T(\epsilon,a)$
of the present paper is 4 times the corresponding quantity of
\cite{Rem1}, \cite{Rem2} and \cite{PieMaRem}.
}:
\be
T(\epsilon,a) = \left( \frac{a}{\mu_{0}^{2}} \right) ^{- \epsilon} 
\frac{a}{\epsilon (\epsilon -1)} \, , 
\label{defT} 
\ee
so that 
\be
\left\{ d^{D}k \right\} = \frac{d^{D}k}{\pi^{\frac{D}{2}} \Gamma 
\left( 3 - \frac{D}{2} \right) } \, . 
\label{defdDk} 
\ee
with $D=4-2\epsilon,\ \epsilon = (4-D)/2$. 

All the results can be downloaded as an input file for FORM in \cite{webpage}.

\subsection{Topologies with $t=3$ \label{3den}}

\bea
\parbox{30mm}{\begin{fmfgraph*}(15,15)
\fmfleft{i}
\fmfright{o}
\fmf{photon}{i,v1}
\fmf{photon}{v2,o}
\fmf{plain,tension=.15,left}{v1,v2}
\fmf{photon,tension=.15}{v1,v2}
\fmf{plain,tension=.15,right}{v1,v2}
\end{fmfgraph*} 
}  \, \, \, & = & \mu_{0}^{2(4-D)} 
\int \{ d^{D}k_{1} \} \{ d^{D}k_{2} \}
\frac{1}{{\mathcal D}_{2} {\mathcal D}_{6} {\mathcal D}_{16} } \\
\, \, \, & = & \left( \frac{a}{\mu_{0}^{2}} \right) ^{-2 \epsilon} 
\sum_{i=-2}^{0} \epsilon^{i} A_{i} + {\mathcal O} \left( 
\epsilon \right) , 
\label{e1} \\
\parbox{30mm}{\begin{fmfgraph*}(15,15)
\fmfleft{i}
\fmfright{o}
\fmf{photon}{i,v1}
\fmf{photon}{v2,o}
\fmflabel{$(k_{1} \cdot k_{2})$ }{o}
\fmf{plain,tension=.15,left}{v1,v2}
\fmf{photon,tension=.15}{v1,v2}
\fmf{plain,tension=.15,right}{v1,v2}
\end{fmfgraph*} 
}  \, \, \, & = & \mu_{0}^{2(4-D)} 
\int \{ d^{D}k_{1} \} \{ d^{D}k_{2} \}
\frac{k_{1} \cdot k_{2}}{{\mathcal D}_{2} {\mathcal D}_{6} {\mathcal D}_{16} } \\
\, \, \, & = & \left( \frac{a}{\mu_{0}^{2}} \right) ^{-2 \epsilon} 
\sum_{i=-2}^{0} \epsilon^{i} B_{i} + {\mathcal O} \left( 
\epsilon \right) . 
\label{e2} 
\eea 
As already said above, from now on we write for short $A_i, B_i$ instead
of $A_i(a,Q^2),\ $ and $B_i(a,Q^2)$. Referring to~\cite{PhDThesis} for 
more details, we find 
\bea
\frac{A_{-2}}{a} & = & - 1 \, ,  
\label{e3} \\
\frac{A_{-1}}{a} & = & - \frac{5}{2} - \frac{1}{4} \left[ x + 
\frac{1}{x} \right] \, ,  
\label{e4} \\
\frac{A_{0}}{a} & = & - \frac{11}{4} - \frac{13}{8} \left[ 
\frac{1}{x} + x \right] - \left[ 1 + \frac{1}{2} \Bigl( 
\frac{1}{x} - x  \Bigr) - \frac{2}{(1-x)} \right] \, H(0,x) \nn\\ 
& & + 2 \, \left[ 1 - \frac{1}{(1-x)} + \frac{1}{(1-x)^{2}} 
\right] \, H(0,0,x) \, ; 
\label{e5} \\
\frac{B_{-2}}{a^{2}} & = & - \frac{1}{4} \left[ x + \frac{1}{x} 
\right] \, , 
\label{e6} \\
\frac{B_{-1}}{a^{2}} & = & -  \frac{1}{24} \left[ \frac{1}{x^{2}} + 
x^{2} \right] - \frac{11}{24} \left[ \frac{1}{x} + x \right] \, , 
\label{e7} \\
\frac{B_{0}}{a^{2}}  & = & - \frac{11}{12} - \frac{1}{48} \left[ 
13 \Bigl( \frac{1}{x^{2}} \! + x^{2} \Bigr) \!
- \! 11 \Bigl( \frac{1}{x} \! + x \Bigr) \right] \! - \frac{1}{2} 
\left[ 1 \! -  \! \frac{2}{(1-x)} \right] H(0,x) \nn\\
& & \! - \frac{1}{12} \left[ \Bigl( \frac{1}{x^{2}} - x^{2} \Bigr) - 7 
\Bigl( \frac{1}{x} - x \Bigr)
\right] \, H(0,x) + \frac{1}{2} \Biggl[ \frac{1}{x} + x - 
\frac{2}{(1-x)} \nn\\
& & \! + \frac{2}{(1-x)^{2}} \Biggr] \, H(0,0,x)  \, .
\label{e8} 
\eea

\bea
\parbox{20mm}{\begin{fmfgraph*}(15,15)
\fmfleft{i}
\fmfright{o}
\fmf{plain}{i,v1}
\fmf{plain}{v2,o}
\fmf{plain,tension=.15,left}{v1,v2}
\fmf{photon,tension=.15}{v2,v1}
\fmf{photon,tension=.15,right}{v1,v2}
\end{fmfgraph*} 
}  & = & \mu_{0}^{2(4-D)} 
\int \{ d^{D}k_{1} \} \{ d^{D}k_{2} \}
\frac{1}{{\mathcal D}_{1} {\mathcal D}_{2} {\mathcal D}_{15} } \\
& = & \left( \frac{a}{\mu_{0}^{2}} \right) ^{-2 \epsilon} 
\sum_{i=-2}^{0} \epsilon^{i} C_{i} + {\mathcal O} \left( 
\epsilon \right) ,
\label{e9} \\
\eea
As mentioned in section \ref{Results}, we calculated this MI directely 
by means of Feynman parameters. We found:
\bea
\frac{C_{-2}}{a} & = & - \frac{1}{2} \, , 
\label{e10} \\
\frac{C_{-1}}{a} & = & - \frac{5}{4} \, ,  
\label{e11} \\
\frac{C_{0}}{a} & = & - \frac{11}{8} - 2 \zeta(2) \, . 
\label{e12} 
\eea

\bea
\parbox{20mm}{\begin{fmfgraph*}(15,15)
\fmfleft{i}
\fmfright{o}
\fmf{photon}{i,v1}
\fmf{photon}{v2,o}
\fmf{plain,tension=.22,left}{v1,v2}
\fmf{plain,tension=.22,right}{v1,v2}
\fmf{plain,right=45}{v2,v2}
\end{fmfgraph*} 
}  & = & \mu_{0}^{2(4-D)} 
\int \{ d^{D}k_{1} \} \{ d^{D}k_{2} \}
\frac{1}{{\mathcal D}_{6} {\mathcal D}_{7} {\mathcal D}_{12} } \\
& = & \left( \frac{a}{\mu_{0}^{2}} \right) ^{-2 \epsilon} 
\sum_{i=-2}^{0} \epsilon^{i} E_{i} + {\mathcal O} \left( 
\epsilon \right) ,
\label{e13} 
\eea
where:
\bea
\frac{E_{-2}}{a}   & = &   - 1 \, , 
\label{e14} \\
\frac{E_{-1}}{a}   & = &   - 3 + \left[ 
1 - \frac{2}{(1-x)} \right] \, H(0,x) \, , 
\label{e15} \\
\frac{E_{0}}{a}   & = &   - 7 - \left[ 
1 - \frac{2}{(1-x)} \right] \Biggl\{ \zeta(2) - 3 H(0,x) - H(0,0,x) 
\nn\\
& &   + 2 H(-1,0,x) \Biggr\} \, .
\label{e16} 
\eea

\bea
\parbox{20mm}{\begin{fmfgraph*}(15,15)
\fmfleft{i}
\fmfright{o}
\fmf{plain}{i,v1}
\fmf{plain}{v2,o}
\fmf{plain,tension=.15,left}{v1,v2}
\fmf{plain,tension=.15}{v1,v2}
\fmf{plain,tension=.15,right}{v1,v2}
\end{fmfgraph*} } & = & \mu_{0}^{2(4-D)} 
\int \{ d^{D}k_{1} \} \{ d^{D}k_{2} \}
\frac{1}{{\mathcal D}_{6} {\mathcal D}_{7} {\mathcal D}_{15} } \\
& = & \left( \frac{a}{\mu_{0}^{2}} \right) ^{-2 \epsilon} 
\sum_{i=-2}^{0} \epsilon^{i} F_{i} + {\mathcal O} \left( 
\epsilon \right) ,
\label{e17} 
\eea
where \cite{PieMaRem}:
\bea
\frac{F_{-2}}{a}   & = &   - \frac{3}{2} \, , 
\label{e18} \\
\frac{F_{-1}}{a}   & = &   - \frac{17}{4} \, , 
\label{e19} \\
\frac{F_{0}}{a}   & = &   - \frac{59}{8} \, .
\label{e20} 
\eea

\subsection{Topologies with $t=4$ \label{4den}}

\bea
\parbox{20mm}{\begin{fmfgraph*}(15,15)
\fmfleft{i}
\fmfright{o}
\fmf{photon}{i,v1}
\fmf{photon}{v3,o}
\fmf{plain,tension=.2,left}{v1,v2}
\fmf{plain,tension=.2,right}{v1,v2}
\fmf{plain,tension=.2,left}{v2,v3}
\fmf{plain,tension=.2,right}{v2,v3}
\end{fmfgraph*} } & = & \mu_{0}^{2(4-D)} 
\int \{ d^{D}k_{1} \} \{ d^{D}k_{2} \}
\frac{1}{{\mathcal D}_{6} {\mathcal D}_{7} {\mathcal D}_{12}  {\mathcal D}_{13}} \\
& = & \left( \frac{a}{\mu_{0}^{2}} \right) ^{-2 \epsilon} 
\sum_{i=-2}^{0} \epsilon^{i} G_{i} + {\mathcal O} \left( 
\epsilon \right) ,
\label{e21} 
\eea
where:
\bea
G_{-2} & = & 1 , 
\label{e22} \\
G_{-1} & = & 4 - 2 \left[ 1 - \frac{2}{(1-x)} 
\right] H(0,x) \, , 
\label{e23} \\
G_{0} & = & 12 + 2 \left[ 1 - \frac{2}{(1-x)} 
\right] \Biggl\{ \zeta(2) - 4 H(0,x) + 2 H(-1,0,x) \nn\\
& & -  \frac{2}{(1-x)} H(0,0,x) \Biggr\} .
\label{e24} 
\eea

\bea
\parbox{20mm}{\begin{fmfgraph*}(15,15)
\fmfleft{i1,i2}
\fmfright{o}
\fmf{plain}{i1,v1}
\fmf{plain}{i2,v2}
\fmf{photon}{v3,o}
\fmf{plain,tension=.3}{v2,v3}
\fmf{plain,tension=.3}{v1,v3}
\fmf{photon,tension=0,right=.5}{v2,v1}
\fmf{photon,tension=0,right=.5}{v1,v2}
\end{fmfgraph*} 
} & = & \mu_{0}^{2(4-D)} 
\int \{ d^{D}k_{1} \} \{ d^{D}k_{2} \}
\frac{1}{{\mathcal D}_{1} {\mathcal D}_{2} {\mathcal D}_{14} {\mathcal D}_{15}} \\
& = & \left( \frac{a}{\mu_{0}^{2}} \right) ^{-2 \epsilon} 
\sum_{i=-2}^{0} \epsilon^{i} I_{i} + {\mathcal O} \left( 
\epsilon \right) ,
\label{e25} 
\eea
where:
\bea
I_{-2} & = & \frac{1}{2} \, , 
\label{e26} \\
I_{-1} & = & \frac{5}{2} - \, \left[ 1 - \frac{2}{(1-x)} 
\right] H(0,x) , 
\label{e27} \\
I_{0} & = & \frac{19}{2} + 2 \zeta(2) + \left[ 
1 - \frac{2}{(1-x)} \right] \bigl\{ 2 \zeta(2) - 5 H(0,x) - 2 H(0,0,x) 
\nn\\
& & + 4 H(-1,0,x) \bigr\}  .
\label{e28} 
\eea

\bea
\parbox{20mm}{\begin{fmfgraph*}(15,15)
\fmfleft{i1,i2}
\fmfright{o}
\fmf{plain}{i1,v1}
\fmf{plain}{i2,v2}
\fmf{photon}{v3,o}
\fmf{plain,tension=.3}{v2,v3}
\fmf{plain,tension=.3}{v1,v3}
\fmf{photon,tension=0}{v2,v1}
\fmf{photon,tension=0,right=.5}{v2,v3}
\end{fmfgraph*} 
} & = & \mu_{0}^{2(4-D)} 
\int \{ d^{D}k_{1} \} \{ d^{D}k_{2} \}
\frac{1}{{\mathcal D}_{1} {\mathcal D}_{2} {\mathcal D}_{10} {\mathcal D}_{14}} \\
& = & \left( \frac{a}{\mu_{0}^{2}} \right) ^{-2 \epsilon} 
\sum_{i=-2}^{0} \epsilon^{i} J_{i} + {\mathcal O} \left( 
\epsilon \right) ,
\label{e29} 
\eea
where:
\bea
J_{-2}(x) & = & \frac{1}{2} \, , 
\label{e30} \\
J_{-1}(x) & = & \frac{5}{2} \, , 
\label{e31} \\
J_{0}(x)  & = & \frac{19}{2} + 2 \biggl[ \frac{1}{(1-x)} - 
\frac{1}{(1+x)} \biggr] \Bigl\{ \zeta(2) H(0,x) + H(0,0,0,x) 
\Bigr\} \nn\\
& & - H(0,0,x) .
\label{e32} 
\eea


\bea
\parbox{20mm}{\begin{fmfgraph*}(15,15)
\fmfleft{i1,i2}
\fmfright{o}
\fmf{plain}{i1,v1}
\fmf{plain}{i2,v2}
\fmf{photon}{v3,o}
\fmf{plain,tension=.3}{v2,v3}
\fmf{photon,tension=.3}{v1,v3}
\fmf{plain,tension=0}{v2,v1}
\fmf{plain,tension=0,right=.5}{v2,v3}
\end{fmfgraph*} }  & = & \mu_{0}^{2(4-D)} 
\int \{ d^{D}k_{1} \} \{ d^{D}k_{2} \}
\frac{1}{{\mathcal D}_{2} {\mathcal D}_{6} {\mathcal D}_{11} {\mathcal D}_{16}} \\
& = & \left( \frac{a}{\mu_{0}^{2}} \right) ^{-2 \epsilon} 
\sum_{i=-2}^{0} \epsilon^{i} K_{i} + {\mathcal O} \left( 
\epsilon \right) , 
\label{e33} \\
\parbox{20mm}{\begin{fmfgraph*}(15,15)
\fmfleft{i1,i2}
\fmfright{o}
\fmfforce{0.22w,0.5h}{v4}
\fmf{plain}{i1,v1}
\fmf{plain}{i2,v2}
\fmf{photon}{v3,o}
\fmf{plain,tension=.3}{v2,v3}
\fmf{photon,tension=.3}{v1,v3}
\fmf{plain,tension=0}{v2,v1}
\fmf{plain,tension=0,right=.5}{v2,v3}
\fmfv{decor.shape=circle,decor.filled=full,decor.size=.1w}{v4}
\end{fmfgraph*} }  & = & \mu_{0}^{2(4-D)} 
\int \{ d^{D}k_{1} \} \{ d^{D}k_{2} \}
\frac{1}{{\mathcal D}_{2} {\mathcal D}_{6} {\mathcal D}^{2}_{11} {\mathcal D}_{16}} \\
& = & \left( \frac{a}{\mu_{0}^{2}} \right) ^{-2 \epsilon} 
\sum_{i=-2}^{0} \epsilon^{i} L_{i} + {\mathcal O} \left( 
\epsilon \right) , 
\label{e34} 
\eea
where: 
\bea
K_{-2} & = & \frac{1}{2} \, , 
\label{e35} \\
K_{-1} & = & \frac{5}{2} \, , 
\label{e36} \\
K_{0} & = & \frac{19}{2} - 2 \zeta(2) + 2
\Biggl[ \frac{1}{(1-x)} - \frac{1}{(1+x)} \Biggr] \bigl[ \zeta(2) H(0,x)
+ H(0,0,0,x) \bigr] \nn\\
& & -  H(0,0,x) \, . 
\label{e37} \\
a L_{-2} & = & - \frac{1}{2} \, , 
\label{e38} \\
a L_{-1} & = & -1  + \frac{1}{2} \Biggl[ 1 - \frac{2}{(1-x)} \Biggr] 
H(0,x) \, , 
\label{e39} \\
a L_{0} & = & - \! 2  + \! \frac{1}{2} \Biggl[ 1 \! - \! 
\frac{2}{(1-x)} \Biggr] \Biggl\{ \zeta(2) + 2 H(0,x) + 4 H(0,0,x) + 
2 H(1,0,x) \nn\\
& & - 6 H(-1,0,x) \Biggr\} + \frac{3}{2} H(0,0,x)
\, .
\label{e40} 
\eea

\bea
\parbox{30mm}{\begin{fmfgraph*}(15,15)
\fmfleft{i1,i2}
\fmfright{o}
\fmf{plain}{i1,v1}
\fmf{plain}{i2,v2}
\fmf{photon}{v3,o}
\fmf{plain,tension=.3}{v2,v3}
\fmf{plain,tension=.3}{v1,v3}
\fmf{plain,tension=0,right=.5}{v2,v1}
\fmf{plain,tension=0,right=.5}{v1,v2}
\end{fmfgraph*} }  & = & \mu_{0}^{2(4-D)} 
\int \{ d^{D}k_{1} \} \{ d^{D}k_{2} \}
\frac{1}{{\mathcal D}_{6} {\mathcal D}_{7} {\mathcal D}_{14} {\mathcal D}_{15}} \\
& = & \left( \frac{a}{\mu_{0}^{2}} \right) ^{-2 \epsilon} 
\sum_{i=-2}^{0} \epsilon^{i} M_{i} + {\mathcal O} \left( 
\epsilon \right) , 
\label{e41} \\
\parbox{30mm}{\begin{fmfgraph*}(15,15)
\fmfleft{i1,i2}
\fmfright{o}
\fmf{plain}{i1,v1}
\fmf{plain}{i2,v2}
\fmf{photon}{v3,o}
\fmflabel{$(p_{2} \cdot k_{1})$ }{o}
\fmf{plain,tension=.3}{v2,v3}
\fmf{plain,tension=.3}{v1,v3}
\fmf{plain,tension=0,right=.5}{v2,v1}
\fmf{plain,tension=0,right=.5}{v1,v2}
\end{fmfgraph*} }  & = & \mu_{0}^{2(4-D)} 
\int \{ d^{D}k_{1} \} \{ d^{D}k_{2} \}
\frac{p_{2} \cdot k_{1}}{{\mathcal D}_{6} {\mathcal D}_{7} {\mathcal D}_{14} 
{\mathcal D}_{15}} \\
& = & \left( \frac{a}{\mu_{0}^{2}} \right) ^{-2 \epsilon} 
\sum_{i=-2}^{0} \epsilon^{i} N_{i} + {\mathcal O} \left( 
\epsilon \right) , 
\label{e42} 
\eea
where: 
\bea
M_{-2} & = & \frac{1}{2} \, ,  
\label{e43} \\
M_{-1} & = & \frac{5}{2} - \Biggl[ 1 - \frac{2}{(1-x)} \Biggr] H(0,x) \,
,  
\label{e44} \\
M_{0} & = & \frac{19}{2} + \zeta(2) + 
\Biggl[ 1 - \frac{2}{(1-x)} \Biggr] \bigl[ \zeta(2) - 5 H(0,x) + 2 
H(-1,0,x)  \bigr] \nn\\
& &  + \frac{2}{(1-x)} H(0,0,x) + \Biggl[ \frac{1}{(1-x)} - \frac{1}{
(1+x)} \Biggr] \bigl[ \zeta(2) H(0,x) \nn\\
& & + H(0,0,0,x) \bigr]   \, .  
\label{e45} \\
\frac{N_{-2}}{a} & = & \frac{1}{8} + \frac{1}{16} \Biggl[ x + 
\frac{1}{x} \Biggr] \, ,  
\label{e46} \\
\frac{N_{-1}}{a} & = & \frac{9}{32} \Biggl[ 2 + x + \frac{1}{x} \Biggr]
- \frac{1}{8} \Biggl[ 4 + x - \frac{1}{x} \Biggr] H(0,x) +
\frac{1}{(1-x)} H(0,x) \, ,  
\label{e47} \\
\frac{N_{0}}{a} & = & \frac{63}{32} + \frac{\zeta(2)}{2} + \frac{63}{64}
\Biggl[ \left( 1 + \frac{16}{63} \zeta(2) \right) x + \frac{1}{x} \Biggr] -
\frac{\zeta(2)}{(1-x)} - \frac{1}{16} \Biggl[ 32 + 9 x \nn\\
& & - \frac{9}{x} \Biggr] H(0,x) + 
\frac{(16 + \zeta(2))}{4(1-x)} H(0,x) - \frac{\zeta(2)}{4(1+x)} H(0,x) 
- \frac{1}{4} \Biggl[ 2 \! - \! \frac{1}{x} \nn\\
& &  - \! 
\frac{4}{(1-x)} \Biggr] H(0,0,x) + \frac{1}{4} \Biggl[ 4 \! + \! x \! - 
\! \frac{1}{x} - \frac{8}{(1-x)} \Biggr] H(\! -1, \! 0,x) \nn\\
& & + \frac{1}{4} \Biggl[ \frac{1}{(1-x)} -  \frac{1}{(1+x)}
\Biggr] H(0,0,0,x) 
\label{e48} 
\, .
\eea

\subsection{Topologies with $t=5$ \label{5den}}

\bea
\parbox{20mm}{\begin{fmfgraph*}(15,15)
\fmfleft{i1,i2}
\fmfright{o}
\fmf{plain}{i1,v1}
\fmf{plain}{i2,v2}
\fmf{photon}{v4,o}
\fmf{plain,tension=.4}{v2,v3}
\fmf{plain,tension=.2}{v3,v4}
\fmf{plain,tension=.15}{v1,v4}
\fmf{photon,tension=0}{v2,v1}
\fmf{photon,tension=0}{v1,v3}
\end{fmfgraph*} 
}  & = & \mu_{0}^{2(4-D)} 
\int \{ d^{D}k_{1} \} \{ d^{D}k_{2} \}
\frac{1}{{\mathcal D}_{1} {\mathcal D}_{2} {\mathcal D}_{9} {\mathcal D}_{14}
{\mathcal D}_{15}} \\
& = & \left( \frac{a}{\mu_{0}^{2}} \right) ^{-2 \epsilon} 
P_{0} + {\mathcal O} \left( 
\epsilon \right) , 
\label{e49} \\
\parbox{20mm}{\begin{fmfgraph*}(15,15)
\fmfleft{i1,i2}
\fmfright{o}
\fmfforce{0.5w,0.3h}{v5}
\fmf{plain}{i1,v1}
\fmf{plain}{i2,v2}
\fmf{photon}{v4,o}
\fmf{plain,tension=.4}{v2,v3}
\fmf{plain,tension=.2}{v3,v4}
\fmf{plain,tension=.15}{v1,v4}
\fmf{photon,tension=0}{v2,v1}
\fmf{photon,tension=0}{v1,v3}
\fmfv{decor.shape=circle,decor.filled=full,decor.size=.1w}{v5}
\end{fmfgraph*} 
}  & = & \mu_{0}^{2(4-D)} 
\int \{ d^{D}k_{1} \} \{ d^{D}k_{2} \}
\frac{1}{{\mathcal D}_{1} {\mathcal D}_{2} {\mathcal D}_{9} {\mathcal D}_{14}
{\mathcal D}^{2}_{15}} \\
& = & \left( \frac{a}{\mu_{0}^{2}} \right) ^{-2 \epsilon} 
\sum_{i=-1}^{0} \epsilon^{i} Q_{i} + {\mathcal O} \left( 
\epsilon \right) , 
\label{e50} 
\eea
where: 
\bea
a P_{0} & = & \frac{1}{2} \left[ \frac{1}{(1-x)} - \frac{1}{(1+x)} 
\right] \biggl\{ \frac{17 \zeta^{2}(2)}{10} - 4 \zeta(3) H(0,x) + 
\zeta(2) H(0,0,x) \nn\\
& & + 4 \zeta(2) H(1,0,x) + 4 H(1,0,0,0,x) - 2 H(0,1,0,0,x) \nn\\
& & - 2 H(0,0,-1,0,x) + 4 H(0,-1,0,0,x) 
\biggr\} \, , 
\label{e51} \\
a^{2} Q_{-1} & = & \frac{1}{4} \left[ \frac{1}{(1-x)} - 
\frac{1}{(1-x)^{2}} + \frac{1}{(1+x)} - \frac{1}{(1+x)^{2}} \right] 
H(0,0,x) \, , 
\label{e52} \\
a^{2} Q_{0} & = & \frac{1}{4} \left[ \frac{1}{(1-x)} \! - \! 
\frac{1}{(1-x)^{2}} \! + \! \frac{1}{(1+x)} \! - \! \frac{1}{(1+x)^{2}} 
\right] \! \Bigl\{ \zeta(3) + \! 5 H(0,0,0,x) \nn\\
& & + 2 H(0,1,0,x) - 4 H(0,-1,0,x) - 4 H(-1,0,0,x) \Bigr\} \, .
\label{e53} 
\eea

\subsection{Topologies with $t=6$ \label{6den}}

\bea
\parbox{30mm}{\begin{fmfgraph*}(15,15)
\fmfleft{i1,i2}
\fmfright{o}
\fmf{plain}{i1,v1}
\fmf{plain}{i2,v2}
\fmf{photon}{v5,o}
\fmf{plain,tension=.3}{v2,v3}
\fmf{plain,tension=.3}{v3,v5}
\fmf{plain,tension=.3}{v1,v4}
\fmf{plain,tension=.3}{v4,v5}
\fmf{photon,tension=0}{v2,v4}
\fmf{photon,tension=0}{v1,v3}
\end{fmfgraph*} 
}  & = & \mu_{0}^{2(4-D)} 
\int \{ d^{D}k_{1} \} \{ d^{D}k_{2} \}
\frac{1}{{\mathcal D}_{1} {\mathcal D}_{2} {\mathcal D}_{9} 
{\mathcal D}_{11} {\mathcal D}_{14} {\mathcal D}_{15} } \\
& = & \left( \frac{a}{\mu_{0}^{2}} \right) ^{-2 \epsilon} 
\sum_{i=-1}^{0} \epsilon^{i} R_{i} + {\mathcal O} \left( 
\epsilon \right) , 
\label{e54} \\
\parbox{30mm}{\begin{fmfgraph*}(15,15)
\fmfleft{i1,i2}
\fmfright{o}
\fmf{plain}{i1,v1}
\fmf{plain}{i2,v2}
\fmf{photon}{v5,o}
\fmflabel{$(k_{1} \cdot k_{2})$}{o}
\fmf{plain,tension=.3}{v2,v3}
\fmf{plain,tension=.3}{v3,v5}
\fmf{plain,tension=.3}{v1,v4}
\fmf{plain,tension=.3}{v4,v5}
\fmf{photon,tension=0}{v2,v4}
\fmf{photon,tension=0}{v1,v3}
\end{fmfgraph*} 
}  & = & \mu_{0}^{2(4-D)} 
\int \{ d^{D}k_{1} \} \{ d^{D}k_{2} \}
\frac{k_{1} \cdot k_{2}}{{\mathcal D}_{1} {\mathcal D}_{2} {\mathcal D}_{9} 
{\mathcal D}_{11} {\mathcal D}_{14} {\mathcal D}_{15} } \\
& = & \left( \frac{a}{\mu_{0}^{2}} \right) ^{-2 \epsilon} 
S_{0} + {\mathcal O} \left( 
\epsilon \right) , 
\label{e55} 
\eea
where: 
\bea
a^{2} R_{-1} & = & - \frac{1}{4} \left[ \frac{1}{(1-x)} \! - \!
\frac{1}{(1-x)^{2}} \! + \! \frac{1}{(1+x)} \! - \! \frac{1}{(1+x)^{2}} 
\right] \bigl[ \zeta(3) \! + \! \zeta(2) H(0,x) \nn\\
& & + 2 H(0,0,0,x) + 2 H(0,1,0,x) - 2 H(0,-1,0,x) \bigr] \, , 
\label{e56} \\
a^{2} R_{0} & = & - \frac{1}{4} \left[ \frac{1}{(1-x)} \! - \! 
\frac{1}{(1-x)^{2}} \! + \! \frac{1}{(1+x)} \! - \! \frac{1}{(1+x)^{2}} 
\right] \! \Biggl[ \frac{37 \zeta^{2}(2)}{10} \! + \! H(0,x) \nn\\
& & - 4 H(-1,x) + 4 \zeta(3) H(1,x) - 2 \zeta(2) H(0,0,x) \nn\\
& & - 4 \zeta(2) H(-1,0,x) - 2 \zeta(2) H(0,-1,x) - 2 \zeta(2) H(0,1,x)
\nn\\
& & + 4 \zeta(2) H(1,0,x) + 12 H(0,0,0,0,x) + 8 H(-1,0,-1,0,x) \nn\\
& & - 8 H(-1,0,0,0,x) - 8 H(-1,0,1,0,x) + 20 H(0,-1,-1,0,x) \nn\\
& & - 16 H(0,-1,0,0,x) - 12 H(0,-1,1,0,x) - 24 H(0,0,-1,0,x) \nn\\
& & + 16 H(0,0,1,0,x) - 12 H(0,1,-1,0,x) + 8 H(0,1,0,0,x) \nn\\
& & + 4 H(0,1,1,0,x) - 8 H(1,0,-1,0,x) + 8 H(1,0,0,0,x) \nn\\
& & + 8 H(1,0,1,0,x) \Biggr] \, , 
\label{e57} \\
a S_{0} & = & \left[ \frac{1}{(1+x)} - \frac{1}{(1-x)} \right] \Biggl\{ 
\frac{\zeta^{2}(2)}{10} - \zeta(3) H(0,x) + \zeta(2) \bigl( 2 H(1,0,x)
\nn\\
& & + 3 H(0,-1,x) \bigr) + \frac{1}{2}  H(0,0,0,0,x) + H(0,-1,0,0,x) 
\nn\\
& & + H(0,0,-1,0,x) + H(0,1,0,0,x) + 2 H(1,0,0,0,x) \Biggr\} \, .
\label{e58} 
\eea

\section{The 6-denominator reducible diagrams \label{Results2}}

The other diagrams of Fig. (\ref{fig1})
are all reducible diagrams. We give in this section their result.

\bea
\parbox{20mm}{\begin{fmfgraph*}(15,15)
\fmfleft{i1,i2}
\fmfright{o}
\fmf{plain}{i1,v1}
\fmf{plain}{i2,v2}
\fmf{photon}{v5,o}
\fmf{plain,tension=.3}{v2,v3}
\fmf{plain,tension=.3}{v3,v5}
\fmf{plain,tension=.3}{v1,v4}
\fmf{plain,tension=.3}{v4,v5}
\fmf{photon,tension=0}{v2,v1}
\fmf{photon,tension=0}{v4,v3}
\end{fmfgraph*} } 
& = & \mu_{0}^{2(4-D)} 
\int \{ d^{D}k_{1} \} \{ d^{D}k_{2} \}
\frac{1}{{\mathcal D}_{1} {\mathcal D}_{2} {\mathcal D}_{9} 
{\mathcal D}_{10} {\mathcal D}_{14} {\mathcal D}_{15} } \\
& = & \left( \frac{a}{\mu_{0}^{2}} \right) ^{-2 \epsilon} 
\sum_{i=-2}^{0} \epsilon^{i} F^{(1)}_{i} + {\mathcal O} \left( 
\epsilon \right) , 
\label{e59} 
\eea
where:
\bea
a^{2} F^{(1)}_{-2}  & = &  - \frac{1}{
4} \!  \left[  \frac{1}{(1-x)} \! -  \! \frac{1}{(1-x)^{2}}  \! +  \! 
\frac{1}{(1+x)} \! - \! \frac{1}{(1+x)^{2}}  \right]  \! H(0,0,x) , 
\label{e60} \\
a^{2} F^{(1)}_{-1}  & = &  \frac{1}{4} \!  \left[ \frac{1}{(1-x)} \! - 
\!\frac{1}{(1-x)^{2}} \! + \! \frac{1}{(1+x)} \! - \! 
\frac{1}{(1+x)^{2}}
\right] \Bigl\{ \zeta(3) \! + \! 2 \zeta(2) \! H(0,x) \nn\\
& &   - H(0,0,0,x) + 4 H(-1,0,0,x) + 2 H(0,1,0,x) \Bigr\} \, ,  
\label{e61} \\
a^{2} F^{(1)}_{0}  & = &  - \frac{1}{4} \biggl[ \frac{1}{(1-x)} - 
\frac{1}{(1-x)^{2}} + \frac{1}{(1+x)} - \frac{1
}{(1+x)^{2}} \biggr] \Bigl\{ \frac{12}{5}\zeta^{2}(2) \nn\\
& & - \zeta(3) [ H(0,x) - 4 H(-1,x)] + \zeta(2) [ 7 H(0,0,x) + 2 
H(0,1,x) \nn\\
& &  + 8 H(-1,0,x) + 4 H(0,-1,x) ] + 5 H(0,0,0,0,x) \nn\\
& &  + 16 H(-1,-1,0,0,x) - 4 H(-1,0,0,0,x) + 8 H(-1,0,1,0,x) \nn\\
& &  - 16 H(0,-1,-1,0,x) + 6 H(0,-1,0,0,x) + 12 H(0,-1,1,0,x) \nn\\
& &  + 14 H(0,0,-1,0,x) - 12 H(0,0,1,0,x) + 12 H(0,1,-1,0,x) \nn\\
& &  - 8 H(0,1,0,0,x) - 4 H(0,1,1,0,x) \Bigr\} \, . 
\label{e62} 
\eea

\bea
\parbox{20mm}{\begin{fmfgraph*}(15,15)
\fmfleft{i1,i2}
\fmfright{o}
\fmfforce{0.2w,0.93h}{v2}
\fmfforce{0.2w,0.07h}{v1}
\fmfforce{0.2w,0.5h}{v3}
\fmfforce{0.8w,0.5h}{v5}
\fmf{plain}{i1,v1}
\fmf{plain}{i2,v2}
\fmf{photon}{v5,o}
\fmf{plain,tension=0}{v2,v5}
\fmf{plain,tension=0}{v3,v4}
\fmf{photon,tension=.4}{v1,v4}
\fmf{plain,tension=.4}{v4,v5}
\fmf{plain,tension=0}{v1,v3}
\fmf{photon,tension=0}{v2,v3}
\end{fmfgraph*} }  & = & \mu_{0}^{2(4-D)} 
\int \{ d^{D}k_{1} \} \{ d^{D}k_{2} \}
\frac{1}{{\mathcal D}_{1} {\mathcal D}_{2} {\mathcal D}_{9} 
{\mathcal D}_{10} {\mathcal D}_{11} {\mathcal D}_{15} } \\
& = & \left( \frac{a}{\mu_{0}^{2}} \right) ^{-2 \epsilon} 
\sum_{i=-2}^{0} \epsilon^{i} F^{(2)}_{i} + {\mathcal O} \left( 
\epsilon \right) , 
\label{e63} 
\eea
where:
\bea
a^{2} F^{(2)}_{-2} & = & - \frac{1}{8} \Biggl[ \frac{1}{(1-x)} -
\frac{1}{(1+x)} \Biggr] H(0,x) \, , 
\label{e64} \\
a^{2} F^{(2)}_{-1} & = & \frac{1}{8} \Biggl[ \frac{1}{(1-x)} -
\frac{1}{(1+x)} \Biggr] \bigl[ \zeta(2) + 2 H(0,x) + 2 H(0,0,x) \nn\\
& & - 2 H(-1,0,x) + 2 H(1,0,x) \bigr] \, , 
\label{e65} \\
a^{2} F^{(2)}_{0} & = & - \frac{3 \zeta(2) \ln{2}}{(1+x)} \Biggl[ 1 -
\frac{1}{(1+x)} \Biggr] - \frac{\zeta(2)}{4} \Biggl[ \frac{1}{(1-x)} -
\frac{1}{(1+x)} \Biggr] \nn\\
& & - \frac{\zeta(3)}{8} \Biggl[ \frac{7}{(1-x)} - \frac{1}{(1+x)} - 
\frac{6}{(1+x)^{2}} \Biggr] - \frac{1}{2} \Biggr[ \,
\frac{(1+2 \zeta(2))}{(1-x)} \\
& & - \frac{(1 \! - \! \zeta(2))}{(1+x)} \! - \! 
\frac{3 \zeta(2)}{(1+x)^{2}} \! \Biggr] \! H(0,x) \! - \! 
\frac{\zeta(2)}{4} \! \Biggl[ \! \frac{1}{(1-x)} \! - \!\frac{1}{(1+x)} 
\Biggr] \! H(1,x) \nn\\
& & + \frac{\zeta(2)}{4} \Biggl[ \frac{1}{(1-x)} +
\frac{11}{(1+x)} - \frac{12}{(1+x)^{2}} \Biggr] H(-1,x) \nn\\
& & - \frac{1}{2} \Biggl[ \frac{1}{(1-x)} - \frac{1}{(1+x)} \Biggr]
\bigl[ H(0,0,x) - H(-1,0,x) + H(1,0,x) \bigr] \nn\\
& & + \frac{1}{4} \Biggl[ \frac{1}{(1-x)} - \frac{7}{(1+x)} + 
\frac{6}{(1+x)^{2}} \Biggr] H(0,0,0,x) \nn\\
& & - \frac{1}{2} \Biggl[ \frac{5}{(1-x)} - \frac{7}{(1+x)} + 
\frac{2}{(1+x)^{2}} \Biggr] H(-1,0,0,x) \nn\\
& & - \frac{1}{2} \Biggl[ \frac{5}{(1-x)} - \frac{9}{(1+x)} + 
\frac{4}{(1+x)^{2}} \Biggr] H(0,-1,0,x) \nn\\
& & + \Biggl[ \frac{1}{(1-x)} - \frac{2}{(1+x)} + 
\frac{1}{(1+x)^{2}} \Biggr] H(0,1,0,x) \nn\\
& & + \Biggl[ \frac{1}{(1-x)} - \frac{2}{(1+x)^{2}} \Biggr] H(1,0,0,x)
+ \frac{1}{2} \Biggl[ \frac{1}{(1-x)} \nn\\
& & - \frac{1}{(1+x)} \Biggr] \bigl[ 7 H(-1,-1,0,x) - 3 H(-1,1,0,x) 
- 3 H(1,-1,0,x) \nn\\
& & + H(1,1,0,x) \bigr]  \, .
\label{e66} 
\eea

\bea
\parbox{20mm}{\begin{fmfgraph*}(15,15)
\fmfleft{i1,i2}
\fmfright{o}
\fmfforce{0.2w,0.93h}{v2}
\fmfforce{0.2w,0.07h}{v1}
\fmfforce{0.8w,0.5h}{v5}
\fmf{plain}{i1,v1}
\fmf{plain}{i2,v2}
\fmf{photon}{v5,o}
\fmf{plain}{v2,v3}
\fmf{photon,tension=.25,right}{v3,v4}
\fmf{plain,tension=.25}{v3,v4}
\fmf{plain}{v4,v5}
\fmf{plain}{v1,v5}
\fmf{photon}{v1,v2}
\end{fmfgraph*} }  & = & \mu_{0}^{2(4-D)} 
\int \{ d^{D}k_{1} \} \{ d^{D}k_{2} \}
\frac{1}{{\mathcal D}_{1} {\mathcal D}_{2} {\mathcal D}^{2}_{9} 
{\mathcal D}_{10} {\mathcal D}_{14}} \\
& = & \left( \frac{a}{\mu_{0}^{2}} \right) ^{-2 \epsilon} 
\sum_{i=-2}^{0} \epsilon^{i} F^{(3)}_{i} + {\mathcal O} \left( 
\epsilon \right) , 
\label{e67} 
\eea
where:
\bea
a^{2} F^{(3)}_{-2} & = & \frac{1}{8} \Biggl[ \frac{1}{(1-x)} - 
\frac{1}{(1+x)} \Biggr] H(0,x) \, , 
\label{e68} \\
a^{2} F^{(3)}_{-1} & = & - \frac{\zeta(2)}{8} \Biggl[ \frac{1}{(1-x)} - 
\frac{1}{(1+x)} \Biggr] - \frac{1}{(1+x)} + \frac{1}{(1+x)^{2}}
+ \frac{1}{2} \Biggl[ \frac{1}{(1-x)} \nn\\
& & - \frac{3}{(1+x)^{2}} \! + \! \frac{2}{(1+x)^{3}} \Biggr] \! H(0,x) 
\! - \! \frac{1}{4} \Biggl[ \frac{1}{(1-x)} \! - \! \frac{1}{(1+x)} 
\Biggr] \bigl[ H(0,0,x)\nn\\
& &  - H(-1,0,x) + H(1,0,x) \bigr] \, , 
\label{e69} \\
a^{2} F^{(3)}_{0} & = & - \frac{\zeta(2)}{2} \Biggl[ \frac{1}{(1-x)} \! 
- \! \frac{3}{(1+x)^{2}} \! + \! \frac{2}{(1+x)^{3}} \Biggr] \! + \! 
\frac{7 \zeta(3)}{8} \! \Biggl[ \! \frac{1}{(1-x)} \! - \! 
\frac{1}{(1+x)} \! \Biggr] \nn\\
& & + \Biggl\{ \Biggl[ \frac{1}{(1-x)} - \frac{3}{(1+x)^{2}} + 
\frac{2}{(1+x)^{3}} \Biggr] +  \frac{7 \zeta(3)}{4} 
\Biggl[ \frac{1}{(1-x)} \nn\\
& & - \frac{1}{(1+x)} \Biggr] \! \Biggr\} \! H(0,x) \! - \! 
\frac{\zeta(2)}{4} \Biggl[ \frac{1}{(1-x)} - \frac{1}{(1+x)} 
\Biggr] \bigl[ H(-1,x) \! \nn\\
& & - \! H(1,x) \bigr] + \Biggl\{ \frac{1}{2} \Biggl[ \frac{1}{(1+x)} - 
\frac{3}{(1+x)^{2}} + \frac{2}{(1+x)^{3}} \Biggr] -  \frac{1}{4} 
\Biggl[ \frac{1}{(1-x)} \nn\\
& & - \frac{1}{(1+x)} \! \Biggr] \! \Biggr\} \! H(0,0,x) \! - \! \Biggl[
\frac{1}{(1+x)} \! - \! \frac{3}{(1+x)^{2}} \! + \! 
\frac{2}{(1+x)^{3}} \! \Biggr] \! H(-1,0,x) \nn\\
& & - \frac{1}{2} \! \Biggl[ \! \frac{1}{(1-x)} \! - \! \frac{1}{(1+x)} 
\! \Biggr] \! \bigl[ \! H(1,0,x) \! - \! H(0,0,0,x) \! + \! 7 \! 
H(-1,-1,0,x) \nn\\
& & - \! 5 \! H( \! -1,0,0,x) \! - \! 3 H( \! -1,1,0,x) \! - \! 5 
H(0, \! -1,0,x) \! + \! 2 \! H(0,1,0,x) \nn\\
& & - 3 H(1,-1,0,x) + 2 H(1,0,0,x) + H(1,1,0,x) \bigr] \, .
\label{e70} 
\eea

\bea
\parbox{20mm}{\begin{fmfgraph*}(15,15)
\fmfleft{i1,i2}
\fmfright{o}
\fmfforce{0.2w,0.93h}{v2}
\fmfforce{0.2w,0.07h}{v1}
\fmfforce{0.2w,0.3h}{v3}
\fmfforce{0.2w,0.7h}{v4}
\fmfforce{0.8w,0.5h}{v5}
\fmf{plain}{i1,v1}
\fmf{plain}{i2,v2}
\fmf{photon}{v5,o}
\fmf{plain}{v2,v5}
\fmf{photon}{v1,v3}
\fmf{photon}{v2,v4}
\fmf{plain}{v1,v5}
\fmf{plain,left}{v3,v4}
\fmf{plain,right}{v3,v4}
\end{fmfgraph*} }  & = & \mu_{0}^{2(4-D)} 
\int \{ d^{D}k_{1} \} \{ d^{D}k_{2} \}
\frac{1}{{\mathcal D}^{2}_{1} {\mathcal D}_{7} {\mathcal D}_{8} 
{\mathcal D}_{9} {\mathcal D}_{10}} \\
& = & \left( \frac{a}{\mu_{0}^{2}} \right) ^{-2 \epsilon} 
\sum_{i=-1}^{0} \epsilon^{i} F^{(4)}_{i} + {\mathcal O} \left( 
\epsilon \right) , 
\label{e71} 
\eea
where:
\bea
a^{2} F^{(4)}_{-1} & = & \frac{1}{(1+x)} - \frac{1}{(1+x)^{2}} -
\frac{1}{6} \Biggl[ \frac{2}{(1-x)} + \frac{1}{(1+x)} -
\frac{9}{(1+x)^{2}} \nn\\
& & + \frac{6}{(1+x)^{3}} \Biggr] H(0,x) \, , 
\label{e72} \\
a^{2} F^{(4)}_{0} & = & - \frac{8}{3} \Biggl[ \frac{1}{(1+x)} - 
\frac{1}{(1+x)^{2}} \Biggr] + \frac{\zeta(2)}{3} \Biggl[ \frac{1}{(1-x)}
+ \frac{2}{(1+x)} \nn\\
& & - \frac{12}{(1+x)^{2}} + \frac{15}{(1+x)^{3}} - 
\frac{6}{(1+x)^{4}} \Biggr] \nn\\
& & + \frac{1}{36} \Biggl[ \frac{11}{(1-x)}
- \frac{5}{(1+x)} - \frac{18}{(1+x)^{2}} + \frac{12}{(1+x)^{3}} - 
\Biggr] H(0,x) \nn\\
& & + \frac{1}{3} \Biggl[ \frac{2}{(1-x)}
+ \frac{1}{(1+x)} - \frac{9}{(1+x)^{2}} + \frac{6}{(1+x)^{3}} - 
\Biggr] H(-1,0,x) \nn\\
& & - \frac{1}{3} \Biggl[ \frac{1}{(1-x)} - \frac{2}{(1+x)^{2}} - 
\frac{1}{(1+x)^{3}} + \frac{2}{(1+x)^{4}} \Biggr] H(0,0,x) 
\label{e73} 
 \, .
\eea

\end{fmffile}

\begin{fmffile}{seco2}

\newcommand{\be}{\begin{equation}}
\newcommand{\ee}{\end{equation}}
\newcommand{\nn}{\nonumber}
\newcommand{\bea}{\begin{eqnarray}}
\newcommand{\eea}{\end{eqnarray}}
\newcommand{\bfig}{\begin{figure}}
\newcommand{\efig}{\end{figure}}
\newcommand{\bc}{\begin{center}}
\newcommand{\ec}{\end{center}}

\section{Expansion for $Q^{2} \gg a$ \label{Q2grande}}

We list, in this section, the asymptotic expansion of the 6-denominator 
vertex diagrams given in the previous sections, in order to show their 
behaviour for momentum transfer larger than the mass.

Putting $y=Q^{2}/a\ ,$ $L = \ln{y} $ 
and keeping terms up to the order $(1/y)^5$, we have: 

\begin{eqnarray}
\parbox{30mm}{\begin{fmfgraph*}(15,15)
\fmfleft{i1,i2}
\fmfright{o}
\fmf{plain}{i1,v1}
\fmf{plain}{i2,v2}
\fmf{photon}{v5,o}
\fmf{plain,tension=.3}{v2,v3}
\fmf{plain,tension=.3}{v3,v5}
\fmf{plain,tension=.3}{v1,v4}
\fmf{plain,tension=.3}{v4,v5}
\fmf{photon,tension=0}{v2,v4}
\fmf{photon,tension=0}{v1,v3}
\end{fmfgraph*} }  & \simeq & \left( \frac{a}{\mu_{0}^{2}} \right) ^{-2
\epsilon} \sum_{i=-1}^{0} \epsilon^{i} \biggl[ 
\sum_{j=2}^{5} \frac{A_{(j)}^{(i)}}{y^{j}} \biggr] \, ,
\label{f2} \\
\parbox{30mm}{\begin{fmfgraph*}(15,15)
\fmfleft{i1,i2}
\fmfright{o}
\fmf{plain}{i1,v1}
\fmf{plain}{i2,v2}
\fmf{photon}{v5,o}
\fmflabel{$(k_{1} \cdot k_{2})$}{o}
\fmf{plain,tension=.3}{v2,v3}
\fmf{plain,tension=.3}{v3,v5}
\fmf{plain,tension=.3}{v1,v4}
\fmf{plain,tension=.3}{v4,v5}
\fmf{photon,tension=0}{v2,v4}
\fmf{photon,tension=0}{v1,v3}
\end{fmfgraph*}}  & \simeq & \left( \frac{a}{\mu_{0}^{2}} \right) ^{-2
\epsilon} 
\sum_{j=1}^{5} \frac{B_{(j)}^{(0)}}{y^{j}}  \, ,
\label{f3} 
\end{eqnarray}
where: 
\bea
a^{2} A_{(2)}^{(-1)} & = &    \zeta(3) 
                       - \zeta(2) L
                       - \frac{1}{3} L^3 \, , 
\label{f4} \\
a^{2} A_{(3)}^{(-1)} & = &  -2 \zeta(2) - 4 \zeta(3) + 4 \zeta(2) L 
                             - 2 L^{2} + \frac{4}{3} L^{3} \, , 
\label{f5} \\ 
a^{2} A_{(4)}^{(-1)} & = & - 1 + 11 \zeta(2) + 16 \zeta(3) 
                        - 5 L
                        - 16 \zeta(2) L
                        + 11 L^{2}
                        - \frac{16}{3} L^3 \, , 
\label{f6} \\
a^{2} A_{(5)}^{(-1)} & = & \frac{10}{3} \! - \! \frac{152}{3} \zeta(2) 
                        - \!  64 \zeta(3) \! + \! 36 L 
                        + \! 64 \zeta(2) L
                        - \! \frac{152}{3}  L^{2}
                        + \! \frac{64}{3} L^3  , 
\label{f7} \\
a^{2} A_{(2)}^{(0)} & = & \frac{37}{10} \zeta^{2}(2)
                 - \zeta(3) L
                 - \zeta(2) L^{2}
		 + \frac{1}{2} L^{4} \, ,  
\label{f8} \\
a^{2} A_{(3)}^{(0)} & = & 
          - 4  \zeta(2)
          - 2  \zeta(3)
          - \frac{74}{5}  \zeta^{2}(2)
          - 4 \zeta(2) L
          + 4 \zeta(3) L 
          - 8 L  \nn\\
& & 
          + 4 \zeta(2) L^{2}
          - 4 L^{2}
          + 4 L^{3}
          - 2 L^{4}  \, ,  
\label{f9} \\ 
a^{2} A_{(4)}^{(0)} & = & 
          + 18 \zeta(2)
          + \frac{296}{5}  \zeta^{2}(2)
          + 15  \zeta(3)
	  - 20
          + 18 \zeta(2) L 
          - 16 \zeta(3) L  \nn\\
& & 
          + 31 L 
          - 16 \zeta(2) L^{2}
          + 37 L^{2}
          - \frac{70}{3} L^{3}
          + 8 L^{4} \, ,  
\label{f10} \\
a^{2} A_{(5)}^{(0)} & = & 
          - \frac{724}{9}  \zeta(2)
          - \frac{1184}{5}  \zeta^{2}(2)
          - \frac{248}{3}  \zeta(3)
	  + 138 
          - \frac{208}{3}  \zeta(2) L \nn\\
& & 
          + 64 \zeta(3) L 
          - \frac{2204}{27} L 
          + 64  \zeta(2) L^{2}
          - \frac{1948}{9} L^{2}
          + 112 L^{3} \nn\\
& & 
          - 32 L^{4} \, ,  
\label{f11} \\
a B_{(1)}^{(0)} & = &  - \frac{1}{5} \zeta^2(2) 
                     - 2 \zeta(3) L
                     - \frac{1}{24} L^4  \, ,  
\label{f12} \\
a B_{(2)}^{(0)} & = & - 2 
                  - 2 \zeta(2) 
                  + \frac{2}{5} \zeta^{2}(2) 
                  - 4 \zeta(3)
                  + 4 \zeta(2) L
                  + 4 \zeta(3) L 
		  - 2 L \nn\\
& & 
                  + \frac{1}{3} L^3 
                  + \frac{1}{12} L^4   \, ,  
\label{f13} \\
a B_{(3)}^{(0)} & = & \frac{31}{8} 
                  + \frac{37}{2} \zeta(2) 
                  - \frac{6}{5} \zeta^2(2) 
                  + 14 \zeta(3) 
                  + \frac{33}{4} L  
                  - 14 \zeta(2) L   \nn\\
& & 
		  - 12 \zeta(3) L
                  + \frac{7}{2} L^2
                  - \frac{7}{6} L^3 
                  - \frac{1}{4} L^4   \, ,  
\label{f14} \\ 
a B_{(4)}^{(0)} & = &  - \frac{2195}{324} 
                  - \frac{767}{9} \zeta(2) 
                  + 4 \zeta^{2}(2) 
                  - \frac{148}{3} \zeta(3) 
                  + \frac{148}{3} \zeta(2) L 
                  + 40 \zeta(3) L  \nn\\
& & 
                  - \frac{1237}{54} L
                  - 18 L^{2} 
		  + \frac{37}{9} L^3 
		  + \frac{5}{6} L^4 \, ,  
\label{f15} \\
a B_{(5)}^{(0)} & = &  \frac{146447}{10368}
                  + \frac{25325}{72} \zeta(2) 
                  - 14 \zeta^2(2) 
                  + \frac{533}{3} \zeta(3) 
                  + \frac{52955}{864} L  \nn\\
& &   
                  - \frac{533}{3} \zeta(2) L
                  - 140 \zeta(3) L
                  + \frac{615}{8} L^2
                  - \frac{533}{36} L^3 
                  - \frac{35}{12} L^4  \, . 
\label{f16} 
\eea

\be
\parbox{20mm}{\begin{fmfgraph*}(15,15)
\fmfleft{i1,i2}
\fmfright{o}
\fmf{plain}{i1,v1}
\fmf{plain}{i2,v2}
\fmf{photon}{v5,o}
\fmf{plain,tension=.3}{v2,v3}
\fmf{plain,tension=.3}{v3,v5}
\fmf{plain,tension=.3}{v1,v4}
\fmf{plain,tension=.3}{v4,v5}
\fmf{photon,tension=0}{v2,v1}
\fmf{photon,tension=0}{v4,v3}
\end{fmfgraph*} } \simeq \left( \frac{a}{\mu_{0}^{2}} \right) ^{-2
\epsilon} \sum_{i=-2}^{0} \epsilon^{i} \biggl[ 
\sum_{j=2}^{5} \frac{C_{(j)}^{(i)}}{y^{j}} \biggr] \, ,
\label{f17} 
\ee
where:
\bea
a^{2} C_{(2)}^{(-2)} & = & \frac{L^{2}}{2} \, , 
\label{f18} \\
a^{2} C_{(3)}^{(-2)} & = &  2 L - 2 L^{2} \, ,  
\label{f19} \\
a^{2} C_{(4)}^{(-2)} & = & 2 - 11 L + 8 L^{2}  \, ,  
\label{f20} \\
a^{2} C_{(5)}^{(-2)} & = & - 14 + \frac{152}{3} L - 32 L^{2} \, ,  
\label{f21} \\
a^{2} C_{(2)}^{(-1)} & = & - \zeta(3) + 2 \zeta(2) L 
- \frac{1}{6} L^{3} \, , 
\label{f23} \\
a^{2} C_{(3)}^{(-1)} & = &  4 \zeta(2) + 4 \zeta(3) - 2 L - 
8 \zeta(2) L - 3 L^{2} + \frac{2}{3} L^{3} \, ,  
\label{f24} \\
a^{2} C_{(4)}^{(-1)} & = & -3 - 22 \zeta(2) - 16 \zeta(3) + \frac{7}{2}
L + 32 \zeta(2) L + \frac{37}{2} L^{2} - \frac{8}{3} L^{3}  \, ,  
\label{f25} \\
a^{2} C_{(5)}^{(-1)} & = & \frac{47}{3} + \frac{304}{3} \zeta(2) + 64
\zeta(3) + \frac{70}{9} L - 128 \zeta(2) L - 92 L^{2} \nn\\
& & + \frac{32}{3} L^{3} \, ,  
\label{f26} \\
a^{2} C_{(2)}^{(0)} & = & \frac{12}{5} \zeta^{2}(2) + \zeta(3) L 
+ \frac{7}{2} \zeta(2) L^{2} + \frac{5}{24} L^{4} \, , 
\label{f27001} \\
a^{2} C_{(3)}^{(0)} & = &  - 8 - 2 \zeta(2) - \frac{48}{5} \zeta^{2}(2) 
+ 6 \zeta(3) - 2 L + 6 \zeta(2) L - 4 \zeta(3) L \nn\\
& & 
- 14 \zeta(2) L^{2} + L^{2} + \frac{7}{3} L^{3} - \frac{5}{6} L^{4} ,  
\label{f27002} \\
a^{2} C_{(4)}^{(0)} & = & 48 + \frac{23}{2} \zeta(2) + \frac{192}{5} 
\zeta^{2}(2) 
- 37 \zeta(3) + \frac{89}{4} L - 25 \zeta(2) L + 16 \zeta(3) L \nn\\
& & + \frac{19}{4} L^{2} + 56 \zeta(2) L^{2} - \frac{27}{2} L^{3} + 
\frac{10}{3} L^{4}  \, ,  
\label{f27003} \\
a^{2} C_{(5)}^{(0)} & = & - \frac{36349}{162} - \frac{362}{9} \zeta(2) 
- \frac{768}{5} \zeta^{2}(2) + 184 \zeta(3) - \frac{3395}{27} L 
+ 88 \zeta(2) L \nn\\
& & - 64 \zeta(3) L - \frac{503}{9} L^{2} - 224 \zeta(2) L^{2} + 
\frac{580}{9} L^{3} - \frac{40}{3} L^{4}
\, .
\label{f27004}
\eea

\be
\parbox{20mm}{\begin{fmfgraph*}(15,15)
\fmfleft{i1,i2}
\fmfright{o}
\fmfforce{0.2w,0.93h}{v2}
\fmfforce{0.2w,0.07h}{v1}
\fmfforce{0.2w,0.5h}{v3}
\fmfforce{0.8w,0.5h}{v5}
\fmf{plain}{i1,v1}
\fmf{plain}{i2,v2}
\fmf{photon}{v5,o}
\fmf{plain,tension=0}{v2,v5}
\fmf{plain,tension=0}{v3,v4}
\fmf{photon,tension=.4}{v1,v4}
\fmf{plain,tension=.4}{v4,v5}
\fmf{plain,tension=0}{v1,v3}
\fmf{photon,tension=0}{v2,v3}
\end{fmfgraph*} } \simeq \left( \frac{a}{\mu_{0}^{2}} \right) ^{-2
\epsilon} \sum_{i=-2}^{0} \epsilon^{i} \biggl[ 
\sum_{j=1}^{5} \frac{E_{(j)}^{(i)}}{y^{j}} \biggr] \, ,
\label{f28} 
\ee
where:
\bea
a^{2} E_{(1)}^{(-2)} & = & + \frac{L}{4} \, , 
\label{f29} \\
a^{2} E_{(2)}^{(-2)} & = & \frac{1}{2} - \frac{L}{2} \, , 
\label{f30} \\
a^{2} E_{(3)}^{(-2)} & = & - \frac{7}{4} + \frac{3}{2} L \, ,   
\label{f31} \\
a^{2} E_{(4)}^{(-2)} & = & \frac{37}{6} - 5 L \, , 
\label{f32} \\
a^{2} E_{(5)}^{(-2)} & = & - \frac{533}{24} + \frac{35}{2} L \, ,   
\label{f33} \\
a^{2} E_{(1)}^{(-1)} & = &  \frac{1}{4} \zeta(2) 
                     - \frac{1}{2} L
                     + \frac{1}{4} L^2 , 
\label{f34} \\
a^{2} E_{(2)}^{(-1)} & = & - 1 - \frac{\zeta(2)}{2} + 2 L - 
\frac{L^{2}}{2} \, , 
\label{f35} \\
a^{2} E_{(3)}^{(-1)} & = &  \frac{17}{4} 
                        + \frac{3}{2} \zeta(2) 
                        - 7 L 
                        + \frac{3}{2} L^2  ,  
\label{f36} \\
a^{2} E_{(4)}^{(-1)} & = & - \frac{101}{6} - 5 \zeta(2) + 
\frac{76}{3} L - 5 L^{2} \, , 
\label{f37} \\
a^{2} E_{(5)}^{(-1)} & = &   \frac{3157}{48} 
                        + \frac{35}{2} \zeta(2) 
                        - \frac{281}{3} L
                        + \frac{35}{2} L^2  , 
\label{f38} \\
a^{2} E_{(1)}^{(0)} & = &  - 3 \zeta(2) \ln(2) 
                - \frac{1}{2} \zeta(2) 
                - \frac{5}{2} \zeta(3) 
                + L + \frac{7}{2} \zeta(2) L
                - \frac{1}{2} L^2 \nn\\
& & 
+ \frac{1}{6} L^3 , 
\label{f39} \\
a^{2} E_{(2)}^{(0)} & = &  5 
                 + 12 \zeta(2) \ln(2) 
                 + 11 \zeta(2) 
                 + \frac{13}{2} \zeta(3) 
                 - 3 L - 10 \zeta(2) L 
                 + \frac{3}{2} L^2 \nn\\
& & 
                 - \frac{5}{6} L^3 , 
\label{f40} \\
a^{2} E_{(3)}^{(0)} & = &  - \frac{161}{8} 
                  - 48 \zeta(2) \ln(2) 
                  - \frac{107}{2} \zeta(2) 
                  - \frac{45}{2} \zeta(3) 
                  + \frac{31}{4} L + 36 \zeta(2) L\nn\\
& & 
                  - \frac{31}{4} L^2
                  + \frac{7}{2} L^3 , 
\label{f41} \\
a^{2} E_{(4)}^{(0)} & = & \frac{2677}{36}
                 + 192 \zeta(2) \ln(2) 
                 + \frac{715}{3} \zeta(2) 
                 + 83 \zeta(3) 
                 - \frac{221}{9} L + 136 \zeta(2) L \nn\\
& & 
                 + \frac{113}{3} L^2
                 - \frac{43}{3} L^3 , 
\label{f42} \\
a^{2} E_{(5)}^{(0)} & = &  - \frac{156965}{576} 
                  - 768 \zeta(2) \ln(2) 
                  - \frac{12323}{12} \zeta(2) 
                  - \frac{629}{2} \zeta(3) + \frac{13319}{144} L \nn\\
& & 
                  + 524 \zeta(2) L 
                  - \frac{1387}{8} L^2 + \frac{349}{6} L^3 .
\label{f43} 
\eea

\be
\parbox{20mm}{\begin{fmfgraph*}(15,15)
\fmfleft{i1,i2}
\fmfright{o}
\fmfforce{0.2w,0.93h}{v2}
\fmfforce{0.2w,0.07h}{v1}
\fmfforce{0.2w,0.3h}{v3}
\fmfforce{0.2w,0.7h}{v4}
\fmfforce{0.8w,0.5h}{v5}
\fmf{plain}{i1,v1}
\fmf{plain}{i2,v2}
\fmf{photon}{v5,o}
\fmf{plain}{v2,v5}
\fmf{photon}{v1,v3}
\fmf{photon}{v2,v4}
\fmf{plain}{v1,v5}
\fmf{plain,left}{v3,v4}
\fmf{plain,right}{v3,v4}
\end{fmfgraph*} } \simeq \left( \frac{a}{\mu_{0}^{2}} \right) ^{-2
\epsilon} \sum_{i=-1}^{0} \epsilon^{i} \biggl[ 
\sum_{j=1}^{5} \frac{F_{(j)}^{(i)}}{y^{j}} \biggr] \, ,
\label{f44} 
\ee
where:

\bea
a^{2} F_{(1)}^{(-1)} & = &  1 + \frac{1}{6} L , 
\label{f45} \\
a^{2} F_{(2)}^{(-1)} & = & - \frac{11}{3} + \frac{5}{3} L , 
\label{f46} \\
a^{2} F_{(3)}^{(-1)} & = &  \frac{113}{6} - 11 L , 
\label{f47} \\
a^{2} F_{(4)}^{(-1)} & = & - \frac{809}{9} + \frac{170}{3} L , 
\label{f48} \\
a^{2} F_{(5)}^{(-1)} & = & \frac{14779}{36} - \frac{805}{3} L ,
\label{f49} \\
a^{2} F_{(1)}^{(0)}  & = & - \frac{8}{3} 
                + \frac{2}{3} \zeta(2) 
                - \frac{4}{9} L , 
\label{f50} \\
a^{2} F_{(2)}^{(0)}  & = & \frac{85}{9} 
                 - \frac{7}{3} \zeta(2) 
                 - \frac{1}{9} L 
                 - \frac{3}{2} L^2 , 
\label{f51} \\
a^{2} F_{(3)}^{(0)}  & = & - \frac{1589}{36}
                  + 13 \zeta(2) 
                  - \frac{43}{6} L
                  + \frac{19}{2} L^2 , 
\label{f52} \\
a^{2} F_{(4)}^{(0)}  & = & \frac{1123}{6}
                 - \frac{214}{3} \zeta(2) 
                 + \frac{574}{9} L 
                 - \frac{149}{3} L^2 ,  
\label{f53} \\
a^{2} F_{(4)}^{(0)}  & = & - \frac{109397}{144} 
                  + \frac{1115}{3} \zeta(2) 
                  - \frac{14345}{36} L
                  + \frac{1445}{6} L^2 ,  
\label{f54} 
\eea

\be
\parbox{20mm}{\begin{fmfgraph*}(15,15)
\fmfleft{i1,i2}
\fmfright{o}
\fmfforce{0.2w,0.93h}{v2}
\fmfforce{0.2w,0.07h}{v1}
\fmfforce{0.8w,0.5h}{v5}
\fmf{plain}{i1,v1}
\fmf{plain}{i2,v2}
\fmf{photon}{v5,o}
\fmf{plain}{v2,v3}
\fmf{photon,tension=.25,right}{v3,v4}
\fmf{plain,tension=.25}{v3,v4}
\fmf{plain}{v4,v5}
\fmf{plain}{v1,v5}
\fmf{photon}{v1,v2}
\end{fmfgraph*} } \simeq \left( \frac{a}{\mu_{0}^{2}} \right) ^{-2
\epsilon} \sum_{i=-2}^{0} \epsilon^{i} \biggl[ 
\sum_{j=1}^{5} \frac{G_{(j)}^{(i)}}{y^{j}} \biggr] \, ,
\label{f55} 
\ee
where:

\bea
a^{2} G_{(1)}^{(-2)} & = & - \frac{1}{4} L , \\
a^{2} G_{(2)}^{(-2)} & = & - \frac{1}{2} + \frac{1}{2} L , \\
a^{2} G_{(3)}^{(-2)} & = & \frac{7}{4} - \frac{3}{2} L , \\
a^{2} G_{(4)}^{(-2)} & = & - \frac{37}{6} + 5 L , \\
a^{2} G_{(5)}^{(-2)} & = & \frac{533}{24} - \frac{35}{2} L , \\
a^{2} G_{(1)}^{(-1)} & = & - 1 
                      - \frac{1}{4} \zeta(2) 
                      - \frac{1}{2} L
                      - \frac{1}{4} L^2 , \\
a^{2} G_{(2)}^{(-1)} & = & 3 + \frac{1}{2} \zeta(2) - 2 L + \frac{1}{2}
                      L^{2} ,\\
a^{2} G_{(3)}^{(-1)} & = & - \frac{69}{4} 
                        - \frac{3}{2} \zeta(2) 
                        + 13 L 
                        - \frac{3}{2} L^2 ,\\
a^{2} G_{(4)}^{(-1)} & = & \frac{517}{6} + 5 \zeta(2) - \frac{196}{3} L
                        + 5 L^{2} , \\
a^{2} G_{(5)}^{(-1)} & = & - \frac{19309}{48} 
                        - \frac{35}{2} \zeta(2) 
                        + \frac{911}{3} L 
                        - \frac{35}{2} L^2 , \\
a^{2} G_{(1)}^{(0)} & = & - \frac{1}{2} \zeta(2) 
                + \frac{7}{4} \zeta(3) 
                - L 
                - \frac{7}{2} \zeta(2) L - \frac{1}{2} L^2
                - \frac{1}{6} L^3 , \\
a^{2} G_{(2)}^{(0)}  & = & - 5
                  - 8 \zeta(2)
		  - \frac{7}{2} \zeta(3) 
                  - 4 L
		  + 7 \zeta(2) L 
                  + \frac{5}{2} L^2
		  + \frac{1}{3} L^{3} , \\
a^{2} G_{(3)}^{(0)}   & = & \frac{127}{8}
                 + 34 \zeta(2) 
                 + \frac{21}{2} \zeta(3) 
                 + 39 L
                 - 21 \zeta(2) L 
		 - \frac{53}{4} L^2 
                 - L^3 , \\
a^{2} G_{(4)}^{(0)}   & = &  - \frac{1183}{36}
                  - \frac{418}{3} \zeta(2)
		  - 35 \zeta(3) 
                  - \frac{701}{3} L
		  + 70 \zeta(2) L 
                  + \frac{367}{6} L^2 \nn\\
& & 
		  + \frac{10}{3} L^{3} , \\
a^{2} G_{(5)}^{(0)}   & = &  - \frac{3397}{576} 
                 + \frac{3421}{6} \zeta(2) 
                 + \frac{245}{2} \zeta(3) 
                 + \frac{14389}{12} L 
                 - 245 \zeta(2) L \nn\\
& &  
		 - \frac{6443}{24} L^2
                 - \frac{35}{3} L^3 , 
\end{eqnarray}

\section{Expansion for $Q^{2} \ll a$ \label{Q2piccolo}}

We list, in this section, the expansion of the vertex diagrams around
$Q^{2}=0$.

Putting $y=Q^{2}/a$, and keeping terms up to the order $y^3$, we
have:

\begin{eqnarray}
\parbox{30mm}{\begin{fmfgraph*}(15,15)
\fmfleft{i1,i2}
\fmfright{o}
\fmf{plain}{i1,v1}
\fmf{plain}{i2,v2}
\fmf{photon}{v5,o}
\fmf{plain,tension=.3}{v2,v3}
\fmf{plain,tension=.3}{v3,v5}
\fmf{plain,tension=.3}{v1,v4}
\fmf{plain,tension=.3}{v4,v5}
\fmf{photon,tension=0}{v2,v4}
\fmf{photon,tension=0}{v1,v3}
\end{fmfgraph*} }  & \simeq & \left( \frac{a}{\mu_{0}^{2}} \right) ^{-2
\epsilon} \sum_{i=-1}^{0} \epsilon^{i} \biggl[ 
\sum_{j=0}^{2} A_{(j)}^{(i)} y^{j} \biggr] \, ,
\label{g2} \\
\parbox{30mm}{\begin{fmfgraph*}(15,15)
\fmfleft{i1,i2}
\fmfright{o}
\fmf{plain}{i1,v1}
\fmf{plain}{i2,v2}
\fmf{photon}{v5,o}
\fmflabel{$(k_{1} \cdot k_{2})$}{o}
\fmf{plain,tension=.3}{v2,v3}
\fmf{plain,tension=.3}{v3,v5}
\fmf{plain,tension=.3}{v1,v4}
\fmf{plain,tension=.3}{v4,v5}
\fmf{photon,tension=0}{v2,v4}
\fmf{photon,tension=0}{v1,v3}
\end{fmfgraph*}}  & \simeq & \left( \frac{a}{\mu_{0}^{2}} \right) ^{-2
\epsilon} 
\sum_{j=0}^{2} B_{(j)}^{(0)} y^{j}  \, ,
\label{g3} 
\end{eqnarray}
where: 
\bea
a^{2}A_{(0)}^{(-1)} & = &  - \frac{1}{4} \, , 
\label{g4} \\
a^{2}A_{(1)}^{(-1)} & = &  \frac{5}{72} \, , 
\label{g5} \\ 
a^{2}A_{(2)}^{(-1)} & = & - \frac{377}{21600} \, , 
\label{g6} \\
a^{2}A_{(0)}^{(0)} & = & 1 - \frac{3}{4} \zeta(2) \, ,  
\label{g8} \\
a^{2}A_{(1)}^{(0)} & = & - \frac{49}{216} + \frac{1}{4} \zeta(2) \, ,  
\label{g9} \\ 
a^{2}A_{(2)}^{(0)} & = &  \frac{16717}{324000} - \frac{17}{240} 
\zeta(2) \, ,  
\label{g10} \\
aB_{(0)}^{(0)} & = &  - 2 \zeta(2) - \frac{3}{4} \zeta(3)
                     + 3 \zeta(2) \ln{2} \, ,  
\label{g12} \\
aB_{(1)}^{(0)} & = & - \frac{7}{36} + \frac{29}{72} \zeta(2) 
                     + \frac{1}{8} \zeta(3) - \frac{1}{2} \zeta(2) 
         \ln{2} \, ,  
\label{g13} \\
aB_{(2)}^{(0)} & = & \frac{37}{720} 
                  - \frac{1247}{14400} \zeta(2) 
                  - \frac{1}{40} \zeta(3) 
                  + \frac{1}{10} \zeta(2) \ln{2}  \, ,  
\label{g14} \\ 
\eea

\be
\parbox{20mm}{\begin{fmfgraph*}(15,15)
\fmfleft{i1,i2}
\fmfright{o}
\fmf{plain}{i1,v1}
\fmf{plain}{i2,v2}
\fmf{photon}{v5,o}
\fmf{plain,tension=.3}{v2,v3}
\fmf{plain,tension=.3}{v3,v5}
\fmf{plain,tension=.3}{v1,v4}
\fmf{plain,tension=.3}{v4,v5}
\fmf{photon,tension=0}{v2,v1}
\fmf{photon,tension=0}{v4,v3}
\end{fmfgraph*} } \simeq \left( \frac{a}{\mu_{0}^{2}} \right) ^{-2
\epsilon} \sum_{i=-2}^{0} \epsilon^{i} \biggl[ 
\sum_{j=0}^{2} C_{(j)}^{(i)} y^{j} \biggr] \, ,
\label{g17} 
\ee
where:
\bea
a^{2}C_{(0)}^{(-2)} & = & \frac{1}{8} \, , 
\label{g17000} \\
a^{2}C_{(1)}^{(-2)} & = & - \frac{1}{24} \, , 
\label{g17001} \\
a^{2}C_{(2)}^{(-2)} & = & \frac{17}{1440} \, , 
\label{g18} \\
a^{2}C_{(0)}^{(-1)} & = & \frac{1}{4} \, , 
\label{g22000} \\
a^{2}C_{(1)}^{(-1)} & = & - \frac{1}{9} \, , 
\label{g22} \\
a^{2}C_{(2)}^{(-1)} & = & \frac{797}{21600} \, , 
\label{g23} \\
a^{2}C_{(0)}^{(0)} & = & - \frac{1}{2} + \frac{3}{2} \zeta(2) \, , 
\label{g27} \\
a^{2}C_{(1)}^{(0)} & = & \frac{17}{108} - \frac{1}{2} \zeta(2)  \, , 
\label{g27000} \\
a^{2}C_{(2)}^{(0)} & = & - \frac{2993}{81000} + \frac{17}{120} \zeta(2)
                      \, , 
\label{g27001}
\, .
\eea

\be
\parbox{20mm}{\begin{fmfgraph*}(15,15)
\fmfleft{i1,i2}
\fmfright{o}
\fmfforce{0.2w,0.93h}{v2}
\fmfforce{0.2w,0.07h}{v1}
\fmfforce{0.2w,0.5h}{v3}
\fmfforce{0.8w,0.5h}{v5}
\fmf{plain}{i1,v1}
\fmf{plain}{i2,v2}
\fmf{photon}{v5,o}
\fmf{plain,tension=0}{v2,v5}
\fmf{plain,tension=0}{v3,v4}
\fmf{photon,tension=.4}{v1,v4}
\fmf{plain,tension=.4}{v4,v5}
\fmf{plain,tension=0}{v1,v3}
\fmf{photon,tension=0}{v2,v3}
\end{fmfgraph*} } \simeq \left( \frac{a}{\mu_{0}^{2}} \right) ^{-2
\epsilon} \sum_{i=-2}^{0} \epsilon^{i} \biggl[ 
\sum_{j=0}^{2} E_{(j)}^{(i)} y^{j} \biggr] \, ,
\label{g28} 
\ee
where:
\bea
a^{2}E_{(0)}^{(-2)} & = & \frac{1}{8} \, , 
\label{g29} \\
a^{2}E_{(1)}^{(-2)} & = & - \frac{1}{48} \, , 
\label{g30} \\
a^{2}E_{(2)}^{(-2)} & = &  \frac{1}{240} \, ,   
\label{g31} \\
a^{2}E_{(0)}^{(-1)} & = &  0 \, , 
\label{g34} \\
a^{2}E_{(1)}^{(-1)} & = & \frac{1}{36} \, , 
\label{g35} \\
a^{2}E_{(2)}^{(-1)} & = & - \frac{29}{3600} \,  ,  
\label{g36} \\
a^{2}E_{(0)}^{(0)} & = &  - \frac{1}{2} + \frac{3}{4} \zeta(2) \, , 
\label{g39} \\
a^{2}E_{(1)}^{(0)} & = &  \frac{67}{432} - \frac{1}{32} \zeta(2) \, , 
\label{g40} \\
a^{2}E_{(2)}^{(0)} & = &  - \frac{3407}{108000} - \frac{13}{1280} 
\zeta(2) \, , 
\label{g41} \\
\eea

\be
\parbox{20mm}{\begin{fmfgraph*}(15,15)
\fmfleft{i1,i2}
\fmfright{o}
\fmfforce{0.2w,0.93h}{v2}
\fmfforce{0.2w,0.07h}{v1}
\fmfforce{0.2w,0.3h}{v3}
\fmfforce{0.2w,0.7h}{v4}
\fmfforce{0.8w,0.5h}{v5}
\fmf{plain}{i1,v1}
\fmf{plain}{i2,v2}
\fmf{photon}{v5,o}
\fmf{plain}{v2,v5}
\fmf{photon}{v1,v3}
\fmf{photon}{v2,v4}
\fmf{plain}{v1,v5}
\fmf{plain,left}{v3,v4}
\fmf{plain,right}{v3,v4}
\end{fmfgraph*} } \simeq \left( \frac{a}{\mu_{0}^{2}} \right) ^{-2
\epsilon} \sum_{i=-1}^{0} \epsilon^{i} \biggl[ 
\sum_{j=0}^{2} F_{(j)}^{(i)} y^{j} \biggr] \, ,
\label{g44} 
\ee
where:

\bea
a^{2}F_{(0)}^{(-1)} & = &  \frac{7}{12} \, , 
\label{g45} \\
a^{2}F_{(1)}^{(-1)} & = & - \frac{13}{72} \, , 
\label{g46} \\
a^{2}F_{(2)}^{(-1)} & = &  \frac{19}{360} \, , 
\label{g47} \\
a^{2}F_{(0)}^{0}  & = & - \frac{35}{36} \, , 
\label{g50} \\
a^{2}F_{(1)}^{0}  & = & \frac{79}{432}  
                + \frac{1}{32} \zeta(2) \, , 
\label{g51} \\
a^{2}F_{(2)}^{0}  & = & - \frac{61}{2160}
                  - \frac{1}{64} \zeta(2) \, , 
\label{g52} \\
\eea

\be
\parbox{20mm}{\begin{fmfgraph*}(15,15)
\fmfleft{i1,i2}
\fmfright{o}
\fmfforce{0.2w,0.93h}{v2}
\fmfforce{0.2w,0.07h}{v1}
\fmfforce{0.8w,0.5h}{v5}
\fmf{plain}{i1,v1}
\fmf{plain}{i2,v2}
\fmf{photon}{v5,o}
\fmf{plain}{v2,v3}
\fmf{photon,tension=.25,right}{v3,v4}
\fmf{plain,tension=.25}{v3,v4}
\fmf{plain}{v4,v5}
\fmf{plain}{v1,v5}
\fmf{photon}{v1,v2}
\end{fmfgraph*} } \simeq \left( \frac{a}{\mu_{0}^{2}} \right) ^{-2
\epsilon} \sum_{i=-2}^{0} \epsilon^{i} \biggl[ 
\sum_{j=0}^{2} G_{(j)}^{(i)} y^{j} \biggr] \, ,
\label{g55} 
\ee
where:

\bea
a^{2}G_{(0)}^{(-2)} & = & - \frac{1}{8} \, , \\
a^{2}G_{(1)}^{(-2)} & = & \frac{1}{48} \, , \\
a^{2}G_{(2)}^{(-2)} & = & - \frac{1}{240} \, , \\
a^{2}G_{(0)}^{(-1)} & = & - 1 \, , \\
a^{2}G_{(1)}^{(-1)} & = & \frac{2}{9} \, , \\
a^{2}G_{(2)}^{(-1)} & = & - \frac{211}{3600} \, , \\
a^{2}G_{(0)}^{(0)} & = & - 1 - \frac{3}{2} \zeta(2) \, , \\
a^{2}G_{(1)}^{(0)}  & = & \frac{19}{216} + \frac{1}{4} \zeta(2) \, , \\
a^{2}G_{(2)}^{(0)}   & = & - \frac{3613}{108000} - \frac{1}{20} 
\zeta(2) \, .
\end{eqnarray}

\section{Summary \label{Summa}}

We have carried out a complete investigation of the scalar integrals 
associated to all the QED 2-loop vertex graphs, for on-shell 
electrons and arbitrary momentum transfer $t=-Q^2$ in the 
$D$-continuous regularization scheme. After identifying all the 
occurring Master Integrals (MI's), we have written the linear, 
non-homogeneous differential equations in $Q^2$ satisfied by the MI's, 
expanded them in $\epsilon=(4-D)/2$ and solved the equations by 
means of the method of the variation of the constants of Euler. 
The method requires the solution of the associated homogeneous 
equations.
It turns out that all the associated homogenous equations are trivial, 
or became trivial after a suitable choice of the MI's for the graph 
topologies involving more than a single MI; typically one had to solve 
a first order homogeneous differential equation with simple rational 
coefficients. 

The repeated integrations implied by Euler's method are immediately 
performed, in close analytic form, 
in terms of Harmonic Polylogarithms of increasing weight; the maximum 
weight occurring in the results presented in this paper was 4 (as in 
the case of the zeroth 
order term in $\epsilon$ of the double cross topology). By further 
iterations of the approach, one could almost mechanically obtain any 
additional term in the $\epsilon$ expansion of the MI's. 

The explicit analytic evaluation of the QED vertex form factors in 
terms of the MI's will be presented elsewhere. 
\vskip 1truecm 

\section{Acknowledgement}

We are grateful to J. Vermaseren for his kind assistance in the use
of the algebra manipulating program {\tt FORM}~\cite{FORM}, by which
all our calculations were carried out.
 
We thank T. Gehrmann, R. Heinesch and Y. Schroder for pointing out several 
misprints in a preliminary version of the manuscript.

R.B. would like to thank the Fondazione Della Riccia for supporting his 
stay at CERN, and the Theory Division of CERN for the hospitality during
a great part of this work.

We are greatful to M. Czakon, J. Gluza, and T. Riemann for 
pointing out misprints in the published version of the present paper
(see \cite{CGR}).

\appendix

\section{Propagators \label{app1}}

We list here the denominators of the integral expressions appeared in
the paper.
\bea
{\mathcal D}_{1} & = & k_{1}^{2} \, , \\
{\mathcal D}_{2} & = & k_{2}^{2} \, , \\
{\mathcal D}_{3} & = & (k_{1}+k_{2})^{2} \, , \\
{\mathcal D}_{4} & = & (p_{1}-k_{1})^{2} \, , \\
{\mathcal D}_{5} & = & (p_{2}-k_{2})^{2} \, , \\
{\mathcal D}_{6} & = & [k_{1}^{2}+a] \, , \\
{\mathcal D}_{7} & = & [k_{2}^{2}+a] \, , \\
{\mathcal D}_{8} & = & [(k_{1}+k_{2})^{2}+a] \, , \\
{\mathcal D}_{9} & = & [(p_{1}-k_{1})^{2}+a] \, , \\
{\mathcal D}_{10} & = & [(p_{2}+k_{1})^{2}+a] \, , \\
{\mathcal D}_{11} & = & [(p_{2}-k_{2})^{2}+a] \, , \\
{\mathcal D}_{12} & = & [(p_{1}+p_{2}-k_{1})^{2}+a] \, , \\
{\mathcal D}_{13} & = & [(p_{1}+p_{2}-k_{2})^{2}+a] \, , \\
{\mathcal D}_{14} & = & [(p_{1}-k_{1}-k_{2})^{2}+a] \, , \\
{\mathcal D}_{15} & = & [(p_{2}+k_{1}+k_{2})^{2}+a] \, , \\
{\mathcal D}_{16} & = & [(p_{1}+p_{2}-k_{1}-k_{2})^{2}+a] \, .
\eea

\section{1-loop ingredients \label{app2}}

We recall in this appendix some useful results about
1-loop diagrams, fundamental ingredients for the 2-loop calculations,
obtained with the method of differential equations. 
They appear in the 2-loop integrals which factorize in the products 
of two 1-loop integral; due to the possible presence of extra powers 
of $1/\epsilon$ in their coefficients, we give the results of 
the $\epsilon$ expansion up to the second order in $\epsilon$. 

The case of the massive bubble is exaustively examined. The differential
equation is presented and solved, as usual in the $\epsilon \rightarrow
0$ expansion. 

\subsubsection{Tadpole}

\bea
\parbox{20mm}{
\begin{fmfgraph*}(15,15)
\fmfleft{i}
\fmfright{o}
\fmfforce{0.5w,0.1h}{v1}
\fmfforce{0.25w,0.62h}{v3}
\fmfforce{0.5w,0.9h}{v7}
\fmfforce{0.74w,0.62h}{v11}
\fmf{plain,left=.1}{v1,v3}
\fmf{plain,left=.5}{v3,v7}
\fmf{plain,left=.5}{v7,v11}
\fmf{plain,left=.1}{v11,v1}
\end{fmfgraph*}} & = & \mu_{0}^{(4-D)} \int \{ d^{D}k \} 
\frac{1}{(k^{2} +a)} \nn\\
& = & \left( \frac{a}{\mu_{0}^{2}} \right) ^{- \epsilon} 
\sum_{i=-1}^{2} \epsilon^{i} A_{i} + {\mathcal O} \left( 
\epsilon^{3} \right) \, ,
\label{appb1}
\eea
where:
\bea
\frac{A_{-1}}{a} & = & - 1 \, , 
\label{appb10001}\\
\frac{A_{0}}{a} & = & - 1 \, , 
\label{appb10002}\\
\frac{A_{1}}{a} & = & - 1 \, , 
\label{appb10003}\\
\frac{A_{2}}{a} & = & - 1  \, .
\label{appb10004}
\eea

\subsubsection{Fully massive bubble}

The topology under consideration has one MI. We choose the scalar
integral itself:
\be
F(\epsilon,a,Q^{2}) = \parbox{15mm}{
\begin{fmfgraph*}(15,15)
\fmfleft{i}
\fmfright{o}
\fmf{photon}{i,v1}
\fmf{photon}{v2,o}
\fmf{plain,tension=.15,left}{v1,v2}
\fmf{plain,tension=.15,left}{v2,v1}
\end{fmfgraph*} }  = \mu_{0}^{(4-D)}
\int \{ d^{D}k \} \frac{1}{
[k^{2} +a] \, [(Q-k)^{2}+a]} \, ,
\label{appb20001}
\ee

The corresponding first-order linear differential equation is the
following:
\bea
\frac{dF(\epsilon,a,Q^{2})}{dQ^{2}} \! & = & \! - \frac{1}{2} \left[ 
\frac{1}{Q^{2}} - \frac{(1 - 2 \epsilon)}{(Q^{2}+4a)} 
\right] F(\epsilon,a,Q^{2}) \nn\\
& & \qquad \qquad - \frac{(1 - \epsilon)}{2a} 
\left[  \frac{1}{Q^{2}} - \frac{1}{(Q^{2}+4a)} \right]
T(\epsilon,a) ,
\label{appb20002}
\eea
where $T(\epsilon,a)$ is the Tadpole.

As in the cases previously discussed, we use our knowledge on the
analytical behaviour of the solution in order to find the initial
condition. In fact, Eq. (\ref{appb20002}) shows two possible 
singularities for the function $F(\epsilon,a,Q^{2})$, for $Q^{2}=0$ and
for $Q^{2}=-4a$. Only the second, nevertheless, is indeed a singularity
for $F$, corresponding to the physical threshold. Multiplying Eq. 
(\ref{appb20002}) for $Q^{2}$ and taking the limit $Q^{2} \rightarrow
0$, we obtain:
\be
F(\epsilon,a,Q^{2}=0) = - \frac{(1- \epsilon)}{a} T(\epsilon,a) \, .
\label{appb20003}
\ee

We look for a solution of Eq. (\ref{appb20002}), with initial condition
(\ref{appb20003}), expanded in Laurent series around $\epsilon=0$:
\be
F(\epsilon,a,Q^{2}) = \sum_{i=-1}^{2}
\epsilon^{i} F_{i}(a,Q^{2}) + {\mathcal O} ( \epsilon^{3} ) \, .
\label{appb20004}
\ee

The associated homogeneous equation at $\epsilon=0$ is 
\be
\frac{df(a,y)}{dy} \, = \, - \frac{1}{2} \left[ 
\frac{1}{y} - \frac{1}{(y+4a)} \right] f(a,y) \, ,
\label{appb20005}
\ee
which has the following solution:
\be
f(\epsilon,a,y) = k \, \sqrt{1+ \frac{4a}{y}} \, .
\label{appb20006}
\ee

We can find the solution of the non-homogeneous equation, order by
order in $\epsilon$, by means of the Euler's method of the variation of 
the constant $k$. We have:
\bea
F_{i}(a,Q^{2}) & = & \sqrt{1+ \frac{4a}{Q^{2}}} \, \, \Biggl\{ 
\int^{Q^{2}} \frac{dy}{\sqrt{1+ \frac{4a}{y}}} \Biggl[ \frac{1}{(y+4a)}
F_{i-1}(a,y) - \frac{1}{2a} \Biggl( \frac{1}{y} \nn\\
& & - \frac{1}{(y+4a)}
\Biggr) [ A_{i} - A_{i-1} ] \Biggr] + k_{i} \Biggr\} \, , 
\label{appb20007}
\eea
where the coefficients $A_{i}$ are those of Eqs. 
(\ref{appb10001}--\ref{appb10004}). The determination of the constants
of integration $k_{i}$ is made by imposing that the solution satisfies
Eq. (\ref{appb20003}).

In terms of the variable $x$, defined in Eq. (\ref{Q2tox}), the solution
reads:

\bea
\! \! \! \! \! \! \parbox{20mm}{
\begin{fmfgraph*}(15,15)
\fmfleft{i}
\fmfright{o}
\fmf{photon}{i,v1}
\fmf{photon}{v2,o}
\fmf{plain,tension=.15,left}{v1,v2}
\fmf{plain,tension=.15,left}{v2,v1}
\end{fmfgraph*} } & = & \mu_{0}^{(4-D)}
\int \{ d^{D}k \} \frac{1}{
[k^{2} +a] \, [(Q-k)^{2}+a]} \nn\\
\! \! \! \! \! \! & = & \left( \frac{a}{\mu_{0}^{2}} \right) ^{
- \epsilon} \sum_{i=-1}^{2} \epsilon^{i} B_{i} 
+ {\mathcal O} \left( \epsilon^{3} \right) \, , 
\label{appb2}
\eea
where:
\bea
\! \! \! \! \! \! B_{-1} & = & 1 \, ,  \\
\! \! \! \! \! \! B_{0} & = &  2 - 2 \left[ \frac{
1}{2} - \frac{1}{(1-x)} \right]  H(0,x) \, , \\
\! \! \! \! \! \! B_{1} & = &  4 - 4 \left[ 
\frac{1}{2} - \frac{1}{(1-x)} \right] \Biggl\{ - \frac{\zeta(2)}{2} + 
H(0,x) + \frac{1}{2} H(0,0,x) \nn\\
\! \! \! \! \! \! & & - H(-1,0,x) \Biggr\} \, , \\
\! \! \! \! \! \! B_{2} & = & 4 + 4 \, \left[ 
\frac{1}{2} - \frac{1}{(1-x)} \right] \Biggl\{ \frac{\zeta(2)}{2} 
\Bigl[ 1+ \frac{1}{2} H(0,x) - H(-1,x) \Bigr] + \frac{\zeta(3)}{2} 
\nn\\
\! \! \! \! \! \! & & - \frac{1}{2} \Bigl[ H(0,x) + \frac{1}{2} 
H(0,0,x) - H(-1,0,x) + \frac{1}{4} H(0,0,0,x) \nn\\
\! \! \! \! \! \! & & - \frac{1}{2} H(-1,0,0,x) - H(0,-1,0,x) - 
H(-1,-1,0,x) \Bigr] \Biggr\} .
%
\eea

\subsubsection{Bubble on the mass-shell}

\bea
\parbox{20mm}{
\begin{fmfgraph*}(15,15)
\fmfleft{i}
\fmfright{o}
\fmf{plain}{i,v1}
\fmf{plain}{v2,o}
\fmf{plain,tension=.15,left}{v1,v2}
\fmf{photon,tension=.15,left}{v2,v1}
\end{fmfgraph*} } & = & \left. \mu_{0}^{(4-D)}
\int \{ d^{D}k \} \frac{1}{k^{2} \, [(Q-k)^{2}+a]} \, \, 
\right| _{Q^{2}=-a} \nn\\
& = & - \frac{(1- \epsilon)}{a (1-2 \epsilon)} \, 
\parbox{15mm}{
\begin{fmfgraph*}(15,15)
\fmfleft{i}
\fmfright{o}
\fmfforce{0.5w,0.1h}{v1}
\fmfforce{0.25w,0.62h}{v3}
\fmfforce{0.5w,0.9h}{v7}
\fmfforce{0.74w,0.62h}{v11}
\fmf{plain,left=.1}{v1,v3}
\fmf{plain,left=.5}{v3,v7}
\fmf{plain,left=.5}{v7,v11}
\fmf{plain,left=.1}{v11,v1}
\end{fmfgraph*}} \nn\\
& = & \left( \frac{a}{\mu_{0}^{2}} \right) ^{- \epsilon} 
\sum_{i=-1}^{2} \epsilon^{i} E_{i} + {\mathcal O} \left( 
\epsilon^{3} \right) \, ,
\label{appb3}
\eea
where:
\bea
E_{-1} & = & 1 \, , \\
E_{0} & = & 2 \, , \\
E_{1} & = & 4 \, , \\
E_{2} & = & 8 
\, .
\eea

\subsubsection{Scalar Vertex at 1 loop}

\bea
\! \! \! \! \! \! \parbox{20mm}{\begin{fmfgraph*}(15,15)
\fmfleft{i1,i2}
\fmfright{o}
\fmf{plain}{i1,v1}
\fmf{plain}{i2,v2}
\fmf{photon}{v3,o}
\fmf{plain,tension=.3}{v2,v3}
\fmf{plain,tension=.3}{v1,v3}
\fmf{photon,tension=0}{v2,v1}
\end{fmfgraph*} } & = & \mu_{0}^{(4-D)}
\int \frac{d^{D}k}{(2 \pi)^{(D-2)}} \frac{1}{
k^{2} \, [(p_{1}-k)^{2}+a] \, [(p_{2}+k)^{2}+a]} \nn\\
\! \! \! \! \! \! & = & \frac{(1-2 \epsilon)}{\epsilon} \, 
\frac{1}{(Q^{2}+4a)}
\, \, \parbox{15mm}{\begin{fmfgraph*}(15,15)
\fmfleft{i}
\fmfright{o}
\fmf{photon}{i,v1}
\fmf{photon}{v2,o}
\fmf{plain,tension=.15,left}{v1,v2}
\fmf{plain,tension=.15,right}{v1,v2}
\end{fmfgraph*}}  - \frac{(1- \epsilon)}{a \, \epsilon} \, 
\frac{1}{(Q^{2}+4a)}  
\parbox{15mm}{
\begin{fmfgraph*}(15,15)
\fmfleft{i}
\fmfright{o}
\fmfforce{0.5w,0.1h}{v1}
\fmfforce{0.25w,0.62h}{v3}
\fmfforce{0.5w,0.9h}{v7}
\fmfforce{0.74w,0.62h}{v11}
\fmf{plain,left=.1}{v1,v3}
\fmf{plain,left=.5}{v3,v7}
\fmf{plain,left=.5}{v7,v11}
\fmf{plain,left=.1}{v11,v1}
\end{fmfgraph*} } \nn\\
\! \! \! \! \! \! & = & \left( \frac{a}{\mu_{0}^{2}} \right) ^{-
\epsilon} \sum_{i=-1}^{2} \epsilon^{i} F_{i} + {\mathcal O} 
\left( \epsilon^{3} \right) ,
\label{appb4}
\eea
where:
\bea
\! \! \! \! a F_{-1} & = & - \frac{1}{2} \, \left[ \frac{1}{(1+x)} - 
\frac{1}{(1-x)} \right] H(0,x) \, , \\
\! \! \! \! a F_{0} & = & \left[ \frac{1}{(1+x)} - 
\frac{1}{(1-x)} \right] \Biggl\{ \frac{\zeta(2)}{2} - \frac{1}{2} 
H(0,0,x) + H(-1,0,x) \Biggr\} \, , \\
\! \! \! \! a F_{1} & = & - 2 \left[ \frac{1}{(1+x)} - 
\frac{1}{(1-x)} \right] \, \Biggl\{ - \frac{\zeta(3)}{2}
- \frac{\zeta(2)}{2} \left[ \frac{1}{2} H(0,x) - H(-1,x) 
\right] \nn\\
\! \! \! \! & & + \frac{1}{4} H(0,0,0,x) + H(-1,-1,0,x) - \frac{1}{2} 
H(-1,0,0,x) \nn\\
\! \! \! \! & & - \frac{1}{2} H(0,-1,0,x) \Biggr\}  \, , \\
\! \! \! \! a F_{2} & = & \left[ \frac{1}{(1+x)} - 
\frac{1}{(1-x)} \right] \, \Biggl\{ - \frac{9 \zeta^{2}(2)}{20}
+ \zeta(3)\left[ H(0,x) - 2 H(-1,x) \right] \nn\\
\! \! \! \! & & + \zeta(2) \left[ \frac{1}{2} H(0,0,x) + 2 H(-1,-1,x) 
- H(-1,0,x) - H(0,-1,x) \right] \nn\\
\! \! \! \! & & - \frac{1}{2} H(0,0,0,0,x) + 4 H(-1,-1,-1,0,x) -
2 H(-1,-1,0,0,x) \nn\\
\! \! \! \! & & - 2 H(-1,0,-1,0,x) - 2 H(0,-1,-1,0,x) + 4
H(-1,0,0,0,x) \nn\\
\! \! \! \! & & + H(0,-1,0,0,x) + H(0,0,-1,0,x) \Biggr\} \, .
\eea

\section{Reducible 2-loop diagrams \label{app3}}

In this appendix we give the expressions of the reducible diagrams of
Figs. (\ref{fig2}--\ref{fig4}).

\bea
\parbox{20mm}{\begin{fmfgraph*}(15,15)
\fmfleft{i1,i2}
\fmfright{o}
\fmf{plain}{i1,v1}
\fmf{plain}{i2,v2}
\fmf{photon}{v4,o}
\fmf{plain,tension=.4}{v2,v3}
\fmf{plain,tension=.2}{v3,v4}
\fmf{plain,tension=.15}{v1,v4}
\fmf{photon,tension=0}{v2,v1}
\fmf{photon,tension=0,left=.5}{v4,v3}
\end{fmfgraph*} }  & = & \mu_{0}^{2(4-D)} 
\int \{ d^{D}k_{1} \} \{ d^{D}k_{2} \}
\frac{1}{{\mathcal D}_{1} {\mathcal D}_{2} {\mathcal D}_{9} {\mathcal D}_{10} 
{\mathcal D}_{14} } \\
& = & \left( \frac{a}{\mu_{0}^{2}} \right) ^{-2 \epsilon} 
\sum_{i=-2}^{0} \epsilon^{i} E^{(1)}_{i} + {\mathcal O} \left( 
\epsilon \right) , 
\eea
where:
\bea
a E^{(1)}_{-2} & = & \frac{1}{2} \left[ \frac{1}{(1-x)} - 
\frac{1}{(1+x)} \right] H(0,x) \, , \\
a E^{(1)}_{-1} & = & - \frac{1}{2} \left[ \frac{1}{(1-x)} - 
\frac{1}{(1+x)} \right] \bigl[ \zeta(2) - 2 H(0,x) - H(0,0,x) \nn\\
& & + 2 H(-1,0,x) \bigr] \\
a E^{(1)}_{0} & = & - \frac{1}{2} \left[ \frac{1}{(1-x)} - 
\frac{1}{(1+x)} \right] \bigl[ 2 \zeta(2) + 2 \zeta(3) - (4+3 \zeta(2)) 
H(0,x) \nn\\ 
& & - 2 \zeta(2) H(-1,x) - 2 H(0,0,x) + 4 H(-1,0,x) - 5 H(0,0,0,x) 
\nn\\
& & + 2 H(0,-1,0,x) + 2 H(-1,0,0,x) - 4 H(-1,-1,0,x) \bigr] \, .
\eea

\bea
\parbox{20mm}{\begin{fmfgraph*}(15,15)
\fmfleft{i1,i2}
\fmfright{o}
\fmf{plain}{i1,v1}
\fmf{plain}{i2,v2}
\fmf{photon}{v4,o}
\fmf{photon,tension=.4}{v2,v3}
\fmf{plain,tension=.2}{v3,v4}
\fmf{plain,tension=.15}{v1,v4}
\fmf{plain,tension=0}{v2,v1}
\fmf{plain,tension=0}{v1,v3}
\end{fmfgraph*} }  & = & \mu_{0}^{2(4-D)} 
\int \{ d^{D}k_{1} \} \{ d^{D}k_{2} \}
\frac{1}{{\mathcal D}_{4} {\mathcal D}_{6} {\mathcal D}_{7} {\mathcal D}_{14} 
{\mathcal D}_{15} } \\
& = & \left( \frac{a}{\mu_{0}^{2}} \right) ^{-2 \epsilon} 
E^{(2)}_{0} + {\mathcal O} \left( 
\epsilon \right) , 
\eea
where:
\bea
a E^{(2)}_{0} & = & - \frac{1}{2} \left[ \frac{1}{(1-x)} - 
\frac{1}{(1+x)} \right] \Biggl\{ \frac{\zeta^{2}(2)}{5} + 2 \zeta(3)
H(0,x) - \zeta(2) H(0,0,x) \nn\\
& & - 2 \zeta(2) H(1,0,x) - 4 H(0,0,-1,0,x) + 2 H(0,0,1,0,x) \nn\\
& & - 4 H(0,1,0,0,x) - 2 H(1,0,0,0,x) \Biggr\} \, .
\eea

\bea
\parbox{20mm}{\begin{fmfgraph*}(15,15)
\fmfleft{i1,i2}
\fmfright{o}
\fmfforce{0.2w,0.93h}{v2}
\fmfforce{0.2w,0.07h}{v1}
\fmfforce{0.2w,0.5h}{v3}
\fmfforce{0.8w,0.5h}{v4}
\fmf{plain}{i1,v1}
\fmf{plain}{i2,v2}
\fmf{photon}{v4,o}
\fmf{photon,tension=0}{v1,v3}
\fmf{plain,tension=0}{v3,v4}
\fmf{photon,tension=0}{v2,v4}
\fmf{plain,tension=0}{v2,v3}
\fmf{plain,tension=0}{v1,v4}
\end{fmfgraph*} }  & = & \mu_{0}^{2(4-D)} 
\int \{ d^{D}k_{1} \} \{ d^{D}k_{2} \}
\frac{1}{{\mathcal D}_{2} {\mathcal D}_{4} {\mathcal D}_{6} {\mathcal D}_{8} 
{\mathcal D}_{11} } \\
& = & \left( \frac{a}{\mu_{0}^{2}} \right) ^{-2 \epsilon} 
E^{(3)}_{0} + {\mathcal O} \left( 
\epsilon \right) , 
\eea
where:
\bea
a E^{(3)}_{0} & = & \left[ \frac{1}{(1-x)} - 
\frac{1}{(1+x)} \right] \Biggl\{ \frac{27\zeta^{2}(2)}{10} + \zeta(3)
H(0,x) + 3 \zeta(2) H(0,0,x) \nn\\
& & - 6 \zeta(2) H(0,-1,x) + 2 H(0,0,0,0,x) + 2 H(0,1,0,0,x) \nn\\
& & - 2 H(0,-1,0,0,x) \Biggr\} \, .
\eea

\bea
\parbox{20mm}{\begin{fmfgraph*}(15,15)
\fmfleft{i1,i2}
\fmfright{o}
\fmfforce{0.2w,0.93h}{v2}
\fmfforce{0.2w,0.07h}{v1}
\fmfforce{0.2w,0.55h}{v3}
\fmfforce{0.2w,0.13h}{v5}
\fmfforce{0.8w,0.5h}{v4}
\fmf{plain}{i1,v1}
\fmf{plain}{i2,v2}
\fmf{photon}{v4,o}
\fmf{photon}{v2,v3}
\fmf{plain,left}{v3,v5}
\fmf{plain,right}{v3,v5}
\fmf{plain}{v1,v4}
\fmf{plain}{v2,v4}
\end{fmfgraph*} }  & = & \mu_{0}^{2(4-D)} 
\int \{ d^{D}k_{1} \} \{ d^{D}k_{2} \}
\frac{1}{{\mathcal D}_{1} {\mathcal D}_{7} {\mathcal D}_{8} {\mathcal D}_{9} 
{\mathcal D}_{10} } \\
& = & \left( \frac{a}{\mu_{0}^{2}} \right) ^{-2 \epsilon} 
\sum_{i=-2}^{0} \epsilon^{i} E^{(4)}_{i} + {\mathcal O} \left( 
\epsilon \right) , 
\eea
where:
\bea
a E^{(4)}_{-2} & = & \frac{1}{2} \left[ \frac{1}{(1-x)} - 
\frac{1}{(1+x)} \right] H(0,x) \, , \\
a E^{(4)}_{-1} & = & - \frac{1}{2} \left[ \frac{1}{(1-x)} - 
\frac{1}{(1+x)} \right] \bigl[ \zeta(2) - H(0,0,x) + 2 H(-1,0,x) 
\bigr] \\
a E^{(4)}_{0} & = & \frac{1}{(1+x)} \left[ 1 - \frac{1}{(1+x)}
\right] \bigl[ 6 \zeta(2) + 2 H(0,0,x)  \bigr] - \Biggl[ \frac{1}{(1-x)}
\nn\\
& & - \frac{1}{(1+x)} \Biggr] \bigl[ \zeta(3) - 2 H(0,x) 
- \zeta(2) H(-1,x) - H(0,0,0,x) \nn\\
& & - 2 H(-1,-1,0,x) + H(-1,0,0,x) + H(0,-1,0,x) \bigr] \, .
\eea

\bea
\parbox{20mm}{\begin{fmfgraph*}(15,15)
\fmfforce{0.5w,0.2h}{v3}
\fmfforce{0.5w,0.8h}{v2}
\fmfforce{0.2w,0.5h}{v1}
\fmfforce{0.8w,0.5h}{v4}
\fmfleft{i}
\fmfright{o}
\fmf{photon}{i,v1}
\fmf{photon}{v4,o}
\fmf{plain,left=.4}{v1,v2}
\fmf{plain,right=.4}{v1,v3}
\fmf{plain,left=.4}{v2,v4}
\fmf{plain,right=.4}{v3,v4}
\fmf{photon}{v2,v3}
\end{fmfgraph*} }  & = & \mu_{0}^{2(4-D)} 
\int \{ d^{D}k_{1} \} \{ d^{D}k_{2} \}
\frac{1}{{\mathcal D}_{2} {\mathcal D}_{6} {\mathcal D}_{8} {\mathcal D}_{12} 
{\mathcal D}_{16} } \\
& = & \left( \frac{a}{\mu_{0}^{2}} \right) ^{-2 \epsilon} 
E^{(5)}_{0} + {\mathcal O} \left( 
\epsilon \right) , 
\eea
where:
\bea
a E^{(5)}_{0} & = & \frac{2}{(1-x)} \Biggl[ 1 - \frac{1}{(1-x)} \Biggr]
\bigl[ 3 \zeta(3) + 4 H(-1,0,0,x) - 4 H(0,-1,0,x) \nn\\
& & + 2 H(0,1,0,x) - 2 H(1,0,0,x) \bigr] \, .
\eea

\bea
\parbox{20mm}{\begin{fmfgraph*}(15,15)
\fmfforce{0.5w,0.2h}{v3}
\fmfforce{0.5w,0.8h}{v2}
\fmfforce{0.2w,0.5h}{v1}
\fmfforce{0.8w,0.5h}{v4}
\fmfleft{i}
\fmfright{o}
\fmf{plain}{i,v1}
\fmf{plain}{v4,o}
\fmf{plain,left=.4}{v1,v2}
\fmf{photon,right=.4}{v1,v3}
\fmf{photon,left=.4}{v2,v4}
\fmf{plain,right=.4}{v3,v4}
\fmf{plain}{v2,v3}
\end{fmfgraph*} }  & = & \mu_{0}^{2(4-D)} 
\int \{ d^{D}k_{1} \} \{ d^{D}k_{2} \}
\frac{1}{{\mathcal D}_{1} {\mathcal D}_{5} {\mathcal D}_{7} {\mathcal D}_{8} 
{\mathcal D}_{10} } \\
& = & \left( \frac{a}{\mu_{0}^{2}} \right) ^{-2 \epsilon} 
E^{(6)}_{0} + {\mathcal O} \left( 
\epsilon \right) , 
\eea
where:
\bea
a E^{(6)}_{0} & = & 6 \zeta(2) \ln{2} - \frac{3}{2} \zeta(3) \, .
\eea

\bea
\parbox{20mm}{\begin{fmfgraph*}(15,15)
\fmfleft{i1,i2}
\fmfright{o}
\fmf{plain}{i1,v1}
\fmf{plain}{i2,v2}
\fmf{photon}{v4,o}
\fmf{plain,tension=.3}{v2,v3}
\fmf{plain,tension=.3}{v1,v3}
\fmf{photon,tension=0}{v2,v1}
\fmf{plain,tension=.2,left}{v3,v4}
\fmf{plain,tension=.2,right}{v3,v4}
\end{fmfgraph*} }  & = & \mu_{0}^{2(4-D)} 
\int \{ d^{D}k_{1} \} \{ d^{D}k_{2} \}
\frac{1}{{\mathcal D}_{1} {\mathcal D}_{7} {\mathcal D}_{9} {\mathcal D}_{10} 
{\mathcal D}_{13} } \\
& = & \left( \frac{a}{\mu_{0}^{2}} \right) ^{-2 \epsilon} 
\sum_{i=-2}^{0} \epsilon^{i} E^{(7)}_{i} + {\mathcal O} \left( 
\epsilon \right) , 
\eea
where:
\bea
a E^{(7)}_{-2} & = & \frac{1}{2} \left[ \frac{1}{(1-x)} - 
\frac{1}{(1+x)} \right] H(0,x) \, , \\
a E^{(7)}_{-1} & = & - \frac{1}{2} \left[ \frac{1}{(1-x)} - 
\frac{1}{(1+x)} \right] \bigl[ \zeta(2) - 2 H(0,x) - H(0,0,x) \nn\\
& & + 2 H(-1,0,x)  \bigr] + \frac{2}{(1-x)} \Biggl[ 1 - \frac{1}{(1-x)}
\Biggr] H(0,0,x) \, , \\
a E^{(7)}_{0} & = & - \frac{1}{2} \left[ \frac{1}{(1-x)} - 
\frac{1}{(1+x)} \right] \bigl[ 2 \zeta(2) + 2 \zeta(3) - (4 - \zeta(2))
H(0,x) \nn\\
& & - 2 \zeta(2) H(-1,x) + 2 H(0,0,x) - 4 H(-1,0,x) - H(0,0,0,x) \nn\\
& & - 4 H(-1,-1,0,x) + 2 H(0,-1,0,x) + 2 H(-1,0,0,x) \nn\\
& & + \frac{2}{(1-x)} \Biggl[ 1 \! - \! \frac{1}{(1-x)} \Biggr] \bigl[
\zeta(2) H(0,x) \! - \! 2 H(0,0,x) \! + \! 3 H(0,0,0,x) \nn\\
& & - 4 H(-1,0,0,x) - 2 H(0,-1,0,x) \bigr] \, .
\eea

\bea
\parbox{20mm}{\begin{fmfgraph*}(15,15)
\fmfforce{0.5w,0.2h}{v3}
\fmfforce{0.5w,0.8h}{v2}
\fmfforce{0.2w,0.5h}{v1}
\fmfforce{0.8w,0.5h}{v4}
\fmfleft{i}
\fmfright{o}
\fmf{photon}{i,v1}
\fmf{photon}{v4,o}
\fmf{plain,left=.4}{v1,v2}
\fmf{plain,right=.4}{v1,v3}
\fmf{plain,left=.4}{v2,v4}
\fmf{plain,right=.4}{v3,v4}
\fmf{photon,left=.6}{v3,v4}
\end{fmfgraph*} }  & = & \mu_{0}^{2(4-D)} 
\int \{ d^{D}k_{1} \} \{ d^{D}k_{2} \}
\frac{1}{{\mathcal D}_{2} {\mathcal D}_{6} {\mathcal D}_{12} {\mathcal D}_{16} } \\
& = & \left( \frac{a}{\mu_{0}^{2}} \right) ^{-2 \epsilon} 
\sum_{i=-2}^{0} \epsilon^{i} E^{(8)}_{i} + {\mathcal O} \left( 
\epsilon \right) , 
\eea
where:
\bea
E^{(8)}_{-2} & = & \frac{1}{2} \, , \\
E^{(8)}_{-1} & = & \frac{5}{2} - \Bigl[ 1 - \frac{2}{(1-x)} \Bigr]
H(0,x) \, , \\
E^{(8)}_{0} & = & \Bigl[ 1 - \frac{2}{(1-x)} \Bigr]
\bigl[ \zeta(2) - 5 H(0,x) + 2 H(-1,0,x) \bigr] \nn\\
& & + \frac{2}{(1-x)^{2}} H(0,0,x) \, .
\eea

\bea
\parbox{20mm}{\begin{fmfgraph*}(15,15)
\fmfforce{0.5w,0.2h}{v3}
\fmfforce{0.5w,0.8h}{v2}
\fmfforce{0.2w,0.5h}{v1}
\fmfforce{0.8w,0.5h}{v4}
\fmfleft{i}
\fmfright{o}
\fmf{plain}{i,v1}
\fmf{plain}{v4,o}
\fmf{photon,left=.4}{v1,v2}
\fmf{plain,right=.4}{v1,v3}
\fmf{photon,left=.4}{v2,v4}
\fmf{plain,right=.4}{v3,v4}
\fmf{photon,left=.6}{v3,v4}
\end{fmfgraph*} }  & = & \mu_{0}^{2(4-D)} 
\int \{ d^{D}k_{1} \} \{ d^{D}k_{2} \}
\frac{1}{{\mathcal D}_{1} {\mathcal D}_{2} {\mathcal D}_{10} {\mathcal D}_{15} } \\
& = & \left( \frac{a}{\mu_{0}^{2}} \right) ^{-2 \epsilon} 
E^{(9)}_{0} + {\mathcal O} \left( 
\epsilon \right) , 
\eea
where:
\bea
E^{(9)}_{0} & = & \frac{19}{2} - 2 \zeta(2) \, .
\eea

\bea
\parbox{20mm}{\begin{fmfgraph*}(15,15)
\fmfforce{0.5w,0.2h}{v3}
\fmfforce{0.5w,0.8h}{v2}
\fmfforce{0.2w,0.5h}{v1}
\fmfforce{0.8w,0.5h}{v4}
\fmfleft{i}
\fmfright{o}
\fmf{plain}{i,v1}
\fmf{plain}{v4,o}
\fmf{plain,left=.4}{v1,v2}
\fmf{photon,right=.4}{v1,v3}
\fmf{plain,left=.4}{v2,v4}
\fmf{plain,right=.4}{v3,v4}
\fmf{plain,left=.6}{v3,v4}
\end{fmfgraph*} }  & = & \mu_{0}^{2(4-D)} 
\int \{ d^{D}k_{1} \} \{ d^{D}k_{2} \}
\frac{1}{{\mathcal D}_{2} {\mathcal D}_{6} {\mathcal D}_{8} {\mathcal D}_{11} } \\
& = & \left( \frac{a}{\mu_{0}^{2}} \right) ^{-2 \epsilon} 
\sum_{i=-2}^{0} \epsilon^{i} E^{(10)}_{i} + {\mathcal O} \left( 
\epsilon \right) , 
\eea
where:
\bea
E^{(10)}_{-2} & = & \frac{1}{2} \, , \\
E^{(10)}_{-1} & = & \frac{5}{2} \, ,  \\
E^{(10)}_{0} & = & \frac{19}{2} - 4 \zeta(2) \, , 
\eea

\bea
\parbox{20mm}{\begin{fmfgraph*}(15,15)
\fmfforce{0.5w,0.2h}{v3}
\fmfforce{0.5w,0.8h}{v2}
\fmfforce{0.2w,0.5h}{v1}
\fmfforce{0.8w,0.5h}{v4}
\fmfleft{i}
\fmfright{o}
\fmf{plain}{i,v1}
\fmf{plain}{v4,o}
\fmf{plain,left=.4}{v1,v2}
\fmf{photon,right=.4}{v1,v3}
\fmf{plain,left=.4}{v2,v4}
\fmf{photon,right=.4}{v3,v4}
\fmf{photon,left=.6}{v3,v4}
\end{fmfgraph*} }  & = & \mu_{0}^{2(4-D)} 
\int \{ d^{D}k_{1} \} \{ d^{D}k_{2} \}
\frac{1}{{\mathcal D}_{1} {\mathcal D}_{2} {\mathcal D}_{3} {\mathcal D}_{10} } \\
& = & \left( \frac{a}{\mu_{0}^{2}} \right) ^{-2 \epsilon} 
\sum_{i=-2}^{0} \epsilon^{i} E^{(11)}_{i} + {\mathcal O} \left( 
\epsilon \right) , 
\eea
where:
\bea
E^{(11)}_{-2} & = &  \frac{1}{2} \, , \\
E^{(11)}_{-1} & = & \frac{5}{2} \, ,  \\
E^{(11)}_{0} & = & \frac{19}{2} - 4 \zeta(2) \, , 
\eea

\bea
\parbox{20mm}{\begin{fmfgraph*}(15,15)
\fmfleft{i1,i2}
\fmfright{o}
\fmf{plain}{i1,v1}
\fmf{plain}{i2,v2}
\fmf{photon}{v3,o}
\fmf{plain,tension=.3}{v2,v3}
\fmf{plain,tension=.3}{v1,v3}
\fmf{photon,tension=0}{v2,v1}
\fmf{plain,right=45}{v3,v3}
\end{fmfgraph*} }  & = & \mu_{0}^{2(4-D)} 
\int \{ d^{D}k_{1} \} \{ d^{D}k_{2} \}
\frac{1}{{\mathcal D}_{1} {\mathcal D}_{7} {\mathcal D}_{9} {\mathcal D}_{10} } \\
& = & \left( \frac{a}{\mu_{0}^{2}} \right) ^{-2 \epsilon} 
\sum_{i=-2}^{0} \epsilon^{i} E^{(12)}_{i} + {\mathcal O} \left( 
\epsilon \right) , 
\eea
where:
\bea
E^{(12)}_{-2} & = & - \frac{1}{2} \left[ \frac{1}{(1-x)} - 
\frac{1}{(1+x)} \right] H(0,x) \, ,  \\
E^{(12)}_{-1} & = & \frac{1}{2} \left[ \frac{1}{(1-x)} - 
\frac{1}{(1+x)} \right] \bigl[ \zeta(2) - H(0,x) - H(0,0,x) \nn\\
& & + 2 H(-1,0,x)  \bigr] \, ,  \\
E^{(12)}_{0} & = & \frac{1}{2} \left[ \frac{1}{(1-x)} - 
\frac{1}{(1+x)} \right] \bigl[ \zeta(2) + 2 \zeta(3) - (1 - \zeta(2))
H(0,x) \nn\\
& & - \zeta(2) H(-1,x) - H(0,0,x) + 4 H(-1,0,x) - H(0,0,0,x) \nn\\
& & - 4 H(-1,-1,0,x) + 2 H(0,-1,0,x) + 2 H(-1,0,0,x) \bigr] \, .
\eea

\bea
\parbox{20mm}{\begin{fmfgraph*}(15,15)
\fmfbottom{v5}
\fmftop{v4}
\fmfleft{i}
\fmfright{o}
\fmf{plain}{i,v1}
\fmf{photon}{v3,o}
\fmf{plain}{v5,v2} 
\fmf{phantom}{v2,v4} 
\fmf{plain,tension=.2,left}{v1,v2}
\fmf{photon,tension=.2,right}{v1,v2}
\fmf{plain,tension=.2,left}{v2,v3}
\fmf{plain,tension=.2,right}{v2,v3}
\end{fmfgraph*} }  & = & \mu_{0}^{2(4-D)} 
\int \{ d^{D}k_{1} \} \{ d^{D}k_{2} \}
\frac{1}{{\mathcal D}_{1} {\mathcal D}_{7} {\mathcal D}_{10} {\mathcal D}_{13} } \\
& = & \left( \frac{a}{\mu_{0}^{2}} \right) ^{-2 \epsilon} 
\sum_{i=-2}^{0} \epsilon^{i} E^{(13)}_{i} + {\mathcal O} \left( 
\epsilon \right) , 
\eea
where:
\bea
E^{(13)}_{-2} & = & 1 \, , \\
E^{(13)}_{-1} & = & 4 - \Bigl[ 1 - \frac{2}{(1-x)} \Bigr]
H(0,x) \, , \\
E^{(13)}_{0} & = & 12 + \Bigl[ 1 - \frac{2}{(1-x)} \Bigr]
\bigl[ \zeta(2) - 4 H(0,x) - H(0,0,x) \nn\\
& & + 2 H(-1,0,x) \bigr]  \, .
\eea

\bea
\parbox{20mm}{\begin{fmfgraph*}(15,15)
\fmfbottom{v5}
\fmftop{v4}
\fmfleft{i}
\fmfright{o}
\fmf{plain}{i,v1}
\fmf{plain}{v3,o}
\fmf{plain}{v5,v2} 
\fmf{plain}{v2,v4} 
\fmf{plain,tension=.2,left}{v1,v2}
\fmf{photon,tension=.2,right}{v1,v2}
\fmf{photon,tension=.2,left}{v2,v3}
\fmf{plain,tension=.2,right}{v2,v3}
\end{fmfgraph*} }  & = & \mu_{0}^{2(4-D)} 
\int \{ d^{D}k_{1} \} \{ d^{D}k_{2} \}
\frac{1}{{\mathcal D}_{2} {\mathcal D}_{4} {\mathcal D}_{6} {\mathcal D}_{11} } \\
& = & \left( \frac{a}{\mu_{0}^{2}} \right) ^{-2 \epsilon} 
\sum_{i=-2}^{0} \epsilon^{i} E^{(14)}_{i} + {\mathcal O} \left( 
\epsilon \right) , 
\eea
where:
\bea
E^{(14)}_{-2} & = & 1 \, , \\
E^{(14)}_{-1} & = & 4 \, , \\
E^{(14)}_{0} & = & 12 \, .
\eea

\bea
\parbox{20mm}{\begin{fmfgraph*}(15,15)
\fmfleft{i}
\fmfright{o}
\fmf{plain}{i,v1}
\fmf{plain}{v2,o}
\fmf{plain,tension=.22,left}{v1,v2}
\fmf{photon,tension=.22,right}{v1,v2}
\fmf{plain,right=45}{v2,v2}
\end{fmfgraph*} }  & = & \mu_{0}^{2(4-D)} 
\int \{ d^{D}k_{1} \} \{ d^{D}k_{2} \}
\frac{1}{{\mathcal D}_{1} {\mathcal D}_{7} {\mathcal D}_{10} } \\
& = & \left( \frac{a}{\mu_{0}^{2}} \right) ^{-2 \epsilon} 
\sum_{i=-2}^{0} \epsilon^{i} E^{(15)}_{i} + {\mathcal O} \left( 
\epsilon \right) , 
\eea
where:
\bea
\frac{E^{(15)}_{-2}}{a} & = & -1 \, , \\
\frac{E^{(15)}_{-1}}{a} & = & -3 \, , \\
\frac{E^{(15)}_{0}}{a} & = & -7 \, .
\eea

\bea
\parbox{20mm}{\begin{fmfgraph*}(15,15)
\fmfleft{i}
\fmfright{o}
\fmfforce{0.5w,0.1h}{v1}
\fmfforce{0.25w,0.62h}{v3}
\fmfforce{0.5w,0.9h}{v7}
\fmfforce{0.74w,0.62h}{v11}
\fmf{plain,left=.1}{v1,v3}
\fmf{plain,left=.5}{v3,v7}
\fmf{plain,left=.5}{v7,v11}
\fmf{plain,left=.1}{v11,v1}
\fmf{photon}{v1,v7}
\end{fmfgraph*} }  & = & \mu_{0}^{2(4-D)} 
\int \{ d^{D}k_{1} \} \{ d^{D}k_{2} \}
\frac{1}{{\mathcal D}_{2} {\mathcal D}_{6} {\mathcal D}_{8} } \\
& = & \left( \frac{a}{\mu_{0}^{2}} \right) ^{-2 \epsilon} 
\sum_{i=-2}^{0} \epsilon^{i} E^{(16)}_{i} + {\mathcal O} \left( 
\epsilon \right) , 
\eea
where:
\bea
\frac{E^{(16)}_{-2}}{a} & = & -1 \, , \\
\frac{E^{(16)}_{-1}}{a} & = & -3 \, , \\
\frac{E^{(16)}_{0}}{a} & = & -7 \, .
\eea

\end{fmffile}

\end{document}